\documentclass[aps,epsf,preprint]{revtex4}
\usepackage{amssymb}
\usepackage{amsmath}
\usepackage{color}

\def\[{\left[}
\def\]{\right]}
\def\be{\begin{eqnarray}}
\def\ee{\end{eqnarray}}
\def\bm{\begin{pmatrix}}
\def\em{\end{pmatrix}}
\def\ba{\begin{array}}
\def\ea{\end{array}}
\def\bi{\begin{itemize}}
\def\ei{\end{itemize}}
\def\nn{\nonumber}
\def\({\left(}
\def\){\right)}
\def\bk#1{\langle#1\rangle}
\def\eq#1{Eq.(\ref{#1})}
\def\a{\alpha}

\def\e{\epsilon}
\def\D{\Delta}
\def\bD{\bar\Delta}

\def\l{\lambda}
\def\m{\mu}
\def\w{\omega}
\def\x{\times}

\def\n{\nu}
\def\p{\partial}
\def\d{\delta}
\def\labels#1{\label{#1}}
\def\bn{\begin{enumerate}}
\def\i{\item}
\def\en{\end{enumerate}}
\def\b{\beta}
\def\g{\gamma}
\def\ba{\begin{array}}
\def\ea{\end{array}}
\def\bc{\begin{center}}
\def\ec{\end{center}}

\def\.{\!\cdot\!}
\def\igw#1{\includegraphics[width=#1cm]}
\def\igwg#1#2{\igw{#1}{#2.png}} 
\def\+{\!+\!}
\def\-{\!-\!}

\def\h{{1\over 2}}

\def\={\stackrel{.}{=}}
\def\c{{\mathfrak c}}
\def\CC{{\mathbb C}}
\def\ol{\overline}
\def\bn{\bar n}
\def\1{\bar 1}
\def\8{\bar 8}
\def\9{\bar 9}
\def\7{\bar 7}

\usepackage{amssymb}
\usepackage{graphicx}

\begin{document}
\title{Color-Kinematics Relation from the Feynman Diagram Perspective}
\author{C.S. Lam}
\affiliation{Department of Physics, McGill University\\
 Montreal, Q.C., Canada H3A 2T8\\
Department of Physics and Astronomy, University of British Columbia,  Vancouver, BC, Canada V6T 1Z1 \\
Email: Lam@physics.mcgill.ca}

\begin{abstract}
Feynman diagrams for gluon tree amplitudes are studied in the Feynman gauge and in any number of spacetime dimensions. The color-kinematics combinations $\D=n_s-n_t-n_u$ of  numerators 
are explicitly calculated for $N=4,5,6$ gluons to see whether the color-kinematics relation $\D=0$ is satisfied. 
This is a tedious task because of the presence of four-gluon vertices, and the large number of Feynman diagrams, numerators, and $\D$
 combinations involved, especially  when $N=6$. 
For on-shell amplitudes, it is found that $\D=0$ for $N=4$, but $\D\not=0$ for $N=5$ and $N=6$
owing to the presence of the  four-gluon vertex. 
However, a {\it local}  generalized gauge transformation can bring about  $\D=0$ for $N=5$, but not for
$N=6$. This raises the question whether gluon amplitudes 
satisfying the color-kinematics relation contain non-local interactions.
\end{abstract}

\maketitle

\section{Introduction}
The color-kinematics (CK) duality  is a remarkable relation which
allows  graviton  amplitudes to be expressed as a double copy of gluon amplitudes \cite{KLT},
thereby opening up a new avenue  to study gravity. It was discovered for $N=4$ gluons long ago \cite{Zhu}, 
but it was Bern, Carrasco, and Johansson \cite{BCJ08} who generalized it
 to all $N$ and made the 
double copy connection. There are many interesting developments and further generalizations since
then that  can
be found in the comprehensive review in Ref.~\cite{BCCJR19}, especially when supersymmetry
is invoked, and there is also a recent reformulation \cite{CM21}.
This article is confined to the narrow objective of studying the CK relation for  pure gluon amplitudes from
the Feynman diagram perspective.

Much is known about the  CK relation for the gluon amplitude, but its ramification on Feynman diagrams 
with more than four external legs seems not to have been thoroughly studied. This is partly due to
the complication of including the four-gluon vertex systematically, and partly because of the large number
of Feynman diagrams and  partial amplitudes involved. In this article we try to
fill in this gap by carrying out these detailed investigations.

The CK relation relies on the fact that the numerator factors $n_a$ in an $N$-point gluon amplitude  $\sum_a\c_an_a/Q_a$ are not unique.   There are many ways to change from one set of $n_a$ to another set $\tilde n_a=n_a-\d n$ that  keep  all the partial amplitudes the same. Such a change is known as a {\it generalized gauge transformation},
or  {\it gauge transformation} for short. In this expression, 
$\c_a$ is the color factor and $Q_a$
 is the propagator factor in the Feynman gauge. 
Certain triplets of $\c_a$ are related by the Jacobi identity, say in the form $\c_s-\c_t-\c_u=0$. The CK
 relation asserts that suitable gauges can be found so that the same relation
 $n_s-n_t-n_u=0$ holds also for the numerator factors. Following \cite{BC11}, 
 those  $n_a$ satisfying the CK relation are said to be in the {\it BCJ representation}.

The existence of  BCJ representation is guaranteed by
 a general formula relating such $n_a$ to the partial amplitudes. 
 Unfortunately the $n_a$ so obtained are highly non-local. 
Since locality is a fundamental attribute of quantum field theory, it is important to find out whether locality can
be preserved in a BCJ representation. A necessary condition to be local
is for every $n_a$ to be a polynomial function
of the polarization vectors $\e_i$ and the external momenta $k_i$. The $n_a$ obtained 
from the general formula  are given by  rational functions, not polynomials, hence non-local. However, the numerators $n_a$
in the Cachazo-He-Yuan (CHY) gluon amplitude \cite{CHY1, CHY2, CHY3} are polynomial functions that also 
satisfy the CK relation,
so according to this criterion local BCJ representation does exist.
However, that may no longer be true when a more stringent criterion of locality is imposed. 
 
There are actually several versions of CHY gluon numerators \cite{LY3, DT17, PD, L19a, ET20}, all yielding
the same partial amplitudes.
They will be collectively denoted as $n'_a$.  A Mathematica program is given in Ref.~\cite{ET20} to calculate
$n'_a$ in one of these versions,  for the coupling
constant $g=-\h$. All explicit calculations of $n'_a$  in this article are carried out  using
this program.

A more stringent condition for locality is to start from 
the numerator factors $\bn_a$ computed from Feynman rules and Feynman diagrams.
Their direct connection with the Yang-Mills field theory makes $\bn_a$  strictly local.  
A more stringent criterion for $n_a$ to be local is to ask 
 $\d n_a=\bn_a-n_a$ to be local, {\it i.e.}, given by polynomials of $\e_i$ and $k_i$. 
We shall find that for $N=4$, $\bn_a$ are already in the BCJ representation, so it is local. For $N=5$, 
the CK relation is not
satisfied by $\bn_a$, but there are such local gauge transformations bringing them into
a BCJ representation. For $N=6$, $\bn_a$ are not in the BCJ representation, and there is no local gauge transformation that can render it in a BCJ representation. In particular, since $n_a'$ for $N=6$ 
satisfy the CK relation, the CHY gluon theory cannot be local according to this more stringent criterion. 
Possible implication of this non-locality will be discussed in Sec.~VII.

To compute $\bn_a$, four-gluon vertices must be converted into cubic vertices. Although there is a unique
way to do it,  their presence greatly complicates the computations. For example, 
 it is necessary to go to at least
$N=6$ before the four-gluon effects can be fully seen.


Sec.~II contains a review of some  known properties of the CK relation for gluons. It also
contains  definitions, conventions, and  other general discussions. Sec.~III explains how the four-gluon vertex should be systematically
incorporated, and the complication that brings along. It also reviews
 the Slavnov-Taylor identity which will  be used  to simplify
calculations. Because of the  rather lengthy calculations to be carried out in  subsequent sections, 
a detailed summary of  the results is also presented  to serve as a guidance of what is to come.

Four-point Feynman amplitudes are discussed in Sec.~IV, five-point in Sec.~V, and six-point in Sec.~VI.
A concluding section can be found in Sec.~VII. 
Lengthy formulas and results are presented in Appendices A, B, C, and D.

\section{Reviews, Notations, and General Discussions}
\subsection{Partial amplitudes}
An $N$-point gluon amplitude $\sum_{\a\in S_{N-2}} \CC_\a A(1\a N)$ consists of a sum of  product of a Del Duca-Dixon-Maltoni (DDM) color factor $\CC_\a$ \cite{DDM} and a color-stripped partial amplitude $A(1\a N)$. $\a$ is a permutation of $\{2,3,\cdots,N\-1\}$, the sum is taken over all the $(N\-2)!$ permutations $\a\in S_{N-2}$, and the external lines $1,\a_2,\a_3,\cdots,\a_{N-1} ,N$ of $A(1\a N)$
are arranged  cyclically in that order in a planar diagram.

Each partial amplitude $A(1\a N)$ is given by a sum of terms of
the form $\pm n_a/Q_a$ that can be computed from Feynman diagrams, or from an $S$-matrix theory such as the 
Cachazo-He-Yuan (CHY) formula. However,
 the numerator factors $\bar n_a$ computed from Feynman diagrams  are { \it not} the same
 as the numerator factors $ n'_a$ computed from the CHY theory, though both necessarily yield the
 same partial amplitudes $A(1\a N)$ on-shell. A change of one set of $n_a$ to another that leaves all $A(1\a N )$ unchanged is known as a  { \it generalized gauge transformation}, or simply  a gauge transformation. 
 
Partial amplitudes depend on the scalar products of  polarization vectors $\e_i$ and outgoing momenta
$k_i$ of the external lines. We shall call a (generalized) gauge transformation {\it local} if $\d n_a$, the difference between two $n_a$'s, is a polynomial of these scalar products for every $a$. Otherwise, it is {\it non-local}. 

The following notations are used in this article: $b_{ij}=\e_i\.\e_j$, $c_{ij}=\e_i\.k_j$, and $s_{ij\cdots \ell}=(k_i\+k_j\+\cdots \+k_\ell)^2:=s_I$, where $I=\{i,j,\cdots,\ell\}$ is the unordered set of subscripts of $s$. 
Note that $b$ and $s$ are symmetric in their subscripts but $c$ is not. Because of momentum conservation, $s_I=s_{I'}$ if
$I'$ is the set of external lines not contained in $I$. To avoid this ambiguity, most of the time we shall use the set of indices not 
containing $N$. Unfortunately, there is no uniuniqueque way to implement  the conservation condition
$\sum_{j=1}^Nc_{ij}=0$, so terms involving $c$ may appear a bit unwieldy.  

Both { \it on-shell} and { \it off-shell} amplitudes will be discussed. An on-shell amplitude obeys $s_i=k_i^2=0$ and $c_{ii}=\e_i\.k_i=0$ for
all $i$. Otherwise it is off-shell. For on-shell amplitudes,  there are
only $D\-2$ independent polarizations $\e_i$  in a spacetime dimension $D$. For off-shell lines, there are $D$ polarization
vectors $\e_i^{(\l)} , \l=1,\cdots,D$, to be normalized so that 
\be (u\.\e_i)(\e_i\.v):=\sum_{\l=1}^D\(u\.\e_i^{(\l)}\)\(\e_i^{(\l)}\.v\)=u\.v\labels{complete1}\ee
for any $i$, and for any vectors $u$ and $v$. The first expression in \eq{complete1} is a short-hand  in
which all polarizations of a repeated off-shell line $i$ are automatically summed. With this convention,
if $i$ is an off-shell line, then
\be b_{pi} b_{iq}=b_{pq} ,\quad b_{pi}c_{iq}=c_{pq} ,\quad c_{ip} c_{iq}=k_p\.k_q=\h\(s_{pq}-s_p-s_q\).\labels{complete2}\ee

The propagator factor $Q_a=s_{I_1}s_{I_2}\cdots s_{I_{N-3}}$ is made up of a product of $N\-3$ Mandelstam variables such that whenever $p<q$, either $I_p\subset I_q$
 or $I_p\cap I_q=\O$. The total number of distinct propagators subject to this constraint
is $(2N\-5)!!$, but only $C_{N-2}$ of them appear in any partial amplitude $A(1\a N)$. For $N=4, 5, 6$ which will be  studied explicitly, the Catalan number $C_{N-2}$ is respectively  $C_{N -2}=2, 5, 14$, while the total number of distinct
propagator factor is $(2N\-5)!!=3, 15, 105$.
Since $C_{N-2}(N-2)!>(2N-5)!!$, the same $n_a/Q_a$ may appear in different partial amplitudes $A(1\a N)$, possibly with different signs.

\subsection{CK relations and CK combinations}
Making use of the explicit form $\sum \pm n_a/Q_a$ for $A(1\a N)$, the gluon amplitude can also be written as $\sum_a\c_an_a/Q_a$,
where $\c_a$ is a color factor obtained from an appropriate combination of the DDM color factors $\CC_\a$.  Jacobi
identity demands certain triplets of these $\c_a$'s to be related, say $\c_s-\c_t-\c_u=0$. The color-kinematics relation (CK relation) asserts
that  (the non-unique) numerators $n_a$ can be 
found  for the { \it on-shell amplitude} so that  $n_s-n_t-n_u=0$ as well. 

The Lie algebra structure constants $f_{ijk}$ from which  $\c_a$ is constructed is antisymmetric in its indices, hence the CK
relation also demands  $n_a$ to have the same antisymmetry as $\c_a$. 

From now on,  partial amplitudes will be studied without the accompanying color factors.
Even without  $\c_a$, the triplet of numerators 
$n_s, n_t, n_u$ can still be identified  from the diagrams, as illustrated in Fig.~1 and Fig.~2.
\bc\igwg{12}{Fig1}\\ Fig.~1.\quad A CK combination. A, B, C, D represent (possibly empty) tree diagrams\ec
\bc\igwg{8}{Fig2}\\ Fig.~2.\quad $n_a$ is antisymmetric when two lines are flipped\ec
These diagrams  resemble, but strictly speaking are not Feynman diagrams. They are diagrams for $n_a$, with
shapes determined by the propagator $Q_a$. More specifically, the distinct components $s_{\a\b}, s_{\b\g}, s_{\a\g}$ in
 the propagator factors
\be
Q_s=s_As_Bs_Cs_Ds_{\a\b} ,\quad Q_t=s_As_Bs_Cs_Ds_{\b\g} ,\quad Q_u=s_As_Bs_Cs_Ds_{\a\g} ,\labels{Qstu}\ee
give  rise to the shapes in Fig.~1. 
The  common factor $P_l:=s_As_Bs_Cs_D$
will be used to label the { \it CK combination} $\D_l=n _s-n_t-n_u$, where $s_A, s_B, s_C, s_D$ represent the products of all inverse propagators in A, B, C, D,  respectively. 
The total number of these common factors is $(2N\-5)!!(N\-3)/3$, so there are 1, 10, 105 $\D_l$'s when $N=4, 5, 6$.

These diagrams contain only cubic vertices which
are antisymmetric when two lines are flipped, as shown in Fig.~2. In addition, each $n_a$ must also contain $N$ factors of $\e$ and $N\-2$ factors of $k$.  Beyond that, $n_a$ could be rather general.

\subsection{Contents of partial amplitudes}
A term $n_a/Q_a$ is contained in  $A(1\a N)$ when $1\a_2\a_3\cdots\a_{N-1}N$ is the order of its external lines. 
Owing to  antisymmetry  at  cubic vertices, some external lines may be flipped, at the cost of a minus sign per  flip. Flips produce different orderings of the external lines, hence
it is possible for a given $n_a/Q_a$ to be contained in several $A(1\a N)$, with a relative sign determined by the number of flips. Its absolute sign will be chosen to be positive in the `smallest' $A(1\a N)$.

The $(N\-2)!$ partial amplitudes $A(1\a N)=A_m$ will be ordered in a way to be discussed later. The `smallest' $A$ refers to the $A_m$ with the smallest $m$.

Note that lines 1 and $N$ are special because they always occupy the two ends of the arguments of every $A_m$. To reflect this, all diagrams will be drawn above a base line with 1 at one end and $N$ at the other end.
As a result, lines 1 and $N$ should never be flipped, and  none of the lines in $\a$ is allowed to flip across the base line connecting 1 and $N$.

Explicit application of these rules will appear in the sections where the amplitudes for $N=4, 5, 6$ are discussed.

\subsection{Dimensional consideration and the variety of terms}
The numerators $n_a$ contain $N$ factors of $\e$ and $N\-2$ factors of $k$. In terms of $b, c, s$, the allowed monomial combinations are
\be
N=4.\qquad &&bcc,\  bbs\nn\\
N=5.\qquad &&bccc,\ bbcs\nn\\
N=6.\qquad &&bcccc,\ bbccs,\  bbbss.\labels{bcs}\ee
More generally, if $\m$ be the largest integer $\le \h N-1$, then the allowed monomial terms are
$b^ic^{N-2i}s^{i-1}$ for $1\le i\le \m$. Terms of the same type can conceivably be combined through momentum
conservation, but terms of different types can never be combined.

On dimensional grounds, it is possible to multiply these forms by a dimensionless rational function,
such as $s_{12}s_{34}/s_{13}s_{23}$, but then the numerator factor $n_a$ will no longer be a polynomial
of $b_{ij}, c_{ij}$, and $s_I$, and whatever interaction that gives rise to this $n_a$ may no longer be local.

For $N=4$, there are six different $b_{ij}$'s, and 3 different $b_{ij}b_{kl}$'s. 
Taking momentum conservation and the on-shell condition into account, there are two $c_{ij}$'s for each $i$,
and two $s_I$. 
So even the four point numerators $n_a$ already contain many different terms each.
For $N=5$, there are 10 different $b$'s, 15 different $bb$'s, three different $c_{ij}$'s for each $i$, and five different
$s_I$.
For $N=6$, there are 15 different $b$'s, 30 different
$bb$'s, 15 different $bbb$'s, four different $c_{ij}$'s for each $i$, and nine different $s_I$.
In short, each $n_a$ contains a vast number of terms, especially for $N=6$.  This makes the computation
of numerators and CK combinations  a very tedious task. Fortunately, considerable simplification
can be obtained by using symmetries of the Feynman diagrams and the Slavnov-Taylor identity.

\subsection{CK relations in $S$-matrix theories}
A short review  on the CK relation  of  gluon tree amplitudes is given in this subsection. 
See Ref.~\cite{BCCJR19} for a more thorough review.
To start with, define
an $(N\-2)!\x(N\-2)!$ { \it propagator matrix} $m$ whose matrix element $m(\a|\b)\equiv m(1\a N|1\b N)$ is
determined by terms common
to $A(1\a N)$ and $A(1\b N)$. 
$m(\a|\b)$ is  equal to those common terms, with all $n_a$ set equal to 1, and with a sign which is the product of
signs of the corresponding terms in the two amplitudes. In other words, up to a sign, it is the sum of the common
propagators of the two amplitudes.
This matrix can also be computed directly using the CHY formula for the biadjoint scalar theory.

With the help of the propagator matrix, the on-shell partial amplitudes 
\be
A(1\a N)=\sum_{\b\in S_{N-2}}m(\a|\b)\n(\b)\labels{Amnu}\ee
can be related to  $(N\-2)!$ parameters $\n(\b)$, to be referred to as the{ \it fundamental numerator factors}. 
Since there are the same number of $A$'s and $\n$'s, one might be inclined to regard
\eq{Amnu}  as the definition of $\n(\b)$
in terms of  the $(N\-2)!$ known $A(1\a N)$'s. Unfortunately such a  definition
is not unique because the partial amplitudes satisfy a set of Bern-Carrasco-Johansson (BCJ) relations
\cite{BCJ08} to be discussed later, making them not linearly independent, and the propagator matrix $m$ not invertible.

Comparing this expression for $A(1\a N)$ with its other
expression $\sum\pm n_a/Q_a$,  a set of relations between $n_a$ and  $\n(\b)$ can be derived
by equating propagators.  To distinguish $n_a$ from
the fundamental numerators $\n(\b)$, $n_a$ are sometimes  referred to as { \it ordinary numerators}. 
A set of ordinary numerators may or my not satisfy the CK relation, but if they are
determined from \eq{Amnu}, then the CK relation will automatically be fulfilled. In other words, the $n_a$
so determined are in the BCJ representation.

Since there are $(2N-5)!!$ ordinary numerators $n_a$ and $(N-2)!$ parameters $\n(\b)$, there must be 
$N_\D=(2N-5)!!-(N-2)!$ on-shell relations between  $n_a$. These are the CK relations. For $N=4, 5, 6$, this number $N_\D$
is 1, 9, 81 respectively. They differ from the numbers 1, 10, 105 given in Sec.~IIB for the number of $\D_l$'s.
The difference of the two sets of numbers, 0, 1, 24, is the number of {\it trivial CK identities}. These are 
identities involving  CK combinations $\D_l$,  valid whatever $n_a$ are. Details will be given in later sections.

The relation between $n_a$ and $\n(\b)$ can be visualized diagrammatically. To do that, start from a half-ladder 
diagram shown in Fig.~3. Since no line in this diagram may be flipped, it belongs to 
 a single $A(1\a N)$ and  appears in a single propagator matrix element $m(\a|\a)$. As a result, its numerator
is equal to a single fundamental numerator: $n_a=\n(\a)$. 
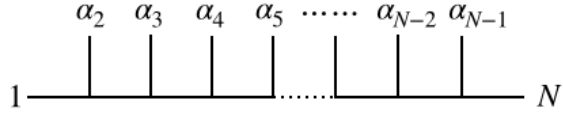
\bc\igwg{8}{Fig3}\\ Fig.~3.\quad A  half-ladder diagram whose propagator
factor is $Q_a=s_{1\a_2}s_{1\a_2\a_3}\cdots s_{1\a_2\a_3\cdots \a_{N-1}}$\ec

Next, consider two  half-ladder diagrams whose external lines $\a$ and $\g$ coincide except for a single neighboring pair. Identifying
the numerators of these two diagrams as $n_s$ and $n_u$, and  using the CK relation shown in Fig~1, one 
obtains a two-term relation
$n_t=n_s-n_u=\n(\a)-\n(\g)$ for $n_t$. In this way, 
starting from  half-ladder diagrams, one can get
progressively to more and more complicated combinations of half-ladder diagrams to express the relations between any $n_a$ and a combination
of  $\n(\b)$'s.

As mentioned above, the on-shell amplitudes satisfy a set of Bern-Carrasco-Johansson (BCJ) relations
\cite{BCJ08, BDV09,S09,BCSV10a}.
There are $(N\-2)!-(N\-3)!=(N\-3)! (N\-3)$ independent relations, given by
\be
0&=&s_{1\b}A(1\b\g_1\cdots\g_{N-3}N)+\cdots +(s_{1\b}\+s_{\g_1\b}\+\cdots\+s_{\g_i\b})A(1\g_1\cdots\g_i\b\cdots\g_{N-3}N)+\cdots\nn\\
&+& (s_{1\b}\+s_{\g_1\b}\+\cdots\+ s_{\g_{N-3}\b})A(1\g_1\cdots\g_{N-3}\b N).\labels{BCJ}\ee
Each relation gives rise to a null vector $u$ for the on-shell propagator matrix \cite{VY07,VY14}. For $N=4, 5, 6$, this number 
of null vectors is 1, 4, 18 respectively.
It follows from \eq{Amnu} that $A(1\a N)$ remains unchanged if we add to the column vector $\n=(\n(\b))$ any linear 
combination of these null vectors $u$. This flexibility of $\n(\b)$ reflects the non-uniqueness of $n_a$ 
in BCJ representations.
Note that the coefficients $x_i$ of these linear combinations do not have to be constants. They can be any function of $b_{ij}, c_{ij}$, and $s_I$, including rational and irrational functions.  A change of $\n(\b)$ causes a gauge transformation of $n_a$. If $x_i$ are constants or polynomials,  the gauge transformation is local. If they are rational or other functions,  the transformation is non-local. 

The  validity of the BCJ relation can be seen diagrammatically in the  half-ladder diagrams of each $A(1\a N)$. In that case, the on-shell kinematical
relation
$s_{1\b}\+s_{\g_1\b}\+\cdots\+s_{\g_i\b}=s_{1\g_1\cdots \g_i\b}-s_{1\g_1\cdots\g_i}$
shows that the factor multiplying each $A(1\a N)$
is just the difference of two consecutive inverse propagators, thereby causing the terms in \eq{BCJ} to cancel pairwise. 

In the presence of null vectors, the propagator matrix $m$ cannot be inverted, so it remains to be shown that there is at least one set of $\n(\b)$ that satisfies \eq{Amnu}.
 Since $m$ has $(N\-3)!(N\-3)$ null vectors and a rank $(N\-3)!$,
only a $(N\-3)!\x(N\-3)!$ sub-matrix of $m$ can be inverted.
If $a=(\a_2\a_3\cdots\a_{N-2})\in S_{N-3}$ is a permutation of the $(N\-3)$ objects
$\{2,3,\cdots,N\-1\}\backslash\a_{N-1}$, then the $(N\-3)!\x(N\-3)!$ matrix $\bar m$, with matrix elements
$\bar m(a|b)=m(1a(N\-1)N|1bN(N\-1))$, has an inverse $\bar m^{-1}:=-S$. If we choose
\be
\n(\b)&=&\sum_{c\in S_{N-3}}S(b|c)A(1cN(N-1)),\qquad {\rm if}\ \b=(b,N\-1),\nn\\
&=&0,\qquad {\rm if}\ \b_{N-1}\not=N\-1,\labels{pnu}\ee
then \eq{Amnu} is satisfied \cite{BDSV10b,CHY3,BCCJR19}, hence it is a solution of \eq{Amnu}.

The matrix element $m(1a(N\-1)N|1bN(N\-1))$ used above is defined similar to the matrix
element  $m(1\a N|1\b N)$. It is given by the common terms in $A(1a(N\-1)N)$
and $A(1bN(N\-1))$, after setting all $n_a=1$ and taking into account the relative signs.
In order for a numerator diagram to be common to these two amplitudes, lines $N$
and $(N\-1)$ must merge into a cubic vertex, thereby producing a $Q_a$
factor $s_{(N\-1)N}$, as shown in Fig.~4. If $o$ is the internal line at that vertex, then
$m(1a(N\-1)N|1bN(N\-1))=-m(1ao|1bo)/s_{(N-1)N}$. Since $o$ is off-shell, the
$(N-3)!\x(N\-3)!$ matrix $m(a|b)\equiv m(1ao|1bo)$ is non-singular and has an inverse.
There is an explicit formula for $S(b|c)$ which will not be displayed here. For our purpose
it is sufficient to know that it is a polynomial of $s_I$ of degree $N\-3$.
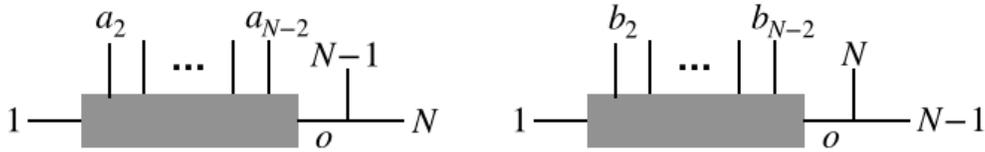
\bc\igwg{14}{Fig4}\\ Fig.~4.\quad The pair of tree diagrams used to compute $m(1a(N\-1)N|1bN(N\-1))$\ec

Equivalently, this solution of $\n(\b)$ can be obtained by using the CK relations to eliminate all but the half-ladder
$n_a$'s, which are equal to some $\n(\a)$. Owing to the BCJ relations, there are
only $(N\-3)!$ independent partial amplitudes, so only $(N\-3)!$ of the  $\n(\a)$'s
can be determined by \eq{Amnu}. The remaining ones can be anything, including zero. The second line
of \eq{pnu} is simply a particular way of choosing what is to be set equal to zero.

 The numerators obtained  in \eq{pnu} is highly non-local. They are rational functions
 of $\e_i$ and $k_i$ rather than polynomials. If we start from a set of local numerators $\bn_a$, say
 obtained from Feynman diagrams, then the gauge transformation $\d n_a$ taking $\bn_a$ into those
 obtained from \eq{pnu} is also non-local. 
 
 \subsection{Local and non-local gauge transformations}
Let $\d n_a$ be a (generalized) gauge transformation, {\it i.e.}, the difference of two sets of $n_a$ that give rise to the same partial amplitudes. If every $\d n_a$ is a polynomial function
 of $b_{ij}, c_{ij},$ and $s_I$, then the  gauge transformation is  {\it local}, otherwise {\it non-local}.
 
 Suppose $Q_a=s_{I_1}s_{I_2}\cdots s_{I_{N-3}}$, with $I_1\prec I_2\prec\cdots\prec I_{N-3}$  (see the next subsection for how $I_a$ is ordered), is the 
 corresponding propagator factor, and $\m$ is the largest integer not exceeding $\h N\-1$. Then  $\d n_a$
 is a {\it local} gauge transformation only if every $\d n_a$ is a multilinear polynomial of $\{s_{I_1}, s_{I_2},\cdots ,
 s_{I_{N-3}}\}$ of degree $\m$, with coefficients that are polynomials of $b_{ij}$ and $c_{ij}$. More precisely, 
 \be
 \d n_a&=&\sum_{i=1}^{N-3} p_a^{(i-1)}s_{I_i}+\sum_{i=1}^{N-3}\sum_{j=i+1}^{N-3} q_a^{(i-1,j-1)}s_{I_i}s_{I_j}
 +\cdots \nn\\
 &+&\sum_{i_1=1}^{N-3}\sum_{i_2=i_1+1}^{N-3}\cdots\sum_{i_\m=i_{\m-1}+1}^{N-3}
 r_a^{(i_1-1,i_2-1,\cdots,i_\m-1)}s_{I_{i_1}}s_{I_{i_2}}\cdots s_{I_{i_\m}},\labels{lgt}\ee
where $p,q,\cdots,r$ are polynomial functions of $b_{ij}$ and $c_{ij}$.
 
With this form, the product of $m$ $s_I$ in  $\d n_a$  cancels $m$ factors of $s$ in $Q_a$, 
leaving behind a term proportional to the $(m\-N\+3)$ power of $s$. To be a gauge transformation, 
one must find other $\d n_b/Q_b$ terms  in the same $A(1\a N)$ with the same $s$ dependence 
to cancel it. This requirement generates a number of  $s$-independent equations for $p, q,\cdots, r$  that must be satisfied. We shall refer to these equations as the {\it gauge constraint equations}.

 An $s$-power larger than $\m$ is not allowed in \eq{lgt} on dimensional grounds. See Sec.~IID. $s_J$ not contained in $Q_a$ is also
 not allowed because then the equations for $p,q,\cdots, r$ etc. will contain a rational function of $s$, making
 the gauge transformation no longer local.

\subsection{Canonical ordering}
For $N>4$, there are quite a few partial amplitudes,  many propagators $Q_a$, and a large
number of CK combinations $\D_l$, so
a systematic method  is needed
to enumerate them. The simplest is to express them as lists, and have the lists ordered canonically.

Each partial amplitude $A(1\a N)$ corresponds to an ordered list $\a$, but $Q_a$ and $\D_l$
are labelled by products of $s_I$'s, with  $I$'s being unordered sets of numbers. The first task is
therefore to convert an unordered set of numbers 
into an ordered list, and then use canonical ordering to order the lists.

An unordered set of three numbers 1, 2, 3 can be written as an  ordered list   $\{1,2,3\}$.
More generally,
an unordered set of $m$ numbers can be turned into an ordered list $\{i_1,i_2,\cdots,i_m\}$
 with $i_1<i_2<\cdots i_m$. 

Two different lists are ordered first
according to their length, and if they have the same length, according the first unequal numbers appearing in the two lists. 
For example,
$\{5,6\}\prec\{1,2,3\}$, and $\{4,7,8,9\}\prec\{4,7,9,8\}$.

Two lists each with several sublists are first ordered by their number of sublists. If they have the same number of sublists, then
they are ordered according to the first unequal sublists. For example, $\{\{5,6\} , \{7,8,9\}\}\prec\{\{1,2\} ,\{3\} ,\{4\}\}$,
and $\{\{3,4,5\} ,\{4,7,8,9\}\}\prec\{\{3,4,5\} ,\{4,7,9,8\}\}$.

Mandelstam variables $s_I$ are ordered according to the list $I$. Thus $s_5\prec s_{23}\prec s_{24}\prec s_{123}$.
Products of $k$ Mandelstam variables $s_{I_1}s_{I_2}\cdots s_{I_k}$ are ordered according to $\{I_1,I_2,\cdots,I_k\}$.
In this way, we can order the inverse propagators (a product of $N\-3$ Mandelstam variables) $Q_a$, and  the  CK combinations (labelled by a product of $N\-4$ Mandelstam variables) $\D_l$. Similarly, the partial amplitudes can be enumerated according to
the list of its arguments. For example,
for $N=5$,  $A_1=A(12345), \ A_2=A(12435),\ A_3=A(13245),\ A_4=A(13425),\ A_5=A(14235)$,\ and $A_6=A(14325)$.

\section{Feynman Diagrams}
\subsection{Three-gluon, four-gluon, and virtual vertices in partial amplitudes}
The main purpose of this article is to investigate how  CK relations  emerge from the Feynman diagrams
of  a Yang-Mills field theory,  and whether the  (generalized) gauge transformation required to accomplish that  is local or not.
Feynman gauge is used in the Feynman diagrams, so propagators are the same as those in scalar theories. Feynman diagrams presented here should be
understood as  diagrams for the numerator  $\bar n_a$
obtained from  three-gluon (3g) and four-gluon (4g) vertex factors alone, without the accompanying propagators $1/Q_a$.

A three-gluon (3g) vertex  $T$ in a color-stripped partial amplitude is given by
\be T(p,q,r)=b_{pq}(c_{rp}-c_{rq})+b_{qr}(c_{pq}-c_{pr})+b_{rp}(c_{qr}-c_{qp}).\labels{T}\ee
These three terms can each be associated with a diagram in the first row of Fig.~5, where $\e$ is depicted by a
black dot, and the difference of two momenta is shown as an arrow.

The 3g vertex is cyclic in its arguments, and antisymmetric when two arguments are exchanged: 
$T(p,q,r)=T(q,r,p)=-T(q,p,r)$.

\bc\igwg{14}{Fig5}\\ Fig.~5.\quad 3g, 4g, and virtual vertices\ec

Since 
four-gluon (4g) vertices are absent  in Fig.~1,  they must 
be converted into  antisymmetric cubic vertices. 
To do so, a 4g vertex $V_4$ should be separated into two terms,
\be V_4(p,q,r,s)&=&2b_{pr}b_{qs}-b_{pq}b_{rs}-b_{ps}b_{qr}={Q(p,q,r,s)s_{pq}\over s_{pq}}+{Q(q,r,s,p)s_{qr}\over s_{qr}} ,\nn\\
Q(p,q,r,s)&=&b_{pr}b_{qs}-b_{ps}b_{qr}=Q(r,s,p,q),\labels{Q}\ee
as shown in Fig.~5. 
Each term is depicted by a pair of `virtual vertices" connected by a dotted line, and each virtual vertex 
has the 
antisymmetry of the 3g vertex when two solid lines are flipped.  Note however that a single virtual vertex does not
really exist, they must come in pairs to form a $Q$.

Each cubic vertex in Fig.~1 can either be a 3g vertex or a virtual vertex. Feynman diagrams
containing virtual vertices can be obtained from purely
3g diagrams by replacing solid propagator lines with dotted lines in all possible ways, 
 subject to the constraint that no two dotted lines may intersect at a vertex. 
 
To save space, commas  between  arguments of functions like $T$ and $Q$ will often be dropped.

\subsection{CK combinations for Feynman diagrams}
With virtual vertices, 
one or two of the four lines $\a, \b, \g, \d$ in Fig.~1 may now be dotted. In terms of the original 4g
vertices, exposing a dotted line is equivalent to cutting through the middle of a 4g vertex.

Fig.~6 shows an example where
$\a$ is dotted, and an example where both $\a$ and $\g$ are dotted. Since two dotted
lines are not allowed to intersect,  the second row of Fig.~6 has only two instead of three diagrams. Also,
for the same reason, diagrams with three or four dotted lines do not exist. 
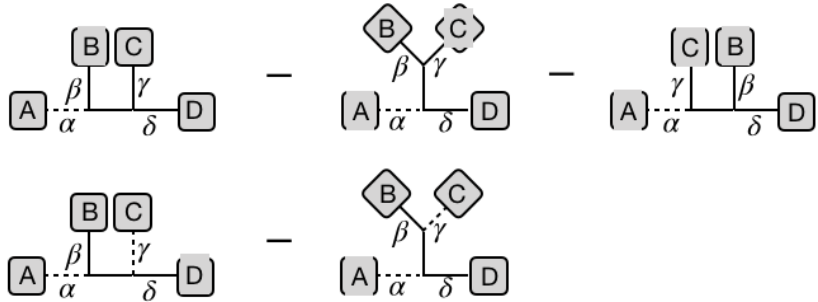
\bc\igwg{12}{Fig6}\\Fig.~6.\quad Virtual vertex in CK combinations\ec
The diagrams in Fig.~6 are merely symbolic, because virtual
vertices cannot exist by themselves. To make sense of Fig.~6, we must pair up all the virtual vertices, as shown
in Fig.~7. This compels us to study Feynman diagrams explicitly also for $N=5$ and $N=6$.

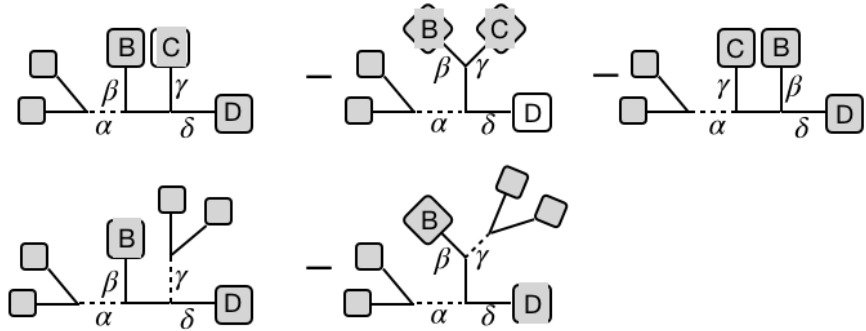
\bc\igwg{12}{Fig7}\\Fig.~7.\quad Completion of the virtual vertices in Fig.~6\ec

These diagrams with dotted lines turn out to be the culprits that cause $\bD_l\not=0$ for $N=5$ and $N=6$.

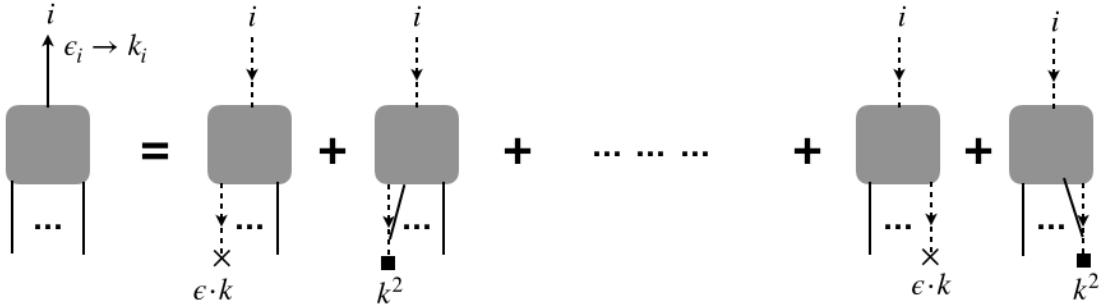
\subsection{Slavnov-Taylor identity}
The Slavnov-Taylor identity will be used to simplify  calculation of the CK relation  for  large $N$. Like the Ward-Takahashi identity
in QED, it is a consequence of local gauge invariance of  the Yang-Mills theory, relating the divergence of a gauge field
to the  gauge and ghost fields. In Feynman gauge, the corresponding Green's function identity is
\be \bk{0|{\p^{\m_1}A_{\m_1}^{a_1}(x_1)A_{\m_2}^{a_2}(x_2)\cdots A_{\m_n}^{a_n}(x_n)}|0}+\sum_{l=2}^n
\bk{0|\bar\w_{a_1}(x_1)A_{\m_2}^{a_2}(x_2)\cdots D_{\m_l}\w^{a_l}(x_l)  \cdots A_{\m_n}^{a_n}(x_n)|0} =0,\nn\\
\labels{STI}\ee
where $A$ is the gluon field, and $\w$, $\bar \w$ are the ghost and anti-ghost fields, whose covariant derivative is given by
 $D_\n\w^a(x)=\p_\n\w^a-gf^{adc}\w^dA_\n^c$.

The diagrammatic expression for \eq{STI} in terms of amplitudes is sketched in Fig.~8. Solid and dashed lines represent gluons and (anti-) ghosts, with
an arrow running from the anti-ghost end to the ghost end. The left hand side,  with $\e_i$ replaced by $k_i$,
is the divergence of the gluon amplitude. The right hand side is given by
covariant derivative terms in momentum space, with a cross ($\x$) representing $\e\.k$ and a
filled box representing $k^2$. For on-shell amplitudes, $\e_i\.k_i$ and $k_i^2$ both vanish,
hence the gluon amplitude is divergenceless.  For off-shell amplitudes, this identity shows how the divergence
of a gluon amplitude is related to amplitudes involving gluons and a single ghost line.
\bc\igwg{16}{Fig8}\\ Fig.~8.\quad Slavnov-Taylor identity. \ec

Let ${\cal M}$ be the gluon amplitude. Singling out the polarization vector $\e_i^{(\l)}$ of its $i$th line,
it can be written as ${\cal M}=\e_{i\m}^{(\l)}{\cal \ol M}^\m=\e_i^{(\l)}\.\ol{\cal M}$. If $i$ is an internal line, the replacement of
$\e_i$ by $k_i$ on the left hand side of Fig.~8 can be accomplished by  $\sum_{\l=1}^D(\e_i^{(\l)}\.k_i)
(\e_i^{(\l)}\.\ol {\cal M})$. In the shorthand notation used in \eq{complete1}, this is just $c_{ii}{\cal M}$.  It is in 
this form that the Slavnov-Taylor identity will be used.

 {\it Numerator Slavnov-Taylor identities}  obtained by equating the residues at various poles will also be used.

\subsection{Summary of results}
In the next few sections, CK combinations and CK relations of  $N$-point Feynman diagrams will be studied
for $N=4,5,6$. Since the formulas tend to be rather lengthy, especially for $N=6$, it might be useful to provide
 a summary here to serve as a guide of what is to come, and what results are to be obtained.

For $N=4$, it has been known for a long time that the CK relation $\bD=\bn_s-\bn_t-\bn_u=0$ computed from Feynman diagrams holds on-shell \cite{Zhu}.
Its off-shell expression which is useful for large $N$ computation will also be computed.

Both the numerator factor $\bn_a$ computed from Feynman diagrams and $n'_a$
computed from the CHY theory satisfy the CK relation, but these two sets of numerators are different.
They will be shown to be related by a {\it local} gauge transformation.

For $N>4$, the CK
combinations $\bD_l$ computed from Feynman numerators $\bn_a$ no longer vanish, even on-shell.
Nevertheless, amplitudes $A(1\a N)$ so computed 
can be used in \eq{pnu} to obtain a set of $\n(\b)$, and from there
a set of $n_a$ satisfying the CK relation. However, such $n_a$'s are highly non-local,  so the  
  gauge transformation $\d n_a=n_a-\bn_a$ is non-local as well.  We would like to know whether
  a local gauge transformation can be found to implement the CK relation. That is, whether a gauge
  transformation of the form \eq{lgt} can cause $\D_l:=\bD_l-\d \D_l=\bD_l-\d(n_s-n_t-n_u)\simeq 0$.
  The notation $\simeq$ is used to indicate equality only for on-shell amplitudes. This requires
  the parameters $p,q,\cdots,r$ in \eq{lgt} to  satisfy a set of equations to
  be referred to as  the {\it CK equations}.
 Since these parameters also need to satisfy the gauge-constraint
  relations, the locality question is equivalent to the question of whether the set of gauge-constraint and CK
  equations have a solution.

For $N=5$, which will  be discussed in Sec.~V,  solutions do exist {\it provided} certain relations between Feynman diagrams with one virtual vertex are satisfied.
Direct calculation shows that these relations are indeed satisfied, so there are local gauge transformations for $N=5$
to implement the CK relation.

The case of $N=6$ to be studied in Sec.~VI is much more complicated. 
The gauge-constraint equations  of the 315 $p$ parameters
can be explicitly solved,
leaving behind 105 free parameters that must still satisfy  210 CK equations. Although there are many more
equations than variables, nevertheless the equations are highly degenerate, so solutions could still exist if
the $\bD_l$ obey a large set of conditions. It turns out that only some of these conditions are obeyed, but not
others, so the $p$ equations have no solution.

There are also 315 $q$ parameters, which must satisfy 144 gauge-constraint relations, and 450 CK equations.
Again it turns out that there are no solutions.

The absence of $p$ solution or $q$ solution tells us that there is no local gauge transformation that can 
implement the CK relation. As an independent check, 
a CHY numerator factor $n'_a$ and the difference  $\d n_a=\bn_a-\hat n_a$ are computed.  
Since $n'_a$ obey the CK
relation, $\d n_a$ must be non-local to be consistent with the $p,q$ conclusions. Indeed it is, because it contains 
$s$ dependences beyond those involved in $Q_a$.

Thus we conclude that, starting from Feynman diagrams, there is no local gauge transformation capable of implementing the CK relation, but we have to go up to $N=6$ to show it. Possible implications of this
result will be discussed in Sec.~VII.

A number of functions are introduced in the calculations. A four-argument function $d$, three five-argument functions
$f, F, D$, and two six-argument functions $g$ and $g_A$. They are used to describe individual and combination of
Feynman diagrams. They possess  symmetries reflecting the antisymmetries of the 3g and the virtual vertices.
Of special interest is the functions $d$ and $D$. Both have a  large amount of symmetries, both vanish on-shell,
and both have  relatively simple expressions off-shell that can be used to facilitate the computation of CK combinations
$\bD_l$ if the Slavnov-Taylor identity is used.

\section{Four-Point Amplitudes}
The numerators of the partial amplitudes
\be
A(1234)&=&{\bar n_s\over s}+{\bar n_t\over t} ,\nn\\
A(1324)&=&-{\bar n_t\over t}+{\bar n_u\over u},\qquad{\rm with}\nn\\
s=s_{12}&=&s_{34} ,\ t=s_{23}=s_{41} ,\ u=s_{13}=s_{24} ,\labels{A1234a}
 \ee
can be computed from the Feynman diagrams in Fig.~9 to be
\be
\bar n_s&=&T(129) T(\bar 934)+Q(1234)s:=P(1234)+Q(1234)s:=A_4(1234),\nn\\
\bar n_t&=&T(419) T(\bar 923)+Q(4123)t=P(4123)+Q(4123)t=A_4(4123),\nn\\
\bar n_u&=&T(139) T(\bar 924)+Q(1324)u=P(1324)+Q(1324)u=A_4(1324).
     \labels{A4}\ee
In \eq{A4} and  in the rest of this article,   high numbers such as 9,8
are used  to designate internal (off-shell) lines  whose polarizations are to be summed over.  
 See \eq{complete1} and \eq{complete2}.   The polarizations of  9 and $\9$ are the same, $\e_9^{(\l)}=\e_{\9}^{(\l)}$, but  
 the momenta  $k_9$ and $k_{\bar 9}$ are opposite to enable all momenta  at
a vertex to be  outgoing. For example, in the $P(1234)=T(129) T(\bar 934)$ term,
 $k_9+k_1+k_2=0$, and $k_{\bar 9}+k_3+k_4=-k_9+k_3+k_4=0$. The minus sign in front of the $\bar n_t/t$
term in $A(1324)$ of \eq{A1234a} is a consequence of antisymmetry at the $\923$ vertex.

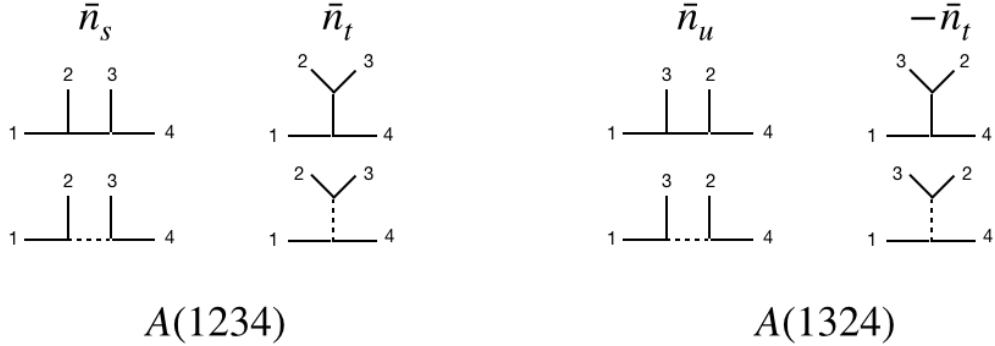
\bc\igwg{14}{Fig9}\\ Fig.9\quad Feynman diagrams for the four-point color-stripped amplitude\ec

The function $P(1234)$ in \eq{A4} can be computed using Feynman rules. The result, valid both on-shell
and off-shell,  is
\be
P(1234)= bcc\ terms\ +\ \h b_{12}b_{34}(s_{13}-s_{14}-s_{23}+s_{24}).\labels{Pformula}\ee
The $bcc$ terms are rather lengthy and will not be displayed here. Suffice to say
 that they are not zero even for on-shell external lines. However,
a direct calculation shows that the CK relation $\bar\D:=\bar n_s-\bar n_t-\bar n_u=0$  is valid on-shell,
implying a complete cancellation of the $bcc$ terms in $\bD$. 
Its off-shell expression, containing only $bcc$ and no $bbs$ terms, 
\be\bar \Delta&:=& \bar n_s-\bar n_t-\bar n_u\nn\\
&=&b_{12}[c_{33}(c_{41}-c_{42})-c_{44}(c_{31}-c_{32})]+b_{13}[c_{44}(c_{21}-c_{23})-c_{22}(c_{41}-c_{43})]\nn\\
&+&b_{14}[c_{22}(c_{31}-c_{34})-c_{33}(c_{21}-c_{24})]+b_{23}[c_{11}(c_{42}-c_{43})-c_{44}(c_{12}-c_{13})]\nn\\
&+&b_{24}[c_{33}(c_{12}-c_{14})-c_{11}(c_{32}-c_{34})]+b_{34}[c_{11}(c_{23}-c_{24})-c_{22}(c_{13}-c_{14})]\nn\\
&:=&d(1234),\labels{D4}\ee
will   be useful later for larger $N$ computations. An equivalent expression which will also be useful later is
\be d(1234)&=&c_{11}T(234)-c_{22}T(341)+c_{33}T(412)-c_{44}T(123).\labels{D4t}\ee

Although the $bcc$ terms from each $\bn_a$ are complicated, their combined expression in \eq{D4}
is relatively simple. The simplicity stems from combinations of terms of the type shown in Fig.~10, yielding
\be \mathbf{c}(\a\b\g\d)=b_{\a\d}\[c_{\b\b}(c_{\g\a}-c_{\g\d})-c_{\g\g}(c_{\b\d}-c_{\b\a})\].\labels{cid}\ee
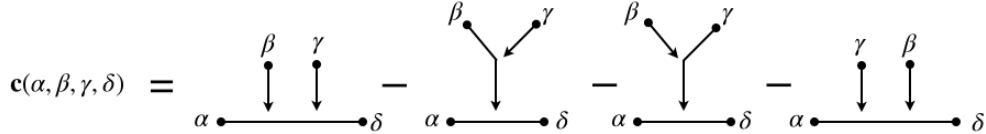
\bc\igwg{13}{Fig10}\\ Fig.~10.\quad Diagrams for the $c$-identity function $\mathbf{c}(\a\b\g\d)$\ec

Either from  the antisymmetry at every vertex or directly from \eq{D4} or \eq{D4t},  $d(1234)$ can be seen to have
 a high degree of symmetry,
\be
d(1234)=-d(2134)=-d(1243)=-d(1324).
\labels{D4sym}\ee

Since $\bn_a$ satisfy the on-shell CK relation, they can be related to the fundamental numerators $\n(\b)$
by using \eq{Amnu} and \eq{A1234a},
\be
\bm A(1234)\\ A(1324)\\ \em&=&\bm 1/s+1/t& -1/t\\ -1/t&1/u+1/t\\ \em \bm \n(23)\\ \n(32)\\ \em=\bm\bar n_s/s+\bar n_t/t\\
\bar n_u/u-\bar n_t/t\\ \em .\labels{A1234b}\ee
The mass matrix $m(\a|\b)$ displayed above is obtained from Fig.~9 and the rules stated at the beginning to Sec.~IIC. The relation between $\bn_a$ and $\n(\b)$ can be read off from \eq{A1234b} to be
\be\bar n_s&=&\n(23)+xu(23),\nn\\
\bar n_u&=&\n(32)+xu(32),\nn\\
\bar n_t&=&\n(23)-\n(32)+x\(u(23)-u(32)\),\labels{nbarnu}\ee
where $x$ is an arbitrary parameter, and $u=\(u(23), u(32)\)^T=(s,-u)^T$ is the null vector of the on-shell mass
matrix $m(\a|\b)$ because $s+t+u\simeq 0$. See Sec.~IIE. 
Since both $\bn_a$ and $n'_a$ are in the BCJ representation, $\d n_a=\bn_a-n'_a$ must be of the form $xu$.
A direct calculation shows that $x=-(b_{13}b_{24}-b_{12}b_{34})$.

\section{Five-Point Amplitude}
\subsection{partial amplitudes and numerators}
As discussed in Sec.~IIA, there are six partial amplitudes for $N=5$,  each containing 
 5 terms, making a total of 30 terms. Since there are only 15 independent propagators $1/Q_a$,  each $n_a/Q_a$  is expected to appear  in several partial amplitudes, with signs determined by the antisymmetry
 of vertices as discussed in Sec.~IIC. 
 
 The explicit enumerations via canonical ordering of $s_I, Q_a=s_{I_1}s_{I_2}$, and $n_a$ 
are given in Appendix A. 
 With those notations,
 the six partial amplitudes are
  \be A_1=A(12345)&=&+{n_1\over s_{12}s_{34}}+{n_2\over s_{12}s_{123}}+{n_{10}\over s_{23}s_{123}}+{n_{11}\over s_{23}s_{234}}+{n_{15}\over s_{34}s_{234}} ,\nn\\
A_2=A(12435)&=&-{n_1\over s_{12}s_{34}}+{n_3\over s_{12}s_{124}}+{n_{12}\over s_{24}s_{124}}+{n_{13}\over s_{24}s_{234}}-{n_{15}\over s_{34}s_{234}} ,\nn\\
 A_3=A(13245)&=&+{n_4\over s_{13}s_{24}}+{n_5\over s_{13}s_{123}}-{n_{10}\over s_{23}s_{123}}-{n_{11}\over s_{23}s_{234}}-{n_{13}\over s_{24}s_{234}} ,\nn\\ 
A_4= A(13425)&=&-{n_4\over s_{13}s_{24}}+{n_6\over s_{13}s_{134}}+{n_{13}\over s_{24}s_{234}}+{n_{14}\over s_{34}s_{134}}-{n_{15}\over s_{34}s_{234}} ,\nn\\
 A_5=A(14235)&=&+{n_{7}\over s_{14}s_{23}}+{n_8\over s_{14}s_{124}}-{n_{11}\over s_{23}s_{234}}-{n_{12}\over s_{24}s_{124}}-{n_{13}\over s_{24}s_{234}} ,\nn\\
 A_6=A(14325)&=&-{n_{7}\over s_{14}s_{23}}+{n_9\over s_{14}s_{134}}+{n_{11}\over s_{23}s_{234}}-{n_{14}\over s_{34}s_{134}}+{n_{15}\over s_{34}s_{234}}.
 \labels{CK5a}\ee
 The canonically ordered $n_a$ used here are different from those
 used in Ref.\cite{BCJ08}. A dictionary relating the two can be found in Table A3 of Appendix A.

 The Feynman diagrams for $A(12345)$ are shown in Fig.~11. Those for other partial amplitudes can be obtained by a relabelling
 of the external lines, together with a change of the numerator designations.
 
 The sign of a $n_a/Q_a$ term in an amplitude $A_m$ can be read out from the Feynman diagrams, as 
discussed in Sec.~IIC.   By convention, the sign of every term in $A_1=A(12345)$ is positive. 
The sign of the $n_1/Q_1$
term in $A_2=A(12435)$ for example is negative because a flip of 3 and 4 is needed  change 12345 to 12435.
 
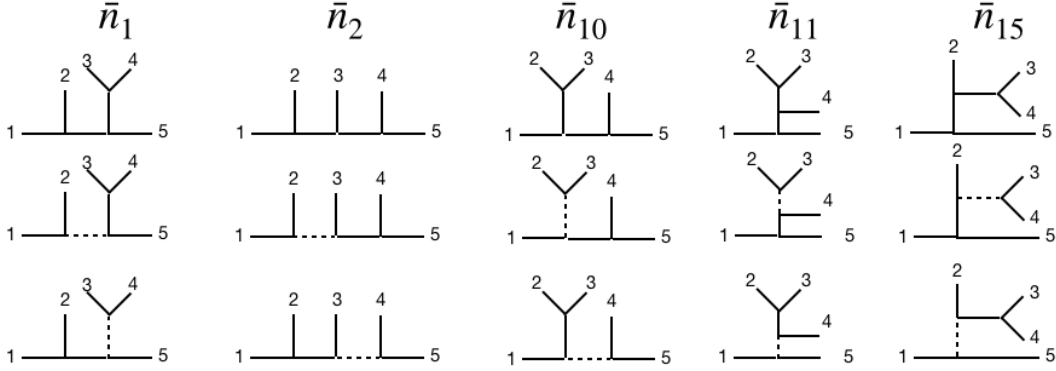
  \bc\igwg{15}{Fig11}\\ Fig.11\quad Feynman diagrams for $A(12345)$\ec

 The numerators $\bar n_a$ read off from Fig.~11 using the Feynman rules are
 \be
 \bar n_1&=&T(129) T(\9\85) T(834)+Q(12\85) T(834) s_{12}+T(129) Q(345\9) s_{34} ,\nn\\ 
 \bar n_2&=&T(129) T(\93\8) T(845)+Q(123\8) T(845) s_{12}+T(129) Q(45\93) s_{123} ,\nn\\
 \bar n_{10}&=&T(239) T(1\9\8) T(845)+Q(23\81) T(845) s_{23}+T(239) Q(451\9) s_{123} ,\nn\\
 \bar n_{11}&=&T(239) T(\94\8) T(851)+Q(234\8) T(851) s_{23}+T(239) Q(51\94) s_{234} ,\nn\\
 \bar n_{15}&=&T(349) T(2\9\8) T(8 51)+Q(34\82) T(851) s_{34}+T(349) Q(2\951) s_{234} .
     \labels{A5a}\ee
The other 10 $\bar n_a$'s,  obtained similarly,  are
\be 
 \bar n_3&=&T(129) T(\94\8) T(835)+Q(124\8) T(835) s_{12}+T(129) Q(35\94) s_{124} ,\nn\\
  \bar n_{12}&=&T(249) T(1\9\8) T(835)+Q(24\81) T(835) s_{24}+T(249) Q(351\9) s_{124} ,\nn\\
 \bar n_{13}&=&T(249) T(\93\8) T(851)+Q(243\8) T(851) s_{24}+T(249) Q(51\93) s_{234} ,\nn\\
 \bar n_4&=&T(139) T(\9\85) T(824)+Q(13\85) T(824) s_{13}+T(139) Q(245\9) s_{24} ,\nn\\ 
  \bar n_5&=&T(139) T(\92\8) T(845)+Q(132\8) T(845) s_{13}+T(139) Q(45\92) s_{123} ,\nn\\
  \bar n_6&=&T(139) T(\94\8) T(825)+Q(134\8) T(825) s_{13}+T(139) Q(25\94) s_{134} ,\nn\\
  \bar n_{14}&=&T(349) T(1\9\8) T(825)+Q(34\81) T(825) s_{34}+T(349) Q(251\9) s_{134} ,\nn\\
 \bar n_7&=&T(149) T(\9\85) T(823)+Q(14\85) T(823) s_{14}+T(149) Q(235\9) s_{23} ,\nn\\
 \bar n_8&=&T(149) T(\92\8) T(835)+Q(142\8) T(835) s_{14}+T(149) Q(35\92) s_{124} ,\nn\\
  \bar n_9&=&T(149) T(\93\8) T(825)+Q(143\8) T(825) s_{14}+T(149) Q(25\93) s_{134} .\labels{A5b}\ee
 Every $\bar n_a$ contains three terms,  $TTT,  QT$, and $TQ$, reflecting the three rows of diagrams in Fig.~11.
 Owing to antisymmetry at the vertices, the last two can all be expressed in terms of a single function $f$
 representing  Fig.~12,
 \be
 f(12345)&:=&Q(123\9) T(945)\nn\\
 &=&b_{2 3} \[b_{4 5} (\-c_{1 4} \+ c_{1 5}) - 
    b_{1 5} (c_{4 4} \+ 2 c_{4 5}) + b_{1 4} (2 c_{5 4} \+ c_{5 5})\] \nn\\
    &+& 
 b_{1 3} \[b_{4 5} (c_{2 4} \- c_{2 5}) + 
    b_{2 5} (c_{4 4} \+ 2 c_{4 5}) - b_{2 4} (2 c_{5 4} \+ c_{5 5})\],
\labels{f1}\ee
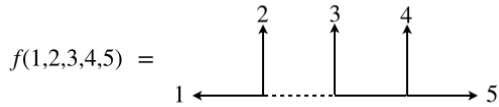
 \bc\igwg{7}{Fig12}\\ Fig.~12.\quad $f(12345)$\ec
which  is antisymmetric in the first two arguments and in the last two arguments:
 \be
 f(12345)=-f(21345)=-f(12354).\labels{f2}\ee

 It is also convenient to introduce another function 
 \be F(12345):=f(12345)+f(12453)+f(12534)\labels{F1}\ee
that  appears in CK combinations. Although $F$ is not zero even on-shell,  it nevertheless possesses a large amount of symmetry,
 \be F(12345)=-F(21345)=-F(12435)=F(12453)=F(12534),\labels{F2}\ee  
needed to simplify calculations.
 
\subsection{CK combinations} 
As discussed in Sec.~IIB, there are ten CK combinations $\bD_l$. Using the canonical ordering 
of double Mandelstam variables specified in Appendix A to label them, these
CK combinations are
\be
\D_1&=& n_2- n_1- n_3,\nn\\
\D_2&=& n_5- n_4- n_6,\nn\\
\D_3&=& n_8- n_7- n_9,\nn\\
\D_4&=& n_{10}- n_7- n_{11} ,\nn\\
\D_5&=& n_{12}- n_4- n_{13} ,\nn\\
\D_6&=& n_1- n_{14}- n_{15} ,\nn\\
\D_7&=& n_2- n_5- n_{10} ,\nn\\
\D_8&=& n_3- n_8- n_{12} ,\nn\\
\D_9&=& n_6- n_9- n_{14} ,\nn\\
\D_{10}&=& n_{11}- n_{13}- n_{15}.\labels{D51}
\ee
Every $ n_a$ occurs twice in these $\D_l$,  enabling
 the following combination of $\D_l$ to be identically zero,
\be \D_1-\D_2+\D_3-\D_4+\D_5+\D_6-\D_7+\D_8-\D_9-\D_{10}=0.\ee
Such a relation will henceforth be referred to as a {\it trivial CK identity}. It is valid on-shell and off-shell
because it is identically zero whatever $n_a$ are. There is only
one trivial identity in $N=5$, but there are many in $N=6$.

Using  \eq{A5a} to \eq{F2}, the CK combinations $\bD_l$ can be calculated to be
\be
\bD_1&=&T(129) d(45\93)+F(12345)s_{12} ,\nn\\
\bD_2&=&T(139) d(45\92)+F(13245)s_{13} ,\nn\\
\bD_3&=&T(149) d(35\92)+F(14235)s_{14} ,\nn\\
\bD_4&=&T(239) d(451\9)+F(23154)s_{23} ,\nn\\
\bD_5&=&T(249) d(351\9)+F(24153)s_{24} ,\nn\\
\bD_6&=&T(349) d(12\95)+F(34512)s_{34} ,\nn\\
\bD_7&=&T(459) d(123\9)+F(45321)s_{123} ,\nn\\
\bD_8&=&T(359) d(124\9)+F(35421)s_{124} ,\nn\\
\bD_9&=&T(259) d(134\9)+F(25431)s_{134} ,\nn\\
\bD_{10}&=&T(159) d(23\94)+F(51432)s_{234}.\labels{D52}
\ee
All has  the form $Td+Fs$. 
The $Td$ terms are of the form $bccc$ (see Sec.~IID), they vanish on-shell on account of the
Slavnov-Taylor identity for the following reason.  Every term in $d$ given by \eq{D4} contains a $c_{ii}$ for some $i$, which is zero unless $i$ is off-shell.  In the expression $Td$, internal line 9 is the only 
one off-shell, hence the only non-vanishing term in $d$ is proportional to $c_{99}$. By setting ${\cal M}$ in Sec.~IIID
to be $T$ and $i$ to be 9, the Slavnov-Taylor identity ensures that the $Td$ terms vanish on-shell. 

The $Fs$ terms are of the form $bbcs$ and are not zero even on-shell, thus  CK combinations $\bD_l$ computed from Feynman diagrams
do not satisfy the CK relation.

Although $\bD_l$ is not quite zero, a large amount of cancellation has already taken place by the
vanishing of the $bccc$ terms. The $bccs$  terms do not vanish,  but since it is linear in $s$,
there is a chance that $\bn_a$ can be converted into a BCJ representation by a {\it local} gauge transformation.

\subsection{Local gauge transformation}
A generalized gauge transformation $\bar n_a\to n_a=\bar n_a-\d n_a$ induces a transformation
$\bD_l\to \D_l=\bD_l-\d \D_l$, with $\d \D_l=\d n_s-\d n_t-\d n_u$. To bring about the CK relation $\D_l=0$ on-shell,  a shift of amount $\d \D_l=\bar \D_l$ is required. Since $\bD_l$ is of the
form $Fs$,  $\d n_a$ is also expected to be linear in $s$. This gives hope that it may be a local gauge transformation.

According to Sec.~IIF, local gauge transformations for $N=5$ must have the form
\be
\d n_a=p_as_a+p'_as'_a,\labels{p51}\ee
where $Q_a=s_as'_a$ is the propagator factor, and by our convention $s_a\prec s'_a$. In other words, 
$\d n_a$ may involve only $s_a$ and $s'_a$, but no other $s$. Note that $s_a$ here is {\it not} a Mandelstam
variable with the canonical subscript $a$. It is defined to be the first Mandelstam factor of $Q_a$.

An examination of \eq{CK5a} shows that every $A(1\a N)$ remains the same if the three parameters in 
every one of the following 10 triplets are equal:
\be
&&t_1=(\-p'_1, p'_2, \-p'_3),\quad t_2=(\-p'_4, p'_5, \-p'_6),\quad t_3=(\-p'_7, p'_8, \-p'_9),\quad t_4=(\-p_7, p'_{10} , \-p'_{11}),\nn\\
&&t_5=(\-p_4, p'_{12} ,\-p'_{13}),\quad
t_6=(p_1, \-p'_{14} , \-p'_{15}),\quad t_7=(p_2, \-p_5, \-p_{10}),\quad t_8=(p_3, \-p_8, \-p_{12}),\nn\\
&& t_9=(p_6, \-p_9, \-p_{14}),\quad
t_{10}= (p_{11} , \-p_{13} , \-p_{15}).\labels{p5e}\ee
For example, take the parameters in the first triplet $t_1$. The only  partial amplitudes affected by these parameters are
\be \d A_1={p'_1\over s_{12}}+{p'_2\over s_{12}} ,\quad \d A_2=-{p'_1\over s_{12}}+{p'_3\over s_{12}} ,\nn\ee
and they are zero if the three parameters in $t_1$ are equal. 

 The $t_1$ triplet contains $p$ parameters for the numerators $(\bar n_1, \bar n_2, \bar n_3)$, 
all members of $\bD_1$. This is why this triplet is labelled $t_1$. 
Similarly, the triplet of parameters in $t_l$
are all related to the numerators involved in $\bD_l$.

Using \eq{p51} and \eq{p5e}, and $t_l$ to represent any of the three $p$ parameters
in its triplet, the shift $\d\D_l=\d n_s-\d n_t-\d n_u$ takes on the form

\be
\d\D_1&=&\d(n_2-n_1-n_{3})= (t_7-t_6-t_8) s_{12}+t_1(s_{34}\+s_{123}\+s_{124}) ,\nn\\
\d\D_2&=&\d(n_{5}-n_{4}-n_6)=(-t_7+t_5-t_9) s_{13}+t_2(s_{24}\+s_{123}\+s_{134}) ,\nn\\
\d\D_3&=&\d(n_{8}-n_7-n_9)=(-t_8+t_4+t_9) s_{14}+t_3(s_{23}\+s_{124}\+s_{134}) ,\nn\\
\d\D_4&=&\d(n_{10}-n_7-n_{11})= (-t_7+t_3-t_{10})s_{23}+t_4 (s_{14}\+s_{123}\+s_{234}),\nn\\
\d\D_5&=&\d(n_{12}-n_{4}-n_{13})=(-t_8+t_2+t_{10})s_{24}+t_5 (s_{13}\+ s_{124}\+s_{234}),\nn\\
\d\D_6&=&\d(n_1-n_{14}-n_{15})=(-t_1+t_9+t_{10})s_{34}+ t_6 (s_{12}\+ s_{134}\+s_{234}),\nn\\
\d\D_7&=&\d(n_2-n_{5}-n_{10})=(t_1-t_2-t_4)s_{123}+t_7 (s_{12}\+ s_{13}\+ s_{23}),\nn\\
\d\D_8&=&\d(n_3-n_8-n_{12})=(-t_1-t_3-t_5)s_{124}+t_8( s_{12}\+ s_{14}\+s_{24}),\nn\\
\d\D_9&=&\d(n_6-n_9-n_{14})= (-t_2+t_3+t_6)s_{134}+t_9 (s_{13}\+ s_{14}\+ s_{34}),\nn\\
\d\D_{10}&=&\d(n_{11}-n_{13}-n_{15})=(-t_4+t_5+t_6) s_{234}+ t_{10}(s_{23}\+ s_{24}+ s_{34}).\labels{dD5a}\ee
These formulas are valid on-shell and off-shell.
Using the kinematical formula
\be s_{I_1I_2}+s_{I_1I_3}+s_{I_2I_3}=s_{I_1}+s_{I_2}+s_{I_3}+s_{I_1I_2I_3}, \labels{stu}\ee
 \eq{dD5a} when on-shell  reduces to

\be
\d\D_1&=& (t_7-t_6-t_8+t_1) s_{12} ,\nn\\
\d\D_2&=&(-t_7+t_5-t_9+t_2) s_{13} ,\nn\\
\d\D_3&=&(-t_8+t_4+t_9+t_3) s_{14} ,\nn\\
\d\D_4&=& (-t_7+t_3-t_{10}+t_4)s_{23} ,\nn\\
\d\D_5&=&(-t_8+t_2+t_{10}+t_5)s_{24} ,\nn\\
\d\D_6&=&(-t_1+t_9+t_{10}+t_6)s_{34} ,\nn\\
\d\D_7&=&(t_1-t_2-t_4+t_7)s_{123} ,\nn\\
\d\D_8&=&(-t_1-t_3-t_5+t_8)s_{124} ,\nn\\
\d\D_9&=& (-t_2+t_3+t_6+t_9)s_{134} ,\nn\\
\d\D_{10}&=&(-t_4+t_5+t_6+t_{10}) s_{234}.\labels{dD5b}\ee
In order for $\d\D_l=\bD_l$, the two sides must have the same $s$ dependence. This can
be verified to be true by
comparing \eq{D52} with \eq{dD5b}.
With this matching of the $s$ dependence, the requirement $\d\D_l=\bD_l$ is reduced to a set of linear equations for the
10 unknowns $t_l$. In matrix form, these CK equations are
\be
\sum_{l'=1}^{10}\tau_{ll'}t_{l'}=d_l,\qquad (1\le l\le 10)\labels{teq}\ee
where $d_l$ is the on-shell expression of $\bD_l$ with the $s$ dependence stripped off, namely,
\be
d_1&=&F(12345),\quad  d_2=F(13245),\quad d_3=F(14235),\quad d_4=F(23154),\nn\\
 d_5&=&F(24153),\quad d_6=F(34512),\quad d_7=F(45321),\quad d_8=F(35421),\nn\\
d_9&=&F(25431),\quad d_{10}=F(51432).\labels{D53}
\ee
The $10\x10$ matrix
{\footnotesize\be
\tau=\bm 
1&0&0&0&0&-1&1&-1&0&0\\
0&1&0&0&1&0&-1&0&-1&0\\
0&0&1&1&0&0&0&-1&1&0\\
0&0&1&1&0&0&-1&0&0&-1\\
0&1&0&0&1&0&0&-1&0&1\\
-1&0&0&0&0&1&0&0&1&1\\
1&-1&0&-1&0&0&1&0&0&0\\
-1&0&-1&0&-1&0&0&1&0&0\\
0&-1&1&0&0&1&0&0&1&0\\
0&0&0&-1&1&1&0&0&0&1\\ \em\ee}
has rank 5, and  has five null vectors
\be
v_1&=&(\-1, 1, 1, \-1, 0, 0, 1, 0, 0, \-1), \nn\\
v_2&=&(\-1, 0, 1, 0, 0, 0, 1, 0, \-1,  0),\nn\\
  v_3&=& (0, \-1, \-1, 0, 0, 0, \-1, \-1, 0, 0), \nn\\
  v_4&=&(\-1, 0, 1, \-1, 0, \-1, 0, 0, 0, 0), \nn\\
  v_5&=&(0, 1, 1, \-1, \-1, 0, 0, 0, 0, 0).\labels{nullv}\ee
 As a result, the inhomogeneous terms $d_l$ must satisfy five sum rules
 $R_x:=\sum_l (v_x)_ld_l=0$ in order for \eq{teq} to have a solution. Explicitly, these sum rules are
 \be
 R_1&=&-d_1+d_2+d_3-d_4+d_7-d_{10}=0,\nn\\
 R_2&=&-d_1+d_3+d_7-d_9=0,\nn\\
 R_3&=&-d_2-d_3-d_7-d_8=0,\nn\\
 R_4&=&-d_1+d_3-d_4-d_6=0,\nn\\
 R_5&=&d_2+d_3-d_4-d_5=0.\labels{R1}\ee
To check whether the sum rules are satisfied, introduce a $D$-function defined by  
  \be
 D(12345):=F(12345)+F(23415)+F(34125)+F(41235).\labels{D1}\ee
 In many ways this five-point function $D$ resembles the four-point function $d$ in \eq{D4}.
It vanishes on-shell, and   has the following expression off-shell,
\be D(12345)&=&c_{11}\(-b_{23}b_{45}+2b_{24}b_{35}-b_{25}b_{34}\)\nn\\
&+&c_{22}\(-b_{34}b_{15}+2b_{13}b_{45}-b_{14}b_{35}\)\nn\\
&+&c_{33}\(-b_{12}b_{45}+2b_{15}b_{24}-b_{14}b_{25}\)\nn\\
&+&c_{44}\(-b_{15}b_{23}+2b_{13}b_{25}-b_{12}b_{35}\)\nn\\
&-&2c_{55}\(-b_{12}b_{34}+2b_{13}b_{24}-b_{14}b_{23}\).\labels{D2}\ee
By construction, it is cyclically symmetric in its first four arguments, but it also possesses the following additional symmetries:
\be D(12345)=D(32145)=D(14325)=D(43215).\labels{D3}\ee
Diagrammatically, $D$ is given by a sum of 12 diagrams of the form Fig~12.

We are now ready to check the sum rules.  
Using \eq{D53} and the symmetry of $F$ given in \eq{F2} , one gets

 \be
 R_1&=&-d_1+d_2+d_3-d_4+d_7-d_{10}\nn\\
 &=&\-F(12345)\+F(13245)\+F(14235)\-F(23154)\+F(45321)\-F(51432)\nn\\
 &=&F(21534)\-F(31452)\-F(14532)\+F(32154)\-F(45312)\+F(16324)\nn\\
&=&\[D(21534)F(53214)\]-\[D(31452)-F(53142)\]=D(21534)-D(31452)\simeq 0,\nn\\
 R_2&=&-d_1+d_3+d_7-d_9=\-F(12345)\+F(14235)\+F(45321)\-F(25431)\nn\\
 &=&F(12543)\+F(41253)\+F(54123)\+F(25413)=D(12543)\simeq 0,\nn\\
 R_3&=&-d_2-d_3-d_7-d_8=\-F(13245)\-F(14235)\-F(45321)\-F(35421)\nn\\
 &=&F(13542)\+F(41352)\+F(54132)\+F(35412) =D(13542)\simeq 0,\nn\\
 R_4&=&-d_1+d_3-d_4-d_6=\-F(12345)\+F(14235)\-F(23154)\-F(34512)\nn\\
 &=&-F(12345)-F(41235)-F(23415)-F(34125)=-D(12345)\simeq 0,\nn\\
 R_5&=&d_2+d_3-d_4-d_5=F(13245)\+F(14235)\-F(23154)\-F(24153)\nn\\
 &=&F(31425)\+F(14235)\+F(23145)\+F(42315)= D(31425)\simeq 0,\labels{R2}\ee
where $\simeq$ means equality on-shell. Thus the sum rules are all true.

With these sum rules satisfied, one can proceed to solve \eq{teq}. Since the rank of $\tau$ is five,
only five $t_l$ can be solved, leaving the other five $t_l$   free. For example, solving for $t_4, t_5, t_6, t_7, t_{10}$, one gets
\be t_4 &=& -d_3 - t_3 + t_8 - t_9, \nn\\
t_5 &=&d_8 - t_1 - t_3 + t_8,\nn\\
t_6 &=& -d_9 + t_2 - t_3 - t_9, \nn\\
t_7&=& -d_3 - d_7 - t_1 + t_2 - t_3 + t_8 - t_9,\nn\\
t_{10}&=&d_4 + d_7 + t_1 - t_2 + t_3,\labels{tsol}\ee 
where \eq{R1} has been used to simplify expressions.     

Let us briefly summarize the results obtained so far for $N=5$. The CK combinations $\bD_l$ computed from Feynman diagrams are not
zero on-shell, though their $bccc$ terms all vanish. A generalized gauge transformation $\bar n_a\to n_a=\bar n_a-\d n_a$ can restore the CK relation, namely,  can bring them to a BCJ representation,
if the induced change $\d\D_l=\d(n_s-n_t-n_u)$ is equal to $\bD_l$ for all $l$. In order for the 
gauge transformation to be local, the latter must be of the form $\d n_a=p_as_a+p'_as'_a$, with the 30 parameters $p_a, p'_a$
obeying 10 triplet identities, thereby leaving only 10 parameters $t_l$ to accomplish the requirement $\d\D_l=\bD_l$.
This requirement has a solution only when the 10 $\bD_l$'s satisfy five sum rules $R_x=0$ shown in \eq{R1}. A detailed
calculation using the explicit expressions of $\bD_l$ shows that these sum rules are indeed satisfied. With that, five of the ten
$t_l$'s can be solved, leaving the other five $t_l$'s as free parameters. If $t_1, t_2, t_3, t_8, t_9$ are chosen as free parameters,
then the solution for $t_4, t_5, t_6, t_7, t_{10}$ is given in \eq{tsol}.

These 10 $t_l$ move $\bar n_a$ to $n_a$, satisfying the ten CK equations $\D_l =0$. The solution 
carries five free $t$ parameters, which is a consequence of the well-known fact that any solution
of the homogeneous equation can be added to a particular solution to get a new solution of the 
inhomogeneous equation.
The homogeneous equation $\sum_{l'}\tau_{ll'}t^0_{l'}=0$ has five independent solutions $v_x$ given by
\eq{nullv}, corresponding to the five free $t_l$ parameters. 
The addition of $v_x$ is therefore a further gauge transformation in the BCJ representation,
 keeping all CK relations intact. According
to the discussion of Sec.~IIE, $v_x$ must therefore be related to the null vectors $u_i$ of the mass matrix $m$,
but there  are five $v_x$
and only four $u_i$. Why don't they match?

\subsection{Null vectors of $\boldsymbol{\tau}$ and  $\boldsymbol{m}$ }
As discussed in Sec.~IIE, the $N=5$ BCJ relations  can be obtained from \eq{BCJ} to be
\be 0&=&A(1324 5) s_{13} + A(12345) (s_{13} +s_{23}) + 
 A(1243 5) (s_{13}+ s_{23} + s_{34}),\nn\\
 0&=& A(1342 5) s_{13} + A(1432 5) (s_{13} +  s_{34}) + 
 A(1423 5) (s_{13} + s_{23} +  s_{34}),\nn\\
 0&=&A(1432 5) s_{14} + A(1342 5) (s_{14} + s_{34}) + 
 A(1324 5) (s_{14}+ s_{24} + s_{34}),\nn\\
 0&=&A(1234 5) s_{12} + A(1324 5) (s_{12} + s_{23}) + 
 A(1342 5) (s_{12} + s_{23} + s_{24}).\labels{mon5}\ee
These relations give rise to 4 null vectors of the on-shell propagator matrix $m$. They are
\be
u_1&=&(s_{13}+s_{23} ,s_{13}+s_{23}+s_{34} ,s_{13} ,0,0,0)=(s_{13}+s_{23} ,-s_{124} ,s_{13} ,0,0,0),\nn\\
u_2&=&(0,0,0,s_{13} ,s_{13}+s_{23}+s_{34} ,s_{13}+s_{34})=(0,0,0,s_{13} ,-s_{124} ,s_{13}+s_{34}),\nn\\
u_3&=&(0,0,s_{14}+s_{24}+s_{34} ,s_{14}+s_{34} ,0,s_{14})=(0,0,-s_{123} ,s_{14}+s_{34} ,0,s_{14}),\nn\\
u_4&=&(s_{12} ,0,s_{12}+s_{23} ,s_{12}+s_{23}+s_{24} ,0,0)=(s_{12} ,0,s_{12}+s_{23} ,-s_{134} ,0,0).\labels{u5}\ee
These null vectors enable a generalized gauge transformation  $\n\to\n-\d\n$ on the fundamental numerator factors in the form
\be
\d\n=\d\bm \n_1\\ \n_2\\ \n_3 \\ \n_4\\ \n_5\\ \n_6\\ \em=
\d\bm \n(234)\\ \n(243)\\ \n(324) \\ \n(342)\\ \n(423)\\ \n(432)\\ \em=
\sum_{i=1}^4x_iu_i=
\bm (x_4-x_1)s_{12}+x_1s_{123}\\
-x_1 s_{124}\\
(x_1-x_4)s_{13}+(x_4-x_3)s_{123}\\
(x_2-x_3)s_{13}+(x_3-x_4)s_{134}\\
-x_2s_{124}\\
(x_3-x_2)s_{14}+x_2s_{134}\\ \em,\labels{delnu}\ee
without altering any of the partial amplitudes given by \eq{Amnu}. In \eq{delnu}, $x_i$ are arbitrary parameters, and on-shell kinematical relations such as $s_{13}+s_{23}=s_{123}-s_{12}$ have been used.
With the relation between $n_a$ and $\n(\b)$ discussed in Sec.~IIE, changes $\d \n(\b)$ give rise to changes in $\d n_a$. Let us examine the basic changes associated with half-ladder diagrams.

Recall from Sec.~IIE that the ordinary numerator $n_a$ for a half-ladder diagram  in $A(1\a N)$ is equal to 
the fundamental numerator $\n(\a)$. With the enumeration scheme given in Appendix A, these 
half-ladder $n_a$ are
\be (n_2, n_3, n_5, n_6, n_8, n_9)=(\n_1,\n_2,\n_3,\n_4,\n_5,\n_6).\labels{nnu5}\ee
Their gauge transformation $\d n_a=p_as_a+p'_as'_a$ expressed in $t_l$ are
\be
\d n_2=p_2s_{12}+p'_2s_{123}&=&  t_7s_{12}+t_1s_{123} ,\nn\\
\d n_3=p_3s_{12}+p'_3s_{124}&=&t_8s_{12}-t_1s_{124} ,\nn\\
\d n_5=p_5s_{13}+p'_5s_{123}&=&-t_7s_{13}+t_2s_{123} ,\nn\\
\d n_6=p_6s_{13}+p'_6s_{134}&=&t_9s_{13}-t_2s_{134} ,\nn\\
\d n_8=p_8s_{14}+p'_8s_{124}&=&-t_8s_{14}+t_3s_{124} ,\nn\\
\d n_9=p_9s_{12}+p'_9s_{134}&=&-t_9s_{12}-t_3s_{134}.\labels{pt}\ee
The homogeneous part $t^0_l$ causes change between the $n_a$'s that satisfy the CK relations. For those $n_a$ which are
equal to some $\n(\a)$, \eq{pt} with $t_l$ replaced by $t_l^0$ can be identified with \eq{delnu}. Therefore
\be t_7^0s_{12}+  t_1^0s_{123}&=& (x_4-x_1)s_{12}+x_1s_{123} ,\nn\\
  t_8^0s_{12}-  t_1^0s_{124}&=&-x_1 s_{124} ,\nn\\
 t_7^0s_{13}+  t_2^0s_{123}&=&(x_1-x_4)s_{13}+(x_4-x_3)s_{123} ,\nn\\
  t_9^0s_{13}-  t_2^0s_{134}&=&(x_2-x_3)s_{13}+(x_3-x_4)s_{134} ,\nn\\
  t_8^0s_{14}+  t_3^0s_{124}&=&-x_2s_{124} ,\nn\\
  t_9^0s_{12}-  t_3^0s_{134}&=&(x_3-x_2)s_{14}+x_2s_{134} .\labels{pxt}\ee
This allows the identification
\be
  t_1^0&=&x_1,\nn\\
  t_2^0&=&x_4-x_3,\nn\\
  t_3^0&=&-x_2,\nn\\
  t_7^0&=&x_4-x_1,\nn\\
  t_8^0&=&0,\nn\\
  t_9^0&=&x_2-x_3.\labels{teqx}\ee
 Let us compare these expressions for $t_l^0$ with those given by  \eq{tsol} after setting all $d_l=0$. 
 There are five free parameters in \eq{tsol}, $t_1^0, t_2^0, t_3^0, t_8^0, t_9^0$, through which the
 other five $t_l^0$ can be obtained. For example, $t_7^0=-t_1^0+t_2^0-t_3^0+t_8^0-t_9^0$, which is
 consistent with \eq{teqx}. The other five $t_l^0$ in \eq{teqx} should all be free, but that equation shows
 only four free parameters related to $x_i$, and $t_8^0=0$ is not free. How come? The reason is that
 $t_8^0$ is really not free. $t_8$ can be made to disappear if we make the replacement
 $t_4'=t_4-t_8, t_5'=t_5-t_8, t_7'=t_7-t_8$ in  \eq{tsol} and \eq{dD5b}.
 
 \subsection{Feynman and CHY numerators}
 Recall the discussion in Sec.~IID about the structure of $n_a$. For $N=5$, it contains terms of the form
 $bccc$ and $bbcs$, each type possessing many different terms that cannot be combined. That observation
of course applies both to $\bn_a$ computed from Feynman diagrams, and $n_a'$ computed from the CHY
theory. Since the CK relation is satisfied by $n_a'$, the difference $\d n_a=\bn_a-n'_a$ must take on the
form $p_as_a+p'_as'_a$ if it is a local gauge transformation. In other words, their $bccc$ terms must
completely cancel, and the difference of their $bbcs$ terms must involve only $s_a$ and $s'_a$, without
any other $s_I$. Conversely, by computing $\bn_a-n'_a$ explicitly, one can verify whether 
it is of the form $p_as_a+p'_as'_a$ and therefore whether $\d n_a$ is
indeed a local gauge transformation.

Using the Mathematica program in Ref.~\cite{ET20} to compute $n'_a$, and then $\bn_a-n'_a$
explicitly, I have verified that the difference is indeed given by a local gauge transformation.
Namely, $\d n_a$ is of the form $p_as_a+p'_as'_a$. The coefficients $p_a$ and $p'_a$ are rather
complicated and will not be reproduced here, but they are polynomials of $b$ and $c$.
This provides an independent confirmation of the general assertion that a local gauge transformation
can bring $\bn_a$ into a set of $n_a$ that satisfy the CK relation.

\section{Six-Point Amplitude}
\subsection{partial amplitudes and numerators}
$N=6$  can be dealt with in much the same way as $N=5$, but everything is much lengthier.
For that reason, many details are  relegated to the appendices.

There are now $4!=24$ partial amplitudes,  listed in Table B1 of Appendix B. There are 25 single Mandelstam variables $s_e$
 listed in
Table B2, 105 double Mandelstam variables used to enumerate $\D_l$, listed in Table B3 , and 
105
triple Mandelstam variables used to enumerate $n_a/Q_a$, listed in Table B4. Of the 105 possible terms $n_a/Q_a$, only
14 are contained in each partial amplitude. For example, the amplitude $A_1$ is given by
\be
A_1=A(123456)&=&{n_{2}\over s_{12}s_{34}s_{345}}+{n_{3}\over s_{12}s_{34}s_{1234}}+{n_{7}\over s_{12}s_{45}s_{123}}+{n_{8}\over s_{12}s_{45}s_{345}}+{n_{10}\over s_{12}s_{123}s_{1234}}\nn\\
&+&{n_{61}\over s_{23}s_{45}s_{123}}+{n_{63}\over s_{23}s_{45}s_{2345}}+{n_{64}\over s_{23}s_{123}s_{1234}}+{n_{66}\over s_{23}s_{234}s_{1234}}+{n_{67}\over s_{23}s_{234}s_{2345}}\nn\\
&+&{n_{90}\over s_{34}s_{234}s_{1234}}+{n_{91}\over s_{34}s_{234}s_{2345}}+{n_{93}\over s_{34}s_{345}s_{2345}}
+{n_{105}\over s_{45}s_{345}s_{2345}}.\labels{A6}\ee
The others can be obtained by a suitable permutation of lines 2,3,4,5, together with the appropriate renaming of $n_a$, and a possible minus sign determined by the number of flips as discussed in Sec.~IIC.

Three-gluon Feynman diagrams for $\bn_a$ are shown in Fig.~B1 of Appendix B.
On top of those, diagrams
containing virtual vertices must be added, like those in the two bottom  rows of Fig.~11. However, this time there are four additional rows,
three containing a single dotted line, and one containing two dotted lines.

$N=6$ is the first time when a 3g vertex with three {\it internal} lines appear. It can be seen  in Fig.~B1
that the 15 $\bn_a$ with
\be a\in I_a=\{3, 6, 9, 18, 21, 24, 33, 36, 39, 48, 51, 54, 63, 72, 81\}\labels{In}\ee
contain such an internal vertex, and the remaining 90 $\bn_a$ with
\be a\in J_a&=&\{1, 2, 4, 5, 7, 8, 10, 11, 12, 13,14, 15, 16, 17, 19,\nn\\
 &&\ 20, 22, 23, 25, 26,27, 28, 29, 30, 31, 32, 34, 35, 37, 38, \nn\\
 &&\ 40, 41, 42, 43, 44, 45, 46, 47, 49, 50, 52, 53, 55, 56, 57,\nn\\
 &&\ 58, 59, 60, 61, 62, 64, 65, 66,67, 68, 69, 70, 71, 73, 74, \nn\\
&&\ 75, 76, 77, 78, 79, 80, 82, 83, 84, 85, 86, 87, 88, 89, 90,\nn\\
&&\ 91, 92, 93, 94, 95, 96, 97, 98, 99, 100, 101, 102, 103, 104, 105\}\labels{cIn}\ee
do not. Since two dotted lines are not allowed to intersect, those $\bn_a$ in $I_a$ do not contain any diagram
with two virtual vertices, or two dotted lines, but those in $J_a$ do.

\subsection{CK combinations and trivial CK identities}
The 105 CK combinations $\D_l=n_s-n_t-n_u$ are listed in Table B5, with $s, t, u$ specified in row B. The number without a bar
on top is $s$, and the other two numbers with a bar on top are $t$ and $u$. The $Q_a$ of those three $n_a$ all contain a common
double Mandelstam variable, listed in row A. For example, $\D_1=n_3-n_2-n_1$, and  $Q_3, Q_2, Q_1$ all contain 
$s_{12}s_{34}=s_1s_8$, a fact that can be checked directly using Tables B1, B2, B3.

The 105 $\D_l$'s can be divided into groups containing a single common Mandelstam variable $s_e$ in every one of
their $n_a$.  These $n_a$ occur twice in each group, enabling the $\D_l$ in the same group to be combined into trivial 
CK identities. These trivial CK identities are valid on-shell
and off-shell because they are identically zero, no matter what $n_a$ are. There is only one such relation for $N=5$, but for $N=6$, 
there are
25 $s_e$'s, so there are 25  trivial CK identities. They are listed below, with the common $s_e$ of
the group shown in parenthesis.

\be 0&=& -\D_{1} + \D_{2} + \D_{3} + \D_{4} - \D_{5} + \D_{6} - \D_{7} - 
     \D_{8} + \D_{9}-\D_{10},\quad (s_{12}) \nn\\
   0&=& -\D_{11} + \D_{12} + \D_{13} + \D_{14} - \D_{15} + \D_{16} - \D_{17} - 
     \D_{18} + \D_{19}-\D_{20},\quad (s_{13})\nn\\
  0&=&
    \D_{22} + \D_{23} + \D_{24} + \D_{26} + 
     \D_{29} - (\D_{21} + \D_{25} + \D_{27} + \D_{28}+\D_{30}), \quad (s_{14})\nn\\
  0&=& -\D_{31} + \D_{32} + \D_{33} + \D_{34} - \D_{35} + \D_{36} - \D_{37} - 
     \D_{38} + \D_{39}-\D_{40},\quad (s_{15}) \nn\\
  0&=&-\D_{21} + \D_{31} + \D_{41} + \D_{42} - \D_{43} - \D_{44} + \D_{45} - 
     \D_{46} + \D_{47}-\D_{48},\quad (s_{23}) \nn\\
  0&=&-(\D_{11} - \D_{32} - \D_{49} - \D_{50} + \D_{51} + \D_{52} - \D_{53} + 
       \D_{54} - \D_{55}+\D_{56}),\quad (s_{24}) \nn\\
  0&=& -\D_{12} + \D_{22} + \D_{57} + \D_{58} - \D_{59} - \D_{60} + \D_{61} - 
     \D_{62} + \D_{63}-\D_{64}, (s_{25}) \nn\\
 0&=& 
    \D_{1} + \D_{33} + \D_{57} + \D_{65} - \D_{66} - \D_{67} + \D_{68} - \D_{69} - 
     \D_{70}-\D_{71},\quad (s_{34})\nn\\
     0&=& 
    \D_{2} + \D_{23} + \D_{49} + \D_{72} - \D_{73} - \D_{74} + \D_{75} - \D_{76} - 
     \D_{77}-\D_{78},\quad (s_{35})\nn\\ 
    0&=&
    \D_{3} - \D_{13} - \D_{41} - \D_{79} + \D_{80} + \D_{81} + \D_{82} + \D_{83} - 
     \D_{84}-\D_{85},\quad(s_{45})\nn\\
    0&=& -\D_{4} + \D_{14} + \D_{42} - \D_{79} + \D_{86}-\D_{87}, \quad (s_{123})\nn\\
  0&=& -\D_{5} + \D_{24} + \D_{50} - \D_{72} + \D_{88}-\D_{89},\quad(s_{124}) \nn\\
  0&=& -\D_{6} + \D_{34} + \D_{58} - \D_{65} + \D_{90}-\D_{91},\quad(s_{125}) \nn\\
 0&=& -\D_{15} + \D_{25} - \D_{59} + \D_{66} + \D_{92}-\D_{93},\quad(s_{134}) \nn\\
 0&=& -\D_{16} + \D_{35} - \D_{51} + \D_{73} + \D_{94}-\D_{95},\quad (s_{135}) \nn\\
0&=& -\D_{26} + \D_{36} - \D_{43} + \D_{80} + \D_{96}-\D_{97},\quad(s_{145}) \nn\\  
0&=&
     - \D_{37} - \D_{44}  + \D_{52}  + 
     \D_{67} +\D_{98}-\D_{99},\quad (s_{234}) \nn\\ 
       0&=& 
  - \D_{27}  - \D_{45} 
     + \D_{60}  + \D_{74} +\D_{100}
     -\D_{101}, \quad(s_{235})\nn\\  
      0&=&  -\D_{17} - \D_{53}  + \D_{61} + 
     \D_{81} + \D_{102} -\D_{103},\quad(s_{245}) \nn\\    
  0&=&\D_7-\D_{68}+\D_{75}+\D_{82}-\D_{104}-\D_{105},\quad(s_{345})\nn\\     
  0&=& 
    \D_{8} - \D_{18} + \D_{28} - \D_{46} + \D_{54} + \D_{69} - \D_{86} + \D_{88} - 
     \D_{92}-\D_{98}, \quad(s_{1234})\nn\\           
   0&=&  \D_{9}  - \D_{19} + \D_{38} - \D_{47} + \D_{62} + 
     \D_{76} - \D_{87} + \D_{90} - \D_{94}-\D_{100}, \quad (s_{1235})\nn\\   
      0&=&\D_{10} - \D_{29}  + \D_{39}-\D_{55} + \D_{63}  + \D_{83}-\D_{89}+\D_{91}  - \D_{96}-\D_{102},\quad(s_{1245} )\nn\\          
 0&=&  \D_{20}  - \D_{30} + \D_{40}- \D_{70}+ \D_{77}  + \D_{84} - \D_{93} + 
     \D_{95} - \D_{97}-\D_{104}, \quad(s_{1345})\nn\\
0&=&  \D_{48}-\D_{56}+\D_{64}-\D_{71}+\D_{78}+\D_{85}-\D_{99}+\D_{101}-\D_{103}-\D_{105}.\quad(s_{2345})\labels{primdel}\ee

\subsection{CK combinations for Feynman diagrams}
Recall from Sec.~IID that the $\bD_l$ terms are in the form $bcccc,\  bbccs,\  bbbss$. We shall see that the $bcccc$
terms are always zero on-shell,  but not the other two kinds.

$\bD_l$ are computed from Feynman diagrams shown in Fig.~B1 of Appendix B, plus similar diagrams containing virtual vertices. 
The computation is similar to those in \eq{A5a}, \eq{A5b}, and \eq{D52},  but much lengthier.
The result, shown in \eq{d1tod105} in Appendix B4,   
comes in two types, depending on whether $\bD_l$
contains an $\bn_a$ in $I_a$ or not. Those that do contain two double-dotted Feynman diagrams, and
have an $l$ index given in the list
\be
I_l&=&\{1, 2, 3, 8, 9, 10, 11, 12, 13, 18, 19, 20, 21, 22, 23,\nn\\
&&\ 28, 29, 30, 31, 32, 33, 38, 39, 40, 41, 46, 47, 48, 49, 54,\nn\\
&&\ 55, 56, 57, 62, 63, 64, 69, 70, 71, 76, 77, 78, 83, 84, 85\},\labels{Il}\ee
and those that do not contain three double-dotted Feynman diagrams, and have an $l$ index given in the list
\be J_l&=&\{4, 5, 6, 7, 14, 15, 16, 17, 24, 25, 26, 27, 34, 35, 36,\nn\\
&&\  37, 42, 43, 44, 45, 50, 51, 52, 53, 58, 59, 60, 61, 65, 66,\nn\\
&&\  67, 68, 72, 73, 74, 75, 79, 80, 81, 82, 86, 87, 88, 89, 90,\nn\\
&&\  91, 92, 93, 94, 95, 96, 97, 98, 99, 100, 101, 102, 103, 104, 105\}.\labels{Jl}\ee
$I_l$ has 45 members and $J_l$ has and 60 members.
They are illustrated below with one example 
of each type.

First consider $\bD_1=\bn_3-\bn_2-\bn_1$, which is an example of type $I_l$,  with Feynman diagrams shown in Fig.~13.
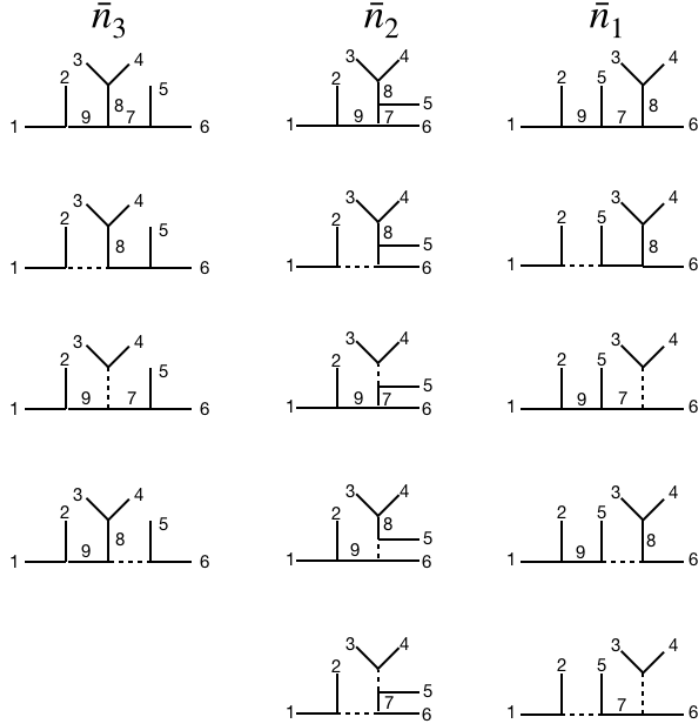
\bc\igwg{10}{Fig13}\\ Fig.~13.\quad Feynman diagrams for $\bD_1$\ec
Using Fig.~9 and \eq{D4} for off-shell $N=4$, the first four rows of Fig.~13 are
\be
&Rows 1\&4.&\quad d(9856)T(\834)T(\912)\nn\\
&Row 2.&\quad T(\834)\[f(12856)-f(12658)-f(12586)\]s_{34}=T(\834)F(12568)s_{34}\nn\\
&Row 3. &\quad T(\912)\[f(34965)-f(34569)+f(34659)\]s_{12}=T(\912)F(34659)s_{12}.\labels{dD11234}\ee

To accommodate the fifth row with two dotted lines,
a new function defined by Fig.~14 is needed. This function 
\be g(123456)=Q(1239)Q(\9456)=b_{23}(b_{16}  b_{4 5} - b_{1 5}  b_{46} )+ 
 b_{1 3} (-b_{2 6} b_{4 5} + b_{2 5} b_{4 6})\labels{g1}\ee
has the symmetry
\be g(123456)=-g(213456)=-g(123465)=g(654321),\labels{g2}\ee
and obeys the sum rule
\be g(123456)+g(123564)+g(123645)=0.\labels{g3}\ee
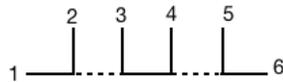
 \bc\igwg{4}{Fig14}\\ Fig.~14.\quad $g(123456)$\ec
What appears in $\bD_l$ is $g_A$, which is $g$ with the two middle arguments anti-symmetrized:
\be g_A(123456)=g(123456)-g(124356).\labels{g4}\ee
 In addition to having the symmetry properties of $g$ shown in \eq{g2}, it also obeys
 \be g_A(123456)=g_A(341256)=g_A(125634).\labels{g5}\ee

Putting these together, we get
\be\bD_1&=&s_{12}F(12568)T(\ol 834)+\[s_{34}F(34659)+d(9856)T(\834)\]T(\ol 912)+s_{12}s_{34}g_A(125634)\nn\\
&\simeq&s_{12}F(12568)T(\ol 834)+s_{34}F(34659)T(\ol 912)+s_{12}s_{34}g_A(125634).\labels{bD1a}\ee
The second line follows  because the $dTT\simeq 0$ on account of the Slavnov-Taylor identity.

Next, consider the type-$J_l$ example $\bD_4=\bn_{10}-\bn_7-\bn_{11}$, whose Feynman diagrams are shown in Fig.~15.
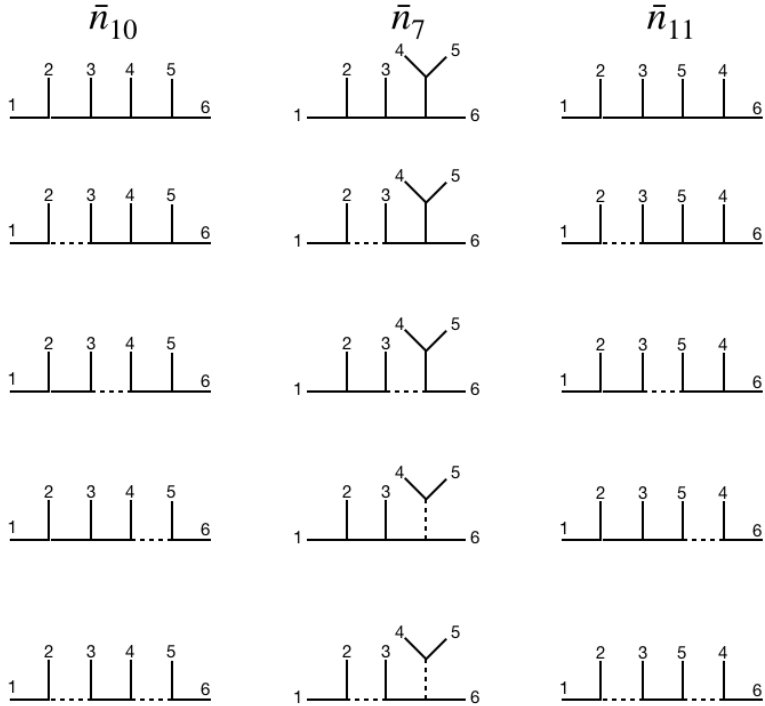
\bc\igwg{11.5}{Fig15}\\ Fig.~15.\quad Feynman diagrams for $\bD_4$\ec
The difference between this type and the $I_l$ type can be spotted in the fifth row of the diagram. There are now three double-dotted diagrams
instead of the previous two in type $I_l$.  

In the present case,  the diagrams in Fig.~15 are
\be
&Rows 1\&4.&\quad d(9456)P({12}3\9)\nn\\
&Row 2\&5.&\quad d(9456)Q(123\9)s_{12}\nn\\
&Row 3. &\quad T(\912)\[f(93456)-f(93654)-f(93546)\]s_{123}=T(\912)F(93456)s_{123},\ee
where $P(1234)$ is given in Fig.~9 and Fig.~16(a), with its $bbs$ terms  shown in Fig.~16(b).
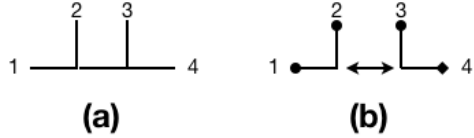
\bc\igwg{7}{Fig16}\\ Fig.~16.\quad (a) the functions $P(1234)$,\ (b) the $bbs$ part of $P(1234)$ \ec
Collecting the results, we get
\be
\bD_4&=&d(9456)\[P(123\9)+Q(123\9)s_{12}\]+T(\912)F(93456)s_{123}\nn\\
&=&d(9456)A_4(123\9)+T(\912)F(93456)s_{123}.\labels{dD412345}\ee
Note that the double dotted diagrams in row 5 have conveniently been combined with those in row 2
to make things simple. As a result, there are no $ss$ terms in \eq{dD412345}. A different
way to explain the absence of the $ss$ terms is to appeal to \eq{g3}, which shows that the three
double dotted diagrams add up to zero.

By a direct calculation, one sees that the $bcccc$ terms of $d(9456)A_4(123\9)$ add up to zero on-shell.
This is expected from the numerator Slavnov-Taylor identity because the $bcccc$ terms of $A_4(123\9)$
are the residue of $A(123\9)$ at the pole $1/s_{12}$. See \eq{A1234a} and \eq{A4}. The remaining on-shell $bbccs$ terms can be computed
with the help of \eq{A4} to \eq{D4t} to be 
\be d(9456)A_4(123\9)&\simeq&c_{99}T(456)A_4(123\9)\simeq -T(456)T(123)s_{12}.\labels{dA4}\ee
The full Slavnov-Taylor identity can  be used to verify the correctness of this relation. The identity asserts that 
$A(1234)=A_4(1234)/s_{12}+A_4(2341)/s_{23}=A_4(1234)/s_{12}-A_4(2314)/s_{23}$ must be zero on-shell when $\e_4$ is replaced by $k_4$. As a consequence, $\e_4\to k_4$ on $A_4(1234)$ must produce a result
proportional to $s_{12}$, with a proportionality constant   invariant under a cyclic permutation
of 1,2,3. This is precisely what \eq{dA4} gives.

 Caution should be exercised in using \eq{dA4}. $T(pqr)$ is given by \eq{T}, but there is no momentum conservation between the three lines $p,q,r$.

Substituting \eq{dA4} into \eq{dD412345}, one gets
\be \bD_4&\simeq&-s_{12}T(123)T(456)+T(\912)F(93456)s_{123}.\labels{dD412345a}\ee

Let $d_es_e+d'_es'_e$ be the on-shell expression of those $\bD_l$ terms linear in $s$, when $\bD_l$ is labelled by the double
Mandelstam variable $P_l=s_es'_e$ with $s_e\prec s'_e$. Then $d_e, d'_e$ can be read off from \eq{d1tod105}
of Appendix B, together with manipulations similar to \eq{bD1a} and \eq{dD412345a}. The result is 
shown in Table B6 of Appendix B.

\subsection{Local gauge transformations}
The vanishing of the $bcccc$ terms in $\bD_l$ leaves it proportional to $s$ and $ss$.
That gives hope that a {\it local} gauge transformation of the type given in Sec.~IIF,
\be \d n_a=p_as_a+p'_as'_a+p''_as''_a+q_as'_as''_a+q'_as_as''_a+q''_as_as'_a,\labels{pq6}\ee
may be able to  to bring the set $\bn_a$ to a new set  $n_a$ satisfying the CK relation $\D_l=\bD_l-\d\D_l=0$.
The purpose of this subsection and the next two is to investigate whether this hope is realized.

As discussed in Sec.~IIF, each of the $p$ and the $q$ parameters 
must satisfy a set of gauge constraint and a set of
CK equations. It turns out that there are many $p$ and many $q$ equations.
We shall find that neither the $p$ equations nor the $q$ equations have consistent solutions,  hence the
gauge transformation used to restore CK relation must be non-local. To verify this conclusion independently,
and to understand 
why that happens, we shall also 
compute the numerators $n'_a$ from the CHY theory. Since the CHY theory is known to satisfy the CK relations.
this gauge transformation $\d n_a=\bn_a-n'_a$ must be non-local in order to agree with the conclusion above.  This is indeed the case because $\d n_a$ contains  $s$ dependences beyond those allowed by \eq{pq6}.

\subsection{$\boldsymbol{p}$ equations}
The $p$ parameters of \eq{pq6} give rise to a change
\be
{\d n_a\over Q_a}={p_a\over s'_as''_a}+{p'_a\over s_as''_a}+{p''_a\over s_as''_a}.\labels{p61}\ee
Similar to the situation in $N=5$, in order to keep all $\d A_m=0$, the $3\x105=315$ $p$ parameters must form 105 triplets $t_l$,
with the $p$ parameters inside each triplet equal. As in $N=5$, each triplet  is a associated with a $\D_l$.
These triplets are
\be
&&t_1=(p''_{3}, -p''_{2}, -p''_{1}),\quad  t_2=(p''_{6}, -p''_{5}, -p''_{4}),\quad t_3=(p''_{7},-p''_{8}, -p''_{9}),
 \quad t_4=(p''_{10}, -p'_{7}, -p''_{11}), \nn\\
&& t_{5}=(p''_{12}, -p'_{4}, -p''_{13}), \quad t_{6}=(p''_{14}, -p'_{1}, -p''_{15}), \quad t_{7}=(p'_{2}, -p'_{5}, -p'_{8}), \quad t_{8}=(p'_{10}, -p'_{3}, -p'_{12}), \nn\\
&& t_{9}=(p'_{11}, -p'_{6}, -p'_{14}), \quad t_{10}=(p'_{13}, -p'_{9}, -p'_{15}), \quad t_{11}=(p''_{18}, -p''_{17}, -p''_{16}),\quad t_{12}= (p''_{21}, -p''_{20}, -p''_{19}), \nn\\
&& t_{13}=(p''_{22}, -p''_{23}, -p''_{24}), \quad t_{14}=(p''_{25}, -p'_{22}, -p''_{26}), \quad t_{15}=(p''_{ 27}, -p'_{19}, -p''_{28}), \quad t_{16}=(p''_{29}, -p'_{16}, -p''_{30}),\nn\\
&& t_{17}=(p'_{17}, -p'_{20}, -p'_{23}),\quad t_{18}= (p'_{25}, -p'_{18}, -p'_{27}), \quad t_{19}=(p'_{26}, -p'_{21}, -p'_{29}), \quad t_{20}=(p'_{28}, -p'_{24}, -p'_{30}),  \nn\\
&& t_{21}=(p''_{33}, -p''_{32}, -p''_{31}),\quad t_{22}=(p''_{36}, -p''_{35}, -p''_{34}), \quad t_{23}=(p''_{37}, -p''_{38}, -p''_{39}), \quad t_{24}=(p''_{40}, -p'_{37}, -p''_{41}), \nn\\
&&t_{25}= (p''_{42}, -p'_{34}, -p''_{43}), \quad t_{26}= (p''_{44}, -p'_{31}, -p''_{45}), \quad t_{27}=(p'_{32}, -p'_{35}, -p'_{38}), \quad t_{28}=(p'_{40}, -p'_{33}, -p'_{42}),  \nn\\
&& t_{29}=(p'_{41}, -p'_{36}, -p'_{44}),\quad t_{30}=(p'_{43}, -p'_{39}, -p'_{45}), \quad t_{31}=(p''_{48}, -p''_{47}, -p''_{46}), \quad t_{32}=(p''_{51}, -p''_{50}, -p''_{49}),  \nn\\
&& t_{33}=(p''_{52}, -p''_{53}, -p''_{54}), \quad t_{34}=(p''_{55}, -p'_{52}, -p''_{56}), \quad t_{35}=(p''_{57}, -p'_{49}, -p''_{58}), \quad t_{36}=(p''_{59}, -p'_{46}, -p''_{60}),  \nn\\
&&t_{37}=(p'_{47}, -p'_{50}, -p'_{53}),\quad t_{38}= (p'_{55}, -p'_{48}, -p'_{57}), \quad t_{39}=(p'_{56}, -p'_{51}, -p'_{59}), \quad t_{40}=(p'_{58}, -p'_{54}, -p'_{60}), \nn\\
&& t_{41}=(p''_{61}, -p''_{62}, -p''_{63}), \quad t_{42}= (p''_{64}, -p'_{61}, -p''_{65}), \quad t_{43}=(p_{31}, -p_{46}, -p'_{62}), \quad t_{44}=(p''_{66}, -p_{47}, -p''_{67}),  \nn\\
&&t_{45}=(p''_{68}, -p_{32}, -p''_{69}), \quad  t_{46}= (p'_{64}, -p_{33}, -p'_{66}), \quad t_{47}=(p'_{65}, -p_{48}, -p'_{68}), \quad t_{48}=(p'_{67}, -p'_{63}, -p'_{69}), \nn\\
&& t_{49}=(p''_{70}, -p''_{71}, -p''_{72}),\quad t_{50}= (p''_{73}, -p'_{70}, -p''_{74}), \quad t_{51}=(p_{16}, -p_{49}, -p'_{71}), \quad t_{52}=(p''_{75}, -p_{50}, -p''_{76}),  \nn\\
&&t_{53}=(p''_{77}, -p_{17}, -p''_{78}),\quad t_{54}= (p'_{73}, -p_{18}, -p'_{75}), \quad t_{55}=(p'_{74}, -p_{51}, -p'_{77}), \quad t_{56}=(p'_{76}, -p'_{72}, -p'_{78}), \nn\\
&& t_{57}=(p''_{79}, -p''_{80}, -p''_{81}),\quad t_{58}= (p''_{82}, -p'_{79}, -p''_{83}), \quad t_{59}=(p_{19}, -p_{34}, -p'_{80}), \quad t_{60}=(p''_{84}, -p_{35}, -p''_{85}), \nn\\
&&t_{61}= (p''_{86}, -p_{20}, -p''_{87}),\quad t_{62}= (p'_{82}, -p_{21}, -p'_{84}), \quad t_{63}=(p'_{83}, -p_{36}, -p'_{86}), \quad t_{64}=(p'_{85}, -p'_{81}, -p'_{87}),\nn\\
&& t_{65}= (p_{1}, -p_{52}, -p_{79}), \quad t_{66}=(p''_{88}, -p_{80}, -p''_{89}), \quad t_{67}=(p''_{90}, -p_{53}, -p''_{91}), \quad t_{68}=(p_{2}, -p''_{92}, -p''_{93}),  \nn\\
&&t_{69}=(p_{3}, -p'_{88}, -p'_{90}),\quad t_{70}=(p'_{89}, -p_{54}, -p'_{92}),\quad t_{71}= (p'_{91}, -p_{81}, -p'_{93}), \quad t_{72}=(p_{4}, -p_{37}, -p_{70}),\nn\\
&&t_{73}= (p''_{ 94}, -p_{71}, -p''_{95}),\quad t_{74}= (p''_{96}, -p_{38}, -p''_{97}), \quad t_{75}=(p_{5}, -p''_{98}, -p''_{99}), \quad t_{76}=(p_{6}, -p'_{94}, -p'_{96}),  \nn\\
&&t_{77}=(p'_{95}, -p_{39}, -p'_{98}),\quad t_{78}=(p'_{97}, -p_{72}, -p'_{99}), \quad t_{79}=(p_{7}, -p_{22}, -p_{61}), \quad t_{80}=(p''_{100}, -p_{62}, -p''_{101}),  \nn\\
&&t_{81}=(p''_{102}, -p_{23}, -p''_{103}), \ t_{82}=(p_{8}, -p''_{104}, -p''_{105}), \ t_{83}=(p_{9}, -p'_{100}, -p'_{102}), \ t_{84}=(p_{24}, -p'_{101}, -p'_{104}),  \nn\\
&&t_{85}=(p_{63}, -p'_{103}, -p'_{105}),\quad t_{86}= (p_{10}, -p_{25}, -p_{64}), \quad t_{87}=(p_{11}, -p_{26}, -p_{65}),\quad t_{88}= (p_{12}, -p_{40}, -p_{73}),\nn\\
&&t_{89}= (p_{13}, -p_{41}, -p_{74}),\quad t_{90}= (p_{14}, -p_{55}, -p_{82}),\quad t_{91}= (p_{15}, -p_{56}, -p_{83}),\quad t_{92}= (p_{27}, -p_{42}, -p_{88}), \nn\ee
\be
&& t_{93}=(p_{28}, -p_{43}, -p_{89}), t_{94}= (p_{29}, -p_{57}, -p_{94}),\quad t_{95}= (p_{30}, -p_{58}, -p_{95}),\quad t_{96}= (p_{44}, -p_{59}, -p_{100}),  \nn\\
&&t_{97}=(p_{45}, -p_{60}, -p_{101}),t_{98}= (p_{66}, -p_{75}, -p_{90}), \quad t_{99}=(p_{67}, -p_{76}, -p_{91}), \quad t_{100}=(p_{68}, -p_{84}, -p_{96}),  \nn\\
&&t_{101}=(p_{69}, -p_{85}, -p_{97}),\ t_{102}= (p_{77}, -p_{86}, -p_{102}),\ t_{103}= (p_{78}, -p_{87}, -p_{103}),\ t_{104}= (p_{92}, -p_{98}, -p_{104}),\nn\\
&&t_{105}= (p_{93}, -p_{99}, -p_{105}).\labels{p6t}\ee
For example, the $t_1$ triplet contains $p$ parameters associated with $n_3, n_2, n_1$, all components of $\D_1$, and the $t_4$ triplet contains $p$ parameters  associated with $n_{10}, n_7, n_{11}$, all  components
of $\D_4$. The equality of  triplet parameters comes from the requirement $\d A_m=0$. 
For example, in order for  $\d A_1=0$, all $\d n_a/Q_a$ terms must  cancel. To have the coefficients
of $1/s_{12}s_{34}$ cancel, $p''_3=-p''_2$ is required. To have the coefficients of $1/s_{12}s_{123}$
to cancel, $p''_{10}=-p'_7$ is required. 

In order to have $\D_l=\bD_l-\d\D_l=0$ on-shell for every $l$, the $p$ parameters must obey a number of CK equations. For example, 
\be
\d\D_1&=&\d(n_3-n_2-n_1)\nn\\
&=&s_{12}(p_3-p_2-p_1)+s_{34}(p'_3-p'_2-p'_1)+(p''_3s_{1234}-p''_2s_{345}-p''_1s_{125})\nn\\
&=&s_{12}(t_{69}-t_{68}-t_{65})+s_{34}(-t_8-t_7+t_6)+t_1(s_{1234}+s_{345}+s_{125})\nn\\
&=&s_{12}(t_{69}-t_{68}-t_{65})+s_{34}(-t_8-t_7+t_6)+t_1(s_{12}+s_{34})\nn\\
&=&s_{12}(t_{69}-t_{68}-t_{65}+t_1)+s_{34}(-t_8-t_7+t_6+t_1),\labels{dd1p6a}\ee
where \eq{p6t} and \eq{stu} have been used to get to the final result.

Excluding the $ss$ term which is associated with the $q$ parameter, the on-shell terms in $\bD_1$ 
shown in \eq{bD1a} are
\be\bD_1\simeq s_{12}F(12569)T(\ol 934)+s_{34}F(34659)T(\ol 912)+O(s^2).\nn\ee 
A matching the $s$-dependent coefficients of \eq{dd1p6a} and \eq{bD1a} yields
\be
t_{69}-t_{68}-t_{65}+t_1&=&F(12569)T(\ol 934):=d_1,\nn\\
-t_8-t_7+t_6+t_1&=&F(34659)T(\ol 912):=d'_1.\labels{dd1p6c}\ee
These are the two equations associated with the requirement $\d\D_1=\bD_1$. 
Similarly,
\be
\d\D_4&=&\d(n_{10}-n_7-n_{11})\nn\\
&=&s_{12}(p_{10}-p_7-p_{11})+s_{123}(p'_{10}-p''_7-p''_{11})+(p''_{10}s_{1234}
-p'_7s_{45}-p'_{11}s_{1235})\nn\\
&=&s_{12}(t_{86}-t_{79}-t_{87})+s_{123}(t_8-t_3-t_9)+t_4(s_{1234}+s_{45}+s_{1235})\nn\\
&=&s_{12}(t_{86}-t_{79}-t_{87})+s_{123}(t_8-t_3-t_9)+t_4s_{123}\nn\\
&=&s_{12}(t_{86}-t_{79}-t_{87}+t_4)+s_{123}(t_8-t_3-t_9).\labels{dd4p6a}\ee
This differs from $\d\D_1$ in that $t_4$ is added to one of the $s$ terms in the last line, but not the other $s$ term. 

The $s$-dependence of $\d\D_4$ matches the $s$-dependence of $\bD_4$ given by
\eq{dD412345a} to be
\be \bD_4\simeq -s_{12}T(123)T(456)+T(\912)F(93456)s_{123}+O(s^2).\nn\ee
As a result,
\be  t_{86}-t_{79}-t_{87}+t_4&=&-T(123)T(456):=d_4,\nn\\
 t_8-t_3-t_9&=&T(\912)F(93456):=d'_4.\labels{dd4p6c}\ee
 
More generally,  let the $\bD_l$ terms linear in $s$ to be designated as $d_es_e+d'_es'_e$,
where $P_l=s_es'_e$ is the double Mandelstam variable specifying $\bD_l$, with $s_e\prec s'_e$.
The $2\x105=210$ equations for  $\d\D_l=\bD_l$ are listed in Appendix C. 
For $l\in I_l$, the equations resemble \eq{dd1p6c}. For $l\in J_l$, they resemble \eq{dd4p6c}.

These equations can written 
in a matrix form similar to \eq{teq}, 
\be \sum_{i'=1}^{105}(\tau)_{ii'}t_{i'}=\ol d_{i},\labels{teq6}\ee
where $i=(2l-1,2l)$ with $1\le l\le 105$, and $\ol d_i=(d_e,d'_e)$.

With 105 unknowns $t_l$ and 210 equations in \eq{teq6}, one might expect to have no solution unless 
the $210\x105$ matrix $\tau$ is highly degenerate. In fact, it is, with  a rank $79$. It has 131 left null vectors
$u_x\ (1\le x\le 131)$ so that $\sum_{i=1}^{210}(u_x)_i(\tau)_{ii'}=0$. As a result, \eq{teq6} has a solution only
when all the following 131 sum rules are satisfied:
\be
R_x=\sum_{i=1}^{210}(u_x)_i\bar d_i=0.\labels{Rx6}\ee
Some of these sum rules have two terms, some four terms,  some six terms, but some have many
more terms. The shorter sum rules are often satisfied, but the longer ones are generally not. As a result, \eq{teq6} has no solution. Details are shown in Appendix C.

\subsection{$\boldsymbol{q}$ equations}
The $q$ parameters of \eq{pq6} give rise to a change
\be
{\d n_a\over Q_a}={q_a\over s_a}+{q'_a\over s'_a}+{q''_a\over s''_a}.\labels{q1}\ee
To be a generalized gauge transformation, it is necessary to keep all $\d A_m=0$, thus the $3\x105=315$ $q$ parameters must satisfy many gauge-constraint equations. 
Unlike the  $p$ parameters, the constraint equations for $q$ are much more complicated.
For example, in order to cancel the $1/s_{12}$ terms in $\d A_1$, it follows from \eq{A6} that one needs
\be q_2+q_3+q_7+q_8+q_{10}=0,\labels{qeq1}\ee
and in order to cancel the $1/s_{345}$ terms in $\d A_1$, one needs
\be q''_2+q''_8+q'_{93}+q'_{105}=0.
\labels{qeq2}\ee
Since each $n_a$ is often contained in several $A_m$'s, these equations are generally coupled.
If we go through the cancellation of every $1/s$ term for every $\d A_m$, then there are 144 such homogeneous
equations for the 315 $q$ parameters to satisfy. These 144 constraint equations are listed in \eq{C1} in Appendix D.

On top of that, $q$ must also satisfy the many CK equations implementing $\d\D_l=\bD_l$. For example, the equation for $\d\D_1=\bD_1$ is
\be \bD_1&=&s_{12}s_{34}g_A(125634):=s_{12}s_{34}g_1\nn\\
&=&\d\D_1=\d(n_3-n_2-n_1)\nn\\
&=&(q''_3-q''_2-q''_1)s_{12}s_{34}+s_{12}(q'_3s_{1234}-q'_2s_{345}-q'_1s_{125})\nn\\
&+&s_{34}(q_3s_{1234}-q_2s_{345}-q_1s_{125}).\labels{dbd1}\ee
For the $s$ dependence of $\d\D_1$ and $\bD_1$ to match, one way is put $q_1=q_1'=q_2=q_2'=q_3=q_3'=0$ so that the
last two terms vanish. There is however potentially a more general solution. 
Using the on-shell version of the kinematical identity \eq{stu},
\be s_{1234}+s_{345}+s_{125}=s_{12}+s_{34},\nn\ee
the last line can be reduced to
\be
(q''_3-q''_2-q''_1)s_{12}s_{34}+q'_3s_{12}(s_{12}+s_{34})
+q_3s_{34}(s_{12}+s_{34}),\nn\ee
provided $q_3'=-q'_2=-q'_1$ and $q_3=-q_2=-q_1$, but a priori neither set has to vanish. Unfortunately
this manipulation does not help because it produces extra terms proportional to
$s_{12}^2$ and $s_{34}^2$, which could be avoided only when $q'_3=q_3=0$. In other words, in order to satisfy \eq{dbd1}, it 
is necessary to have
\be
g_1=g_A(125634)&=&q''_3-q''_2-q''_1,\qquad {\rm and}\labels{dbd1a1}\\
0&=&q'_3=q_3=q'_2=q_2=q'_1=q_1.\labels{dbd1a2}\ee

Although this manipulation does not pan out here, it does work for  $\d\D_4=\bD_4$.
\be
\bD_4&=&0:=s_{12}s_{123}g_4=\d(n_{10}-n_{7}-n_{11})\nn\\
&=&s_{12}s_{123}(q''_{10}-q'_7-q''_{11})+s_{12}(q'_{10}s_{1234}-q''_7s_{45}-q'_{11}s_{1235})\nn\\
&+&s_{123}(q_{10}s_{1234}-q_7s_{45}-q_{11}s_{1235}). \labels{bdb4}\ee
Again one requires  $q'_{10}=-q''_7=-q'_{11}$ and $q_{10}=-q_7=-q_{11}$ to 
be able to combine the extra $s$ terms, but this time
\be 
s_{1234}+s_{45}+s_{1235}=s_{123}\ee
results in one term instead of two, thus avoiding one of the two quadratic $s$ terms appearing above.
 As a result, as long as $q_{10}=-q_7=-q_{11}=0$, the $s$ dependence
would match,  \eq{bdb4} would be satisfied if
\be
g_4=0&=&q''_{10}-q'_7-q''_{11}+q'_{10},\labels{bdb4a1}\\
0&=&q_{10}=-q_7=-q_{11},\labels{bdb4a2}\\
q'_{10}&=&-q''_7=-q'_{11}.\labels{bdb4a3}\ee
The parameters in the last line must be equal but they do not have to vanish.

In general, the $q$ parameters for $\d\D_l=\bD_l$ need to satisfy a relation similar to those satisfied by
$\bD_1$ if $\bD_l$ is of type $I_l$, or a relation similar to those satisfied by $\bD_4$ if $\bD_l$ is type $J_l$. 
These relations
come in three groups: $X,Y,Z$. Group $X$ consists of equations like \eq{dbd1a1} and \eq{bdb4a1},
group $Y$ consists of equations like \eq{bdb4a3}, with equal $q$ parameters but not necessarily zero,
and group $Z$ consists of equations like \eq{dbd1a2} and \eq{bdb4a1}, with zero $q$ parameters.
The number of equations in groups $X,Y,Z$ are respectively (105, 120, 225).
They are listed in Appendix D.

On top of these, remember that there are also 144 equations derived from the gauge constraints
to be satisfied. With so many 
equations  and only $3\x105=315$ unknowns $q_a, q'_a, q''_a$, one might expect to  have no solutions.
That is indeed the case. Details are shown in Appendix D.

\subsection{Non-local gauge transformation between Feynman and CHY numerators}
In order to confirm the conclusion that there is no {\it local} gauge transformation moving the Feynman numerators
$\bn_a$ to a set of $n_a$ satisfying the CK relations, because neither the $p$ equations nor the
$q$ equations have solutions, and in order to understand why this is so,
we  calculate explicitly the 
CHY fundamental numerator factor $\n(2345)$. It is equal to the CHY ordinary numerator factor $n'_{10}$,
so the gauge transformation $\d n_{10}=\bn_{10}-n'_{10}$ should be non-local in order to agree with
the general conclusion.
If local, the terms of $\d n_{10}$ linear in $s$ may depend only 
on $s_{12}, s_{123}=s_{12}+s_{13}+s_{23}$, and $s_{1234}=s_{56}$. Its terms quadratic in $s$ may depend
on the three non-diagonal quadratic  products of these three $s$, and its $s$-independent (or $bcccc$)
terms must be zero. If any of these is violated,  $\d n_{10}$ is non-local.

Using the Mathematica program of Ref.~\cite{ET20} to compute $n'_{10}=\n(2345)$, and Feynman rules to compute
$\bn_{10}$, one indeed gets  no $bcccc$ terms in the resulting $\d n_{10}$. 
However, its linear $s$ dependence and its quadratic $s$ dependence
are much more complicated then those terms allowed above, so $\d n_{10}$
is indeed non-local.

\section{Conclusion}
Numerator factors $\bn_a$ and CK combinations $\bD_l=\bn_s-\bn_t-\bn_u$ have been computed
from the Feynman diagrams of $N$-particle pure gluon amplitudes. To do so, four-gluon
vertices must be converted into a pair of virtual cubic  vertices. The necessity for  virtual vertices to come in pairs implies that
their effect on $\bD_l$ cannot be fully realized until at least $N=6$. 

For $N=4$, the CK relation $\bD=0$ is valid on-shell.

For $N=5$, there is one trivial CK identity
and nine non-trivial CK combinations which are not zero even on-shell. However, a local (generalized)
gauge transformation can convert  every $\bD_l$ into some $\D_l$ that satisfy the CK relation, provided five sum rules
involving Feynman diagrams with one virtual vertex are obeyed. Explicit calculation shows that the sum rules
are fulfilled, thus proving that  $\bD_l$ can obey CK relations through a
local gauge transformation. This result is verified by computing the difference between Feynman and CHY
numerator factors.

For $N=6$, there are 25 trivial CK identities  and 80 non-trivial combinations $\bD_l$ which are not
zero on-shell. After a lengthy
calculation, one concludes  that there is no local (generalized) gauge transformation that can render the CK relations  valid. This conclusion  is confirmed also by computing the difference between the Feynman and CHY
numerator factors.

Locality is a fundamental attribute of quantum field theory that one would like to preserve. For scalar amplitudes,
locality is often deemed to be maintained when a scattering amplitude displays the same propagators
as the Feynman tree amplitude. Strictly speaking
this is correct only when it is also true off-shell.
For a gluon theory,
that is not sufficient, for the Yang-Mills coupling with local gauge invariance are also reflected in the numerator factors. Even with a local redefinition of fields or the insertion of total derivative terms, one
would still expect the numerator factors to be polynomials of $\e_i$ and $k_i$ if the $S$-matrix theory preserves
locality. Since Feynman diagrams derived from the Yang-Mills Lagrangian are by definition local, the generalized gauge
transformation between it and the $S$-matrix theory must also be local. 
The failure to do so
for the CHY gluon theory at $N=6$ raises the possibility that the inherent interaction in the CHY theory may not be local. 

The difficulty of extending the CHY gluon amplitude off-shell  may
be another circumstantial evidence suggesting its non-locality. 
The CHY scalar
theories can be extended off-shell  to agree
with the Feynman  amplitudes, and  to keep its  signature M\"obius invariance intact \cite{LY1,L20},
thus confirming the locality of the CHY scalar amplitudes. 
The CHY gluon amplitude can also be extended
off-shell maintaining M\"obius invariance \cite{L19a}, but that extension does not agree with the 
off-shell Feynman amplitude, and it does not satisfy the Slavnov-Taylor identity.

It is conceivable that the non-locality discussed above indeed  reflects an inherently non-local interaction mechanism in  $S$-matrix theories such as
the CHY theory, but it is perhaps also possible that this non-locality
can be fixed up by an ordinary gauge change, say from the Feynman gauge used here to
a $R_\xi$ gauge. 
Further investigation is required to answer this interesting question and to determine the true meaning of the non-locality found in this article.

I would like to thank York-Peng Yao for helpful discussions.

\appendix
\section{$\boldsymbol{N=5}$}
The canonical ordering of the 10 Mandelstam variables $s_e=s_I$ is shown in Table A1.
These are
all the $s_I$ that  enter into the double Mandelstam-variable expression for $Q_a$, but they
are not all independent. For example, $s_{123}=s_{12}+s_{13}+s_{23}$.

$$\ba{|c||c|c|c|c|c|c|c|c|c|c|}\hline
e&1&2&3&4&5&6&7&8&9&10\\ \hline
s_I&s_{12}&s_{13}&s_{14}&s_{23}&s_{24}&s_{34}&s_{123}&s_{124}&s_{134}&s_{234}\\ 
\hline\ea$$
\bc Table A1.\quad Ordered Mandelstam variables for $N=5$\ec

The canonical ordering of the 15 double Mandelstam variables $Q_a=s_{e_1}s_{e_2}$ is shown in
Table A2.

$$\ba{|c||c|c|c|c|c|}\hline
a&1&2&3&4&5\\ \hline
e_1,e_2&1,6&1,7&1,8&2,5&2,7\\ \hline\hline
a&6&7&8&9&10\\ \hline
e_1,e_2&2,9&3,4&3,8&3,9&4,7\\ \hline\hline
a&11&12&13&14&15\\ \hline
e1,e2&4,10&5,8&5,10&6,9&6,10\\
\hline\ea$$
\bc Table A2.\quad Ordered double Mandelstam variables for $N=5$\ec

A dictionary relating the numerator factor $n_a$ used in \eq{CK5a} and those used in Ref.~\cite{BCJ08} is given in Table A3.
$$\ba{|c|c|c|c|c|c|c|c|c|c|c|c|c|c|c|c|}\hline
\eq{CK5a}&n_1&n_2&n_3&n_4&n_5&n_6&n_7&n_8&n_9&n_{10}&n_{11}&n_{12}&n_{13}&n_{14}&n_{15}\\ \hline
{\rm Ref}.~[3] & n_3&n_1&n_{12}&-n_{10}&n_{15}&n_9&-n_7&n_{14}&n_6&n_{4}&n_{2}&n_{13}&n_{11}&-n_{8}&n_{5}\\
\hline\ea$$
\bc Table A3.\quad $n_a$ correspondence in \eq{CK5a} and in Ref.~\cite{BCJ08}\ec

\section{$\boldsymbol{N=6}$}
\subsection{Canonical ordering}
Table B1 gives the canonical ordering of the partial amplitudes $A_m=A(1\a N)$.
The canonical ordering of the 25 single Mandelstam variables $s_e=s_I$ is shown in Table B2. 
These are
all the $s_I$ that enter into the double Mandelstam-variable expression for $P_l$
and the triple Mandelstam-variable expression for $Q_a$, but they
are not all independent. For example, $s_{123}=s_{12}+s_{13}+s_{23}$.
The canonical ordering of the 105
double Mandelstam variables $P_l=s_{e_1}s_{e_2}$ is shown in Table B3.
The canonical ordering of the 105
triple Mandelstam variables $Q_a=s_{e_1}s_{e_2}s_{e_3}$ is shown in Table B4.

{\footnotesize
$$\ba{|c||c|c|c|c|c|c|c|c|}\hline
m&1&2&3&4&5&6&7&8\\ \hline
A(1\a N)&A(123456)&A(123546)&A(124356)&A(124536)&A(125346)&A(125436)&A(132456)&A(132546)\\ \hline\hline
m&9&10&11&12&13&14&15&16\\ \hline
A(1\a N)&A(134256)&A(134526)&A(135246)&A(135426)&A(142356)&A(142536)&A(143256)&A(143526))\\ \hline\hline
m&17&18&19&20&21&22&23&24\\ \hline
A(1\a N)&A(145236)&A(145326)&A(152346)&A(152436)&A(153246)&A(153426)&A(154236)&A(154326)\\ 
\hline\ea$$}
\bc Table B1.\quad Numbering of the partial amplitudes\ec

{\small
$$\ba{|c||c|c|c|c|c|c|c|c|c|c|c|c|c|c|c|}\hline
e&1&2&3&4&5&6&7&8&9&10&11&12&13&14&15\\ \hline
s_I&s_{12}&s_{13}&s_{14}&s_{15}&s_{23}&s_{24}&s_{25}&s_{34}&s_{35}&s_{45}&s_{123}&s_{124}&s_{125}&s_{134}&s_{135}\\ \hline\hline
e&16&17&18&19&20&21&22&23&24&25&&&&&\\ \hline
s_I&s_{145}&s_{234}&s_{235}&s_{245}&s_{345}&s_{1234}&s_{1235}&s_{1245}&s_{1345}&s_{2345}&&&&&\\
\hline\ea$$}
\bc Table B2.\quad Ordered Mandelstam variables $s_e=s_I$ for $N=6$\ec

{\small
$$\ba{|c||c|c|c|c|c|c|c|c|c|c|c|c|c|c|c|}\hline
l&1&2&3&4&5&6&7&8&9&10&11&12&13&14&15\\ \hline
e_1,e_2&1,8&1,9&1,10&1,11&1,12&1,13&1,20&1,21&1,22&1,23&2,6&2,7&2,10&2,11&2,14\\ \hline\hline
l&16&17&18&19&20&21&22&23&24&25&26&27&28&29&30\\ \hline
e_1,e_2&2,15&2,19&2,21&2,22&2,24&3,5&3,7&3,9&3,12&3,14&3,16&3,18&3,21&3,23&3,24\\ \hline\hline
l&31&32&33&34&35&36&37&38&39&40&41&42&43&44&45\\ \hline
e_1,e_2&4,5&4,6&4,8&4,13&4,15&4,16&4,17&4,22&4,23&4,24&5,10&5,11&5,16&5,17&5,18\\ \hline\hline
l&46&47&48&49&50&51&52&53&54&55&56&57&58&59&60\\ \hline
e_1,e_2&5,21&5,22&5,25&6,9&6,12&6,15&6,17&6,19&6,21&6,23&6,25&7,8&7,13&7,14&7,18\\ \hline\hline
l&61&62&63&64&65&66&67&68&69&70&71&72&73&74&75\\ \hline
e_1,e_2&7,19&7,22&7,23&7,25&8,13&8,14&8,17&8,20&8,21&8,24&8,25&9,12&9,15&9,18&9,20\\ \hline\hline
l&76&77&78&79&80&81&82&83&84&85&86&87&88&89&90\\ \hline
e_1,e_2&9,22&9,24&9,25&10,11&10,16&10,19&10,20&10,23&10,24&10,25&11,21&11,22&12,21&12,23&13,22\\ \hline\hline
l&91&92&93&94&95&96&97&98&99&100&101&102&103&104&105\\ \hline
e_1,e_2&13,23&14,21&14,24&15,22&15,24&16,23&16,24&17,21&17,25&18,22&18,25&19,23&19,25&20,24&20,25\\
\hline\ea$$}
\bc Table B3.\quad Ordered double Mandelstam variables $P_l=s_{e_1}s_{e_2}$ for $N=6$\ec

{\scriptsize\phantom{}\hskip-2.5cm
$\ba{|c||c|c|c|c|c|c|c|c|c|c|c|c|c|c|c|}\hline
a&1&2&3&4&5&6&7&8&9&10&11&12&13&14&15\\ \hline
e_1e_2e_3&1,8,13&1,8,20&1,8,21&1,9,12&1,9,20&1,9,22&1,10,11&1,10,20&1,10,23&1,11,21&1,11,22&1,12,2&1,12,23&1,13,22&1,13,23\\ \hline\hline
a&16&17&18&19&20&21&22&23&24&25&26&27&28&29&30\\ \hline
e_1e_2e_3&2,6,15&2,6,19&2,6,21&2,7,14&2,7,19&2,7,22&2,10,11&2,10,19&2,10,24&2,11,21&2,11,22&2,14,21&2,14,24&2,15,22&2,15,24\\ \hline\hline
a&31&32&33&34&35&36&37&38&39&40&41&42&43&44&45\\ \hline
e_1e_2e_3&3,5,16&3,5,18&3,5,21&3,7,14&3,7,18&3,7,23&3,9,12&3,9,18&3,9,24&3,12,21&3,12,23&3,14,21&3,14,24&3,16,23&3,16,24\\ \hline\hline
a&46&47&48&49&50&51&52&53&54&55&56&57&58&59&60\\ \hline
e_1e_2e_3&4,5,16&4,5,17&4,5,22&4,6,15&4,6,17&4,6,23&4,8,13&4,8,17&4,8,24&4,13,22&4,13,23&4,15,22&4,15,24&4,16,23&4,16,24\\ \hline\hline
a&61&62&63&64&65&66&67&68&69&70&71&72&73&74&75\\ \hline
e_1e_2e_3&{5,10,11}& {5,10,16}& {5,10,25}& {5,11,21}& {5,11,22}& {5,17,21}& {5,17,25}& {5,18,22}& {5,18,25}& {6,9,12}& {6, 9,
15}& {6,9,25}& {6,12,21}& {6,12,23}& {6,17,21}\\ \hline\hline
a&76&77&78&79&80&81&82&83&84&85&86&87&88&89&90\\ \hline
e_1e_2e_3&{6,17,25}& {6,19,23}& {6,19,25}& {7,8,13}& {7,8,14}& {7, 8,
25}& {7,13,22}& {7,13,23}& {7,18,22}& {7,18,25}& {7,19,23}& 
{7,19,25}& {8,14,21}& {8,14,24}& {8,17,21}\\ \hline\hline
a&91&92&93&94&95&96&97&98&99&100&101&102&103&104&105\\ \hline
e_1e_2e_3&{8,17,25}& {8,20,24}& {8,20,25}& {9,15,22}& {9,15,24}& {9,
18,22}& {9,18,25}& {9,20,24}& {9,20,25}& {10,16,23}& {10,
16,24}& {10,19,23}& {10,19,25}& {10,20,24}& {10,20,25}\\
\hline\ea$}
\bc Table B4.\quad Ordered triple Mandelstam variables $Q_a=s_{e_1}s_{e_2}s_{e_3}$  for $N=6$\ec

\subsection{Feynman diagrams}
Fig.~B1 shows the 105 diagrams whose propagators are $1/Q_a$.
\bc\igwg{14}{FigB1a}
\igwg{14}{FigB1b}
\igwg{14}{FigB1c}
\igwg{14}{FigB1d}
\igwg{14}{FigB1e}
\igwg{14}{FigB1f}
\igwg{14}{FigB1g}\\
Fig.~B1.\quad 3g Feynman Diagrams for $N=6$\ec

\subsection{CK combinations}
Table B5 gives the 105 CK combinations $\D_l=n_s-n_t-n_u$. Row A shows the double Mandelstam variable $s_{e_1}s_{e_2}$ common to 
$Q_s, Q_t, Q_u$. Row B shows the  $a$'s of the three $n_a$ making up of $n_s, n_t, n_u$. The one without a bar on top
is $n_s$, the other two are $n_t, n_u$ or $n_u, n_t$.  For example, the Table tells us that
$\D_1=n_3-n_1-n_2$, and that $Q_3, Q_1, Q_2$ all contain the common factor $s_{12}s_{34}$.

{\footnotesize 
$$\phantom{}\hskip-1cm\ba{|c|c|c|c|c|c|c|c|c|c|c|}\hline
&\D_1&\D_2&\D_3&\D_4&\D_5&\D_6&\D_7&\D_8&\D_9&\D_{10}\\ \hline
A&s_{12}s_{34}&s_{12}s_{35}&s_{12}s_{45}&s_{12}s_{123}&s_{12}s_{124}&s_{12}s_{125}&s_{12}s_{345}&s_{12}s_{1234}&s_{12}s_{1235}&s_{12}s_{1245}\\ \hline
B&(\overline1,\ol{2},3)&(\ol{4},\ol{5},6)&(7,\ol{8},\ol{9})&(\ol{7},10,\ol{11})&(\ol{4},12,\overline {13})&(\ol{1},14,\ol{15})&(2,\ol{5},\ol{8})&(\ol{3},10,\ol{12})&(\ol{6},11,\ol{14})&(\ol{9},13,\ol{15})\\ 
\hline\hline
&\D_{11}&\D_{12}&\D_{13}&\D_{14}&\D_{15}&\D_{16}&\D_{17}&\D_{18}&\D_{19}&\D_{20}\\ \hline
A&s_{13}s_{24}&s_{13}s_{25}&s_{13}s_{45}&s_{13}s_{123}&s_{13}s_{134}&s_{13}s_{135}&s_{13}s_{245}&s_{13}s_{1234}&s_{13}s_{1235}&s_{13}s_{1345}\\ \hline
B&(\ol{16},\ol{17},18)&(\ol{19},\ol{20},21)&(22,\ol{23},\ol{24})&(\ol{22},25,\ol{26})&(\ol{19},27,\ol{28})&(\ol{16},29,\ol{30})&(17,\ol{20},\ol{23})&(\ol{18},25,\ol{27})&(\ol{21},26,\ol{29})&(\ol{24},28,\ol{30})\\ \hline
\ea$$}
 
{\footnotesize 
$$\phantom{}\hskip-2.3cm\ba{|c|c|c|c|c|c|c|c|c|c|c|}\hline
&\D_{21}&\D_{22}&\D_{23}&\D_{24}&\D_{25}&\D_{26}&\D_{27}&\D_{28}&\D_{29}&\D_{30}\\ \hline
A&s_{14}s_{23}&s_{14}s_{25}&s_{14}s_{35}&s_{14}s_{124}&s_{14}s_{134}&s_{14}s_{145}&s_{14}s_{235}&s_{14}s_{1234}&s_{14}s_{1245}&s_{14}s_{1345}\\ \hline
B&(\ol{31},\ol{32},33)&(\ol{34},\ol{35},36)&(37,\ol{38},\ol{39})&(\ol{37},40,\ol{41})&(\ol{34},42,\ol{43})&(\ol{31},44,\ol{45})&(32,\ol{35},\ol{38})&(\ol{33},40,\ol{42})&(\ol{36},41,\ol{44})&(\ol{39},43,\ol{45})\\ \hline\hline
&\D_{31}&\D_{32}&\D_{33}&\D_{34}&\D_{35}&\D_{36}&\D_{37}&\D_{38}&\D_{39}&\D_{40}\\ \hline
A&s_{15}s_{23}&s_{15}s_{24}&s_{15}s_{34}&s_{15}s_{125}&s_{15}s_{135}&s_{15}s_{145}&s_{15}s_{234}&s_{15}s_{1235}&s_{15}s_{1245}&s_{15}s_{1345}\\ \hline
B&(\ol{46},\ol{47},48)&(\ol{49},\ol{50},51)&(52,\ol{53},\ol{54})&(\ol{52},55,\ol{56})&(\ol{49},57,\ol{58})&(\ol{46},59,\ol{60})&(47,\ol{50},\ol{53})&(\ol{48},55,\ol{57})&(\ol{51},56,\ol{59})&(\ol{54},58,\ol{60})\\ \hline\hline
&\D_{41}&\D_{42}&\D_{43}&\D_{44}&\D_{45}&\D_{46}&\D_{47}&\D_{48}&\D_{49}&\D_{50}\\ \hline
A&s_{23}s_{45}&s_{23}s_{123}&s_{23}s_{145}&s_{23}s_{234}&s_{23}s_{235}&s_{23}s_{1234}&s_{23}s_{1235}&s_{23}s_{2345}&s_{24}s_{35}&s_{24}s_{124}\\ \hline
B&(61,\ol{62},\ol{63})&(\ol{61},64,\ol{65})&(31,\ol{46},\ol{62})&(\ol{47},66,\ol{67})&(\ol{32},68,\ol{69})&(\ol{33},64,\ol{66})&(\ol{48},65,\ol{68})&(\ol{63},67,\ol{69})
&(70,\ol{71} ,\ol{72})&(\ol{70},73,\ol{74})\\ \hline\hline
&\D_{51}&\D_{52}&\D_{53}&\D_{54}&\D_{55}&\D_{56}&\D_{57}&\D_{58}&\D_{59}&\D_{60}\\ \hline
A&s_{24}s_{135}&s_{24}s_{234}&s_{24}s_{245}&s_{24}s_{1234}&s_{24}s_{1245}&s_{24}s_{2345}&s_{25}s_{34}&s_{25}s_{125}&s_{25}s_{134}&s_{25}s_{235}\\ \hline
B&(16,\ol{49},\ol{71})&(\ol{50},75,\ol{76})&(\ol{17},77,\ol{78})&(\ol{18},73,\ol{75})&(\ol{51},74,\ol{77})&(\ol{72},76,\ol{78})&(79,\ol{80},\ol{81})&(\ol{79},82,\ol{83})&(19,\ol{34},\ol{80})&(\ol{35},84,\ol{85})\\ \hline\hline
&\Delta_{61}&\Delta_{62}&\Delta_{63}&\Delta_{64}&\Delta_{65}&\Delta_{66}&\Delta_{67}&\Delta_{68}&\Delta_{69}&\Delta_{80}\\ \hline
A&s_{25} s_{245}&s_{25} s_{1235}&s_{25} s_{1245}&s_{25} s_{2345}&s_{34} s_{125}&s_{34} s_{134}&s_{34} s_{234}&s_{34} s_{345}&s_{34} s_{1234}&s_{34} s_{1345}\\ \hline
B&(\ol{20},86,\ol{87})&(\ol{21},82,\ol{84})&(\ol{36},83,\ol{86})&(\ol{81},85,\ol{87})&(1,\ol{52},\ol{79})&(\ol{80},88,\ol{89})&(\ol{53},90,\ol{91})&(2,\ol{92},\ol{93})&(3,\ol{88},\ol{90})&(\ol{54},89,\ol{92})\\ \hline\hline
&\Delta_{71}&\Delta_{72}&\Delta_{73}&\Delta_{74}&\Delta_{75}&\Delta_{76}&\Delta_{77}&\Delta_{78}&\Delta_{79}&\Delta_{80}\\
\hline
A&s_{34} s_{2345}&s_{35} s_{124}&s_{35} s_{135}&s_{35} s_{235}&s_{35} s_{345}&s_{35} s_{1235}&s_{35}s_{1345}&s_{35} s_{2345}&s_{45} s_{123}&s_{45} s_{145}\\ \hline
B&(\ol{81},91,\ol{93})&(4,\ol{37},\ol{70})&(\ol{71},94,\ol{95})&(\ol{38},96,\ol{97})&(5,\ol{98},\ol{99})&(6,\ol{94},\ol{96})&(\ol{39},95,\ol{98})&(\ol{72},97,\ol{99})&(7,\ol{22},\ol{61})&(\ol{62},100,\ol{101})\\ \hline\hline
&\Delta_{81}&\Delta_{82}&\Delta_{83}&\Delta_{84}&\Delta_{85}&\Delta_{86}&\Delta_{87}&\Delta_{88}&\Delta_{89}&\Delta_{90}\\ \hline
A&s_{45} s_{245}&s_{45} s_{345}&s_{45} s_{1245}&s_{45} s_{1345}&s_{45} s_{2345}&s_{123} s_{1234}&s_{123} s_{1235}&s_{124} s_{1234}&s_{124} s_{1245}&s_{125} s_{1235}\\ \hline
B&(\ol{23},102,\ol{103})&(8,\ol{104},\ol{105})&(9,\ol{100},\ol{102})&(24,\ol{101},\ol{104})&(63,\ol{103},\ol{105})&(10,\ol{25},\ol{64})&(11,\ol{26},\ol{65})&(12,\ol{40},\ol{73})&(13,\ol{41},\ol{74})&(14,\ol{55},\ol{82})\\ \hline\hline
&\Delta_{91}&\Delta_{92}&\Delta_{93}&\Delta_{94}&\Delta_{95}&\Delta_{96}&\Delta_{97}&\Delta_{98}&\Delta_{99}&\Delta_{100}\\ \hline
A&s_{125} s_{1245}&s_{134} s_{1234}&s_{134} s_{1345}&s_{135} s_{1235}&s_{135} s_{1345}&s_{145} s_{1245}&s_{145} s_{1345}&s_{234} s_{1234}&s_{234} s_{2345}&s_{235} s_{1235}\\ \hline
B&(15,\ol{56},\ol{83})&(27,\ol{42},\ol{88})&(28,\ol{43},\ol{89})&(29,\ol{57},\ol{94})&(30,\ol{58},\ol{95})&(44,\ol{59},\ol{100})&(45,\ol{60},\ol{101})&(66,\ol{75},\ol{90})&(67,\ol{76},\ol{91})&(68,\ol{84},\ol{96})\\ \hline\hline
&\Delta_{101}&\Delta_{102}&\Delta_{103}&\Delta_{104}&\Delta_{105}&&&&&\\ \hline
A&s_{235} s_{2345}&s_{245} s_{1245}&s_{245} s_{2345}&s_{345} s_{1345}&s_{345} s_{2345}&&&&&\\ \hline
B&(69,\ol{85},\ol{97})&(77,\ol{86},\ol{102})&(78,\ol{87},\ol{103})&(92,\ol{98},\ol{104})&(93,\ol{99},\ol{105})&&&&&\\
\hline\ea$$}
\bc Table B5.\quad Composition of the 105 $\D_l$'s\ec

\subsection{$\boldsymbol{\bD_l}$}
Using Table B5 and Fig.~B1, the CK combinations $\bD_l$ for the Feynman diagrams can be calculated and are shown in \eq{d1tod105}. The coefficients $d_e, d'_e$ of their linear $s$ terms are given in Table B6.
{\small
\be
\bD_1&=&(\ol 1,\ol2, 3)=s_{12}F(12569)T(\ol 934)+\[s_{34}F(34659)+d(9856)T(\834)\]T(\ol 912)+s_{12}s_{34}g_A(125634),\nn\\
\bD_2&=&(\ol4,\ol 5,6)=s_{12}F(12469)T(\ol 935)+\[s_{35}F(35649)+d(9846)T(\835)\]T(\ol 912)+s_{12}s_{35}g_A(124635),\nn\\
\bD_3&=&(7,\ol 8,\ol 9)=s_{12}F(12396)T(\ol 945)+\[s_{45}F(45693)+d(9386)T(\845)\]T(\ol 912)+s_{12}s_{45}g_A(123654),\nn\\
\bD_4&=&(\ol 7,10,\ol {11})=s_{12}A_4({12}39)) d(\9456)+s_{123}F(93456)T(\912),\nn\\
\bD_5&=&(\ol 4,12,\ol {13})=s_{12}A_4({12}49) d(\9356)+s_{124}F(94356)T(\912),\nn\\
\bD_6&=&(\ol 1,14,\ol {15})=s_{12}A_4({12}59) d(\9346)+s_{125}F(95346)T(\912),\nn\\
\bD_7&=&(2,\ol 5,\ol 8)=s_{12}A_4({12}96) d(\9345)+s_{345}F(69543)T(\912),\nn\\
\bD_8&=&(\ol 3,10,\ol {12})=s_{12}F(12349)T(\956)+\[s_{1234}F(56439)+d(9348)T(\856)\]T(\912)+s_{12}s_{1234}g_A(123456),\nn\\
\bD_9&=&(\ol 6,11,\ol {14})=s_{12}F(12359)T(\946)+\[s_{1235}F(46539)+d(9358)T(\846)\]T(\912)+s_{12}s_{1235}g_A(123546),\nn\\
\bD_{10}&=&(\ol 9,13,\ol {15})=s_{12}F(12459)T(\936)+\[s_{1245}F(36549)+d(9458)T(\836)\]T(\912)+s_{12}s_{1245}g_A(124536).\nn\\
\bD_{11}&=&(\ol {16},\ol{17}, 18)=s_{13}F(13956)T(\ol 924)+\[s_{24}F(24965)+d(9856)T(\824)\]T(\ol 913)+s_{13}s_{24}g_A(135624),\nn\\
\bD_{12}&=&(\ol{19},\ol {20},21)=s_{13}F(13946)T(\ol 925)+\[s_{25}F(25964)+d(9846)T(\825)\]T(\ol 913)+s_{13}s_{25}g_A(134625),\nn\\
\bD_{13}&=&(22,\ol {23},\ol {24})=s_{13}F(13296)T(\ol 945)+\[s_{45}F(45692)+d(9286)T(\845)\]T(\ol 913)+s_{13}s_{45}g_A(132654),\nn\\
\bD_{14}&=&(\ol {22},25,\ol {26})=s_{13}A_4({13}29) d(\9456)+s_{123}F(92456)T(\913),\nn\\
\bD_{15}&=&(\ol {19},27,\ol {28})=s_{13}A_4({13}49) d(\9256)+s_{134}F(94256)T(\913),\nn\\
\bD_{16}&=&(\ol {16},29,\ol {30})=s_{13}A_4({13}59) d(\9246)+s_{135}F(95246)T(\913),\nn\\
\bD_{17}&=&(17,\ol {20},\ol {23})=s_{13}A_4({13}96) d(\9245)+s_{245}F(69542)T(\913),\nn\\
\bD_{18}&=&(\ol {18},25,\ol {27})=s_{13}F(13249)T(\956)+s_{1234}\[F(56429)+d(9248)T(\856\]T(\913)+s_{13}s_{1234}g_A( 132456),\nn\\
\bD_{19}&=&(\ol {21},26,\ol {29})=s_{13}F(13259)T(\946)+s_{1235}\[F(46529)+d(9258)T(\846)\]T(\913)+s_{13}s_{1235}g_A(132546),\nn\\
\bD_{20}&=&(\ol {24},28,\ol {30})=s_{13}F(13459)T(\926)+s_{1345}\[F(26549)+d(9458)T(\826)\]T(\913)+s_{13}s_{1345}g_A(134526),\nn
\ee
\be
\bD_{21}&=&(\ol {31},\ol{32}, 33)=s_{14}F(14956)T(\ol 923)+\[s_{23}F(23965))+d(9856)T(\823)\]T(\ol 914)+s_{14}s_{23}g_A(145623),\nn\\
\bD_{22}&=&(\ol{34},\ol {35},36)=s_{14}F(14936)T(\ol 925)+\[s_{25}F(25963)+d(9836)T(\825)\]T(\ol 914)+s_{14}s_{25}g_A(143625),\nn\\
\bD_{23}&=&(37,\ol {38},\ol {39})=s_{14}F(14296)T(\ol 935)+\[s_{35}F(35692)+d(9286)T(\835)\]T(\ol 914)+s_{14}s_{35}g_A(142653),\nn\\
\bD_{24}&=&(\ol {37},40,\ol {41})=s_{14}A_4({14}29) d(\9356)+s_{124}F(92356)T(\914),\nn\\
\bD_{25}&=&(\ol {34},42,\ol {43})=s_{14}A_4({14}39) d(\9256)+s_{134}F(93256)T(\914),\nn\\
\bD_{26}&=&(\ol {31},44,\ol {45})=s_{14}A_4({14}59) d(\9236)+s_{145}F(95236)T(\914),\nn\\
\bD_{27}&=&(32,\ol {35},\ol {38})=s_{14}A_4({14}96) d(\9235)+s_{235}F(69532)T(\914),\nn\\
\bD_{28}&=&(\ol {33},40,\ol {42})=s_{14}F(14239)T(\956)+s_{1234}\[F(56329)+d(9238)T(\856)\]T(\ol 914)+s_{14}s_{1234}g_A( 142356),\nn\\
\bD_{29}&=&(\ol {36},41,\ol {44})=s_{14}F(14259)T(\936)+s_{1245}\[F(36529)+d(9258)T(\836)\]T(\ol 914)+s_{14}s_{1245}g_A(142536),\nn\\
\bD_{30}&=&(\ol {39},43,\ol {45})=s_{14}F(14359)T(\926)+s_{1345}\[F(26539)+d(9358)T(\826)\]T(\ol 914)+s_{14}s_{1345}g_A(143526).\nn\\
\bD_{31}&=&(\ol {46},\ol{47}, 48)=s_{15}F(15946)T(\ol 923)+\[s_{23}F(23964)+d(9846)T(\823)\]T(\ol 915)+s_{15}s_{23}g_A(154623),\nn\\
\bD_{32}&=&(\ol{49},\ol {50},51)=s_{15}F(15936)T(\ol 924)+\[s_{24}F(24963)+d(9835)T(\824)\]T(\ol 915)+s_{15}s_{24}g_A(153624),\nn\\
\bD_{33}&=&(52,\ol {53},\ol {54})=s_{15}F(15296)T(\ol 934)+\[s_{34}F(34692)+d(9286)T(\834)\]T(\ol 915)+s_{15}s_{34}g_A(152643),\nn\\
\bD_{34}&=&(\ol {52},55,\ol {56})=s_{15}A_4({15}29) d(\9346)+s_{125}F(92346)T(\915),\nn\\
\bD_{35}&=&(\ol {49},57,\ol {58})=s_{15}A_4({15}39) d(\9246)+s_{135}F(93246)T(\915),\nn\\
\bD_{36}&=&(\ol {46},59,\ol {60})=s_{15}A_4({15}49) d(\9236)+s_{145}F(94236)T(\915),\nn\\
\bD_{37}&=&(47,\ol {50},\ol {53})=s_{15}A_4({15}69) d(\9234)+s_{234}F(69432)T(\915),\nn\\
\bD_{38}&=&(\ol {48},55,\ol {57})=s_{15}F(15239)T(\946)+\[s_{1235}F(46329)+d(9238)T(\846)\]T(\915)+s_{15}s_{1235}g_A(152346),\nn\\
\bD_{39}&=&(\ol {51},56,\ol {59})=s_{15}F(15249)T(\936)+\[s_{1245}F(36429)+d(9248)T(\836)\]T(\915)+s_{15}s_{1245}g_A( 152436),\nn\\
\bD_{40}&=&(\ol {54},58,\ol {60})=s_{15}F(15349)T(\926)+\[s_{1345}F(26439)+d(9348)T(\826)\]T(\915)+s_{15}s_{1345}g_A(153426),\nn\\
\bD_{41}&=&(61,\ol {62},\ol{63})=s_{23}F(23169)T(\ol 945)+\[s_{45}F(45619)+d(1986)T(\845)\]T(\ol 923)+s_{23}s_{45}g_A(231645),\nn\\
\bD_{42}&=&(\ol{61},64,\ol {65})=s_{23}A_4({23}91) d(\9456)+s_{123}F(19456)T(\ol 923),\nn\\
\bD_{43}&=&(31,\ol {46},\ol {62})=s_{23}A_4({23}69) d(145\9)+s_{145}F(96541)T(\ol 923),\nn\\
\bD_{44}&=&(\ol {47},66,\ol {67})=s_{23}A_4({23}49) d(\1\956)+s_{234}F(94165)T(\923),\nn\\
\bD_{45}&=&(\ol {32},68,\ol {69})=s_{23}A_4({23}59) d(1\946)+s_{235}F(95164)T(\923),\nn
\ee

\be
\bD_{46}&=&(\ol {33},64,\ol {66})=s_{23}F(23194)T(\956)+\[s_{1234}F(56491)+d(1948)T(\856)\]T(\923)+s_{23}s_{1234}g_A(231465),\nn\\
\bD_{47}&=&(\ol {48},65,\ol {68})=s_{23}F(23195)T(\946)+\[s_{1235}F(46591)+d(1958)T(\846)\]T(\923)+s_{23}s_{1235}g_A(235164),\nn\\
\bD_{48}&=&(\ol {63},67,\ol {69})=s_{23}F(23459)T(\916)+\[s_{2345}F(61549)+d(9458)T(\861)\]T(\923)+s_{23}s_{2345}g_A(234561),\nn\\
\bD_{49}&=&(70,\ol {71},\ol {72})=s_{24}F(24169)T(\935)+\[s_{35}F(35619)+d(1986)T(\835)\]T(\924)+s_{23}s_{35}g_A(241653),\nn\\
\bD_{50}&=&(\ol {70},73,\ol {74})=s_{24}A_4({24}91) d(\9356)+s_{124}F(19356)T(\924),\nn\\
\bD_{51}&=&(16,\ol {49},\ol{71})=s_{24}A_4({24}69) d(135\9)+s_{135}F(96531)T(\ol 924),\nn\\
\bD_{52}&=&(\ol{50},75,\ol {76})=s_{24}A_4({24}39) d(1\956)+s_{234}F(93165)T(\ol 924),\nn\\
\bD_{53}&=&(\ol {17},77,\ol {78})=s_{24}A_4({24}59) d(1\936)+s_{245}F(95163)T(\ol 924),\nn\\
\bD_{54}&=&(\ol {18},73,\ol {75})=s_{24}F(24193)T(\956)+\[s_{1234}F(56391)+d(1938)T(\856)\]T(\924)+s_{24}s_{1234}g_A(241365),\nn\\
\bD_{55}&=&(\ol {51},74,\ol {77})=s_{24}F(24195)T(\936)+\[s_{1245}F(36591)+d(1958)T(\836)\]T(\924)+s_{24}s_{1245}g_A(241563),\nn\\
\bD_{56}&=&(\ol {72},76,\ol {78})=s_{24}F(24359)T(\916)+\[s_{2345}F(61539)+d(9358)T(\861)\]T(\924)+s_{24}s_{2345}g_A(615324),\nn\\
\bD_{57}&=&(79,\ol {80},\ol {81})=s_{25}F(25169)T(\934)+s_{34}\[F(34619)+d(1986)T(\834)\]T(\925)+s_{25}s_{34}g_A(251634),\nn\\
\bD_{58}&=&(\ol {79},82,\ol {83})=s_{25}A_4({25}91) d(\9346)+s_{125}F(19346)T(\925),\nn\\
\bD_{59}&=&(19,\ol {34},\ol {80})=s_{25}A_4({25}69) d(134\9)+s_{134}F(96431)T(\925),\nn\\
\bD_{60}&=&(\ol {35},84,\ol {85})=s_{25}A_4({25}39) d(1\946)+s_{235}F(93164)T(\925),\nn\\
\bD_{61}&=&(\ol {20},86,\ol{87})=s_{25}A_4({25}49) d(1\936)+s_{245}F(94163)T(\ol 925),\nn\\
\bD_{62}&=&(\ol{21},82,\ol {84})=s_{25}F(25193)T(\946)+\[s_{1235}F(46391)+d(1938)T(\846)\]T(\ol 925)+s_{25}s_{1235}g_A(251364),\nn\\
\bD_{63}&=&(\ol {36},83,\ol {86})=s_{25}F(25194)T(\936))+\[s_{1245}F(36491)+d(1948)T(\836)\]T(\ol 925)+s_{25}s_{1245}g_A(251463),\nn\\
\bD_{64}&=&(\ol {81},85,\ol {87})=s_{25}F(25349)T(\916)+\[s_{2345}F(61439)+d(9348)T(\861)\]T(\925)+s_{25}s_{2345}g_A(614325),\nn\\
\bD_{65}&=&(1,\ol {52},\ol {79})=s_{34}A_4({34}69)  d(125\9)+s_{125}F(96521)T(\934),\nn\\
\bD_{66}&=&(\ol {80},88,\ol {89})=s_{34}A_4({13}49)  d(\9256)+s_{134}F(19256)T(\934),\nn\\
  \bD_{67}&=&(\ol {53},90,\ol {91})=s_{34}A_4({34}92)  d(1\956)+s_{234}F(29165)T(\934),\nn\\
\bD_{68}&=&(2,\ol {92},\ol {93})=s_{34}A_4({34}59)  d(12\96)+s_{345}F(95612)T(\934),\nn\\
\bD_{69}&=&(3,\ol {88},\ol {90})=s_{34}F(34912)T(\956)+\[s_{1234}F(56921)+d(1298)T(\856)\]T(\934)+s_{34}s_{1234}g_A(341256),\nn\\
\bD_{70}&=&(\ol {54},89,\ol {92})=s_{34}F(34195)T(\926)+\[s_{1345}F(26591)+d(1958)T(\826)\]T(\934)+s_{34}s_{1345}g_A(341562),\nn
\ee

\be
\bD_{71}&=&(\ol {81},{91}, \ol{93})=s_{34}F(34295)T(\916)+\[s_{2345}F(61592)+d(9258)T(\816)\]T(\ol 934)+s_{34}s_{2345}g_A(342516),\nn\\
\bD_{72}&=&(4,\ol {37},\ol{70})=s_{35}A_4({35}69) d(124\ol 9)+s_{124}F(96421)T(\ol 935),\nn\\
\bD_{73}&=&(\ol {71},94,\ol {95})=s_{35}A_4({35}91) d(\9246)+s_{135}F(19246)T(\ol 935),\nn\\
\bD_{74}&=&(\ol {38},96,\ol {97})=s_{35}A_4({35}92) d(1\946))+s_{235}F(29164)T(\935),\nn\\
\bD_{75}&=&(5,\ol {98},\ol {99})=s_{35}A_4({35}49) d(12\96)+s_{345}F(94612)T(\935),\nn\\
\bD_{76}&=&(6,\ol {94},\ol {96})=s_{35}F(35912)T(\946)+\[s_{1235} F(46921)+d(1298)T(\846)\]T(\935)+s_{35}s_{1235}g_A(351246),\nn\\
\bD_{77}&=&(\ol {39},95,\ol {98})=s_{35}F(35194)T(\926)+\[s_{1345} F(26491)+d(1948)T(\826)\]T(\935)+s_{35}s_{1345}g_A(351462),\nn\\
\bD_{78}&=&(\ol {72},97,\ol {99})=s_{35}F(35294)T(\916)+\[s_{2345} F(61492)+d(9248)T(\816)\]T(\935)+s_{35}s_{2345}g_A(352416),\nn\\
\bD_{79}&=&(7,\ol {22},\ol {61})=s_{45}A_4({45}69) d(123\9)+s_{123} F(96321)T(\945),\nn\\
\bD_{80}&=&(\ol {62},100,\ol {101})=s_{45}A_4({14}59) d(\9236))+s_{145} F(19236)T(\945),\nn\\
\bD_{81}&=&(\ol {23},{102}, \ol{103})=s_{45}A_4({45}92) d(1\ol 936)+s_{245}F(29163)T(\ol 945),\nn\\
\bD_{82}&=&(8,\ol {104},\ol{105})=s_{45}A_4({45}93) d(12\ol 96)+s_{345}F(39612)T(\ol 945),\nn\\
\bD_{83}&=&(9,\ol {100},\ol {102})=s_{45}F(45912)T(\ol 936)+\[s_{1245}F(36921)+d(1298)T(\836)T\]T(\ol 945)+s_{45}s_{1245}g_A(451236),\nn\\
\bD_{84}&=&(24,\ol {101},\ol {104})=s_{45}F(45913)T(\926)+\[s_{1345}F(26931)+d(1398)T(\826)T\]T(\945)+s_{45}s_{1345}g_A(451326),\nn\\
\bD_{85}&=&(63,\ol {103},\ol {105})=s_{45}F(45923)T(\961)+\[s_{2345}F(61932)+d(2398)T(\861)T\]T(\945)+s_{45}s_{2345}g_A(452361),\nn\\
\bD_{86}&=&(10,\ol {25},\ol {64})=s_{123}F(49321)T(\956)+s_{1234}A_4({94}56) d(123\9),\nn\\
\bD_{87}&=&(11,\ol {26},\ol {65})=s_{123}F(59321)T(\946)+s_{1235}A_4({95}46) d(123\9),\nn\\
\bD_{88}&=&(12,\ol {40},\ol {73})=s_{124}F(93421)T(\956)+s_{1234}A_4({93}56) d(124\9),\nn\\
\bD_{89}&=&(13,\ol {41},\ol {74})=s_{124}F(95421)T(\936)+s_{1245}A_4({95}36) d(124\9),\nn\\
\bD_{90}&=&(14,\ol {55},\ol {82})=s_{125}F(39521)T(\946)+s_{1235}A_4({93}46) d(125\9),\nn\\
\bD_{91}&=&(15,\ol {56}, \ol{83})=s_{125}F(49521)T(\ol 936)+s_{1245}A_4({36}94) d(125\ol 9),\nn\\
\bD_{92}&=&(27,\ol {42},\ol{88})=s_{134}F(29431)T(\ol 956)+s_{1234}A_4({56}92) d(134\ol 9),\nn\\
\bD_{93}&=&(28,\ol {43},\ol {89})=s_{134}F(59431)T(\ol 926)+s_{1345}A_4({26}95) d(134\ol 9),\nn\\
\bD_{94}&=&(29,\ol {57},\ol {94})=s_{135}F(29531)T(\946)+s_{1235}A_4({46}92)  d(135\ol 9),\nn\\
\bD_{95}&=&(30,\ol {58},\ol {95})=s_{135}F(49531)T(\926)+s_{1345}A_4({26}94) d(135\ol 9),\nn
\ee

\be
\bD_{96}&=&(44,\ol {59},\ol {100})=s_{145}F(29541)T(\936)+s_{1245}A_4({36}92) d(145\9),\nn\\
\bD_{97}&=&(45,\ol {60},\ol {101})=s_{145}F(39541)T(\926)+s_{1345}A_4({26}93) d(145\9),\nn\\
\bD_{98}&=&(66,\ol {75},\ol {90})=s_{234}F(91432)T(\956)+s_{1234}A_4({56}19) d(\9234),\nn\\
\bD_{99}&=&(67,\ol {76},\ol {91})=s_{234}F(59432)T(\961)+s_{2345}A_4({61}95) d(\9243),\nn\\
\bD_{100}&=&(68,\ol {84},\ol {96})=s_{235}F(91532)T(\946)+s_{1235}A_4({46}19) d(\9235),\nn\\
\bD_{101}&=&(69,\ol {85}, \ol{97})=s_{235}F(49532)T(\ol 961)+s_{2345}A_4({61}94) d(\9235),\nn\\
\bD_{102}&=&(77,\ol {86},\ol{102})=s_{245}F(91542)T(\ol 936)+s_{1245}A_4({36}19) d(\9245),\nn\\
\bD_{103}&=&(78,\ol {87},\ol {103})=s_{245}F(39542)T(\ol 961)+s_{2345}A_4({61}93) d(\9245),\nn\\
\bD_{104}&=&(92,\ol {98},\ol {104})=s_{345}F(91543)T(\926)+s_{1345}A_4({26}19) d(\9345),\nn\\
\bD_{105}&=&(93,\ol {99},\ol {105})=s_{345}F(92543)T(\961)+s_{2345}A_4({61}29) d(\9345).\labels{d1tod105}
\ee}

$$\ba{|c|c|c||c|c|c|}\hline
l&d_e&d_e'&l&d_e&d_e'\\ \hline
1&F(12569)T(\ol 934)&F(34659)T(\ol 912)&2&F(12469)T(\ol 935)&F(35649)T(\ol 912)\\
3&F(12396)T(\ol 945)&F(45693)T(\ol 912)&4&\-T(123)T(456)&F(93456)T(\ol 912)\\
5&\-T(124)T(356)&F(94356)T(\ol 912)&6&\-T(125)T(346)&F(95346)T(\ol 912)\\
7&\-T(126)T(354)&F(69543)T(\ol 912)&8&F(12349)T(\ol 956)&F(56439)T(\ol 912)\\
9&F(12359)T(\ol 946)&F(46539)T(\ol 912)&10&F(12459)T(\ol 936)&F(36549)T(\ol 912)\\
11&F(13956)T(\ol 924)&F(24965)T(\ol 913)&12&F(13946)T(\ol 925)&F(25964)T(\ol 913)\\
13&F(13296)T(\ol 945)&F(45692)T(\ol 913)&14&\-T(132)T(456)&F(92456)T(\ol 913)\\
15&\-T(134)T(256)&F(94256)T(\ol 913)&16&\-T(135)T(246)&F(95246)T(\ol 913)\\
17&\-T(136)T(254)&F(69542)T(\ol 913)&18&F(13249)T(\ol 956)&F(56429)T(\ol 913)\\
19&F(13259)T(\ol 946)&F(46529)T(\ol 913)&20&F(13459)T(\ol 926)&F(26549)T(\ol 913)\\
21&F(14936)T(\ol 925)&F(23965)T(\ol 914)&22&F(14936)T(\ol 925)&F(25963)T(\ol 914)\\
23&F(14296)T(\ol 935)&F(35692)T(\ol 914)&24&\-T(142)T(356)&F(92356)T(\ol 914)\\
25&\-T(143)T(256)&F(93256)T(\ol 914)&26&\-T(145)T(236)&F(95236)T(\ol 914)\\
27&\-T(146)T(253)&F(69532)T(\ol 914)&28&F(14239)T(\ol 956)&F(56329)T(\ol 914)\\
29&F(14259)T(\ol 936)&F(36529)T(\ol 914)&30&F(14359)T(\ol 926)&F(26539)T(\ol 914)\\
31&F(15946)T(\ol 923)&F(23964)T(\ol 915)&32&F(15936)T(\ol 924)&F(24963)T(\ol 915)\\
33&F(15296)T(\ol 934)&F(34692)T(\ol 915)&34&\-T(152)T(346)&F(92346)T(\ol 915)\\
35&\-T(153)T(246)&F(93246)T(\ol 915)&36&\-T(154)T(236)&F(94236)T(\ol 915)\\
37&\-T(156)T(234)&F(69432)T(\ol 915)&38&F(15239)T(\ol 946)&F(46329)T(\ol 915)\\
39&F(15249)T(\ol 936)&F(36429)T(\ol 915)&40&F(15349)T(\ol 926)&F(26439)T(\ol 915)\\
41&F(23169)T(\ol 945)&F(45619)T(\ol 923)&42&\-T(231)T(564)&F(19456)T(\ol 923)\\
43&\-T(236)T(541)&F(96541)T(\ol 923)&44&\-T(234)T(165)&F(94165)T(\ol 923)\\
45&\-T(235)T(164)&F(95164)T(\ol 923)&46&F(23194)T(\ol 956)&F(56491)T(\ol 923)\\
47&F(23195)T(\ol 946)&F(46591)T(\ol 923)&48&F(23459)T(\ol 916)&F(61549)T(\ol 923)\\
49&F(24169)T(\ol 935)&F(35619)T(\ol 924)&50&\-T(241)T(365)&F(19356)T(\ol 924)\\
51&\-T(246)T(531)&F(96531)T(\ol 924)&52&\-T(243)T(165)&F(93165)T(\ol 924)\\
53&\-T(245)T(163)&F(96153)T(\ol 924)&54&F(24193)T(\ol 956)&F(56391)T(\ol 924)\\
\hline\ea$$

$$\ba{|c|c|c||c|c|c|}\hline
l&d_e&d_e'&l&d_e&d_e'\\ \hline
55&F(24195)T(\ol 936)&F(36591)T(\ol 924)&56&F(24359)T(\ol 916)&F(61539)T(\ol 924)\\
57&F(25169)T(\ol 934)&F(34619)T(\ol 925)&58&\-T(251)T(364)&F(19346)T(\ol 925)\\
59&\-T(256)T(134)&F(96431)T(\ol 925)&60&\-T(253)T(164)&F(93164)T(\ol 925)\\
61&\-T(254)T(163)&F(94163)T(\ol 925)&62&F(25193)T(\ol 946)&F(46391)T(\ol 925)\\
63&F(25194)T(\ol 936)&F(36491)T(\ol 925)&64&F(25349)T(\ol 916)&F(61439)T(\ol 925)\\
65&\-T(346)T(521)&F(96521)T(\ol 934)&66&\-T(341)T(265)&F(19256)T(\ol 934)\\
67&\-T(342)T(165)&F(29165)T(\ol 934)&68&\-T(345)T(612)&F(95612)T(\ol 934)\\
69&F(34912)T(\ol 956)&F(56921)T(\ol 934)&70&F(34195)T(\ol 926)&F(26591)T(\ol 934)\\
71&F(34295)T(\ol 916)&F(61592)T(\ol 934)&72&\-T(356)T(421)&F(96421)T(\ol 935)\\
73&\-T(351)T(264)&F(19246)T(\ol 935)&74&\-T(352)T(146)&F(29164)T(\ol 935)\\
75&\-T(354)T(612)&F(94612)T(\ol 935)&76&F(35912)T(\ol 946)&F(46921)T(\ol 935)\\
77&F(35194)T(\ol 926)&F(26491)T(\ol 935)&78&F(35294)T(\ol 916)&F(61492)T(\ol 935)\\
79&\-T(456)T(321)&F(96321)T(\ol 945)&80&\-T(451)T(263)&F(19236)T(\ol 945)\\
81&\-T(452)T(136)&F(29163)T(\ol 945)&82&\-T(453)T(621)&F(39612)T(\ol 945)\\
83&F(45912)T(\ol 936)&F(36921)T(\ol 945)&84&F(45913)T(\ol 926)&F(26931)T(\ol 945)\\
85&F(45923)T(\ol 961)&F(61932)T(\ol 945)&86&F(49321)T(\ol 956)&\-T(654)T(321)\\
87&F(59321)T(\ol 946)&\-T(645)T(321)&88&F(93421)T(\ol 956)&\-T(653)T(421)\\
89&F(95421)T(\ol 936)&\-T(635)T(421)&90&F(39521)T(\ol 946)&\-T(643)T(521)\\
91&F(49521)T(\ol 936)&\-T(364)T(512)&92&F(29431)T(\ol 956)&\-T(562)T(413)\\
93&F(59431)T(\ol 926)&\-T(625)T(431)&94&F(29531)T(\ol 946)&\-T(642)T(531)\\
95&F(49531)T(\ol 926)&\-T(624)T(531)&96&F(29541)T(\ol 936)&\-T(632)T(541)\\
97&F(39541)T(\ol 926)&\-T(623)T(541)&98&F(91432)T(\ol 956)&\-T(561)T(234)\\
99&F(59432)T(\ol 961)&\-T(165)T(243)&100&F(91532)T(\ol 946)&\-T(461)T(235)\\
101&F(49532)T(\ol 961)&\-T(164)T(235)&102&F(91542)T(\ol 936)&\-T(361)T(245)\\
103&F(39542)T(\ol 961)&\-T(163)T(245)&104&F(91543)T(\ol 926)&\-T(261)T(345)\\
105&F(92543)T(\ol 961)&\-T(612)T(345)&&&\\
\hline\ea$$
\bc Table B6.\quad A list of $d_e$ and $d'_e$ in $\bD_l=d_es_e+d'_es'_e+O(s^2)$, with $s_e\prec s'_e$ \ec

\section{$\boldsymbol{p}$ equations for $\boldsymbol{N=6}$} 
\subsection{$\boldsymbol{\d\D_l=\bD_l}$ equations}
The 210 equations for parameters $p$, or $t_l$, can be obtained from  the
 list $E$ below by setting $E_{2l\-1}=d_l$ and $E_{2l}=d'_l,\ (1\le l\le 105)$. Each $t$ combination in $E_i$
is  expressed as a signed list of $m$. For example, $E_1=(1, \-65, \-68, 69)$
and $E_2=(1, 6, \-7, \-8)$ corresponds to \eq{dd1p6c}
\be t_1-t_{65}-t_{68}+t_{69}&=&d_1,\nn\\
t_1+t_6-t_7-t_8&=&d'_1.\nn\ee
The list $E$ with 210 members is 
{\small\be
&&E=\{\nn\\
&&(1, -65, -68, 69),  (1, 6, -7, -8),  (2, -72, -75, 76),  (2, 5, 7, -9),  (3, 79, -82, -83),  (3, -4, 7, 10),\nn\\
&&  (-79, 86, -87),  (-3, 4, 8, -9), (-72, 88, -89),  (2, 5, -8, -10),  (-65, 90, -91),  (1, 6, -9, 10),\nn\\
&&  (68, -75, -82),  (-1, 2, 3, 7),  (8, -69, 86, -88),  (-1, 4, -5, 8),  (9, -76, 87, -90),  (-2, -4, -6, 9),\nn\\
&&  (10, \-83, 89, -91),  (3, \-5, 6, 10),  (11, \-51, 53, \-54),  (11, 16, \-17, \-18),  (12, \-59, 61, \-62),(12, 15, 17, \-19), \nn\\
&&   (13, -79, 81, -84),  (13, -14, 17, 20),  (79, -86, 87),  (-13, 14, 18, -19), (-59, 92, -93),  (12, 15, -18, -20),\nn\\
&&   (-51, 94, -95),  (11, 16, -19, 20),  (-53, 61, 81),  (-11, 12, 13, 17), (18, 54, -86, -92), (-11, 14, -15, 18),\nn\\
&&   (19, 62, \-87, \-94),  (\-12, \-14, \-16, 19),  (20, \-84, 93, \-95),(13, \-15, 16, 20),  (21, \-43, 45, \-46), (21, 26, \-27, \-28),  \nn\\
&&   (22, 59, 60, -63),  (22, 25, 27, -29),(23, -72, 74, 77),  (23, -24, 27, 30),(72, -88, 89),  (-23, 24, 28, -29), \nn\\
&&     (59, -92, 93),  (22, 25, -28, -30),  (-43, 96, -97),  (21, 26, -29, 30), (-45, 60, 74), (-21, 22, 23, 27),\nn\\
&&   (28, 46, \-88, 92),  (\-21, 24, \-25, 28), (29, 63, \-89, \-96),  (\-22, \-24, \-26, 29),(30, 77, \-93, \-97),  (23, \-25, 26, 30),\nn\\
&&  (31, 43, 44, -47), (31, 36, -37, -38),  (32, 51, 52, -55),(32, 35, 37, -39),  (33, -65, 67, 70),  (33, -34, 37, 40), \nn\\
&&   (65, -90, 91),  (-33, 34, 38, -39), (51, -94, 95),  (32, 35, -38, -40),  (43, -96, 97),  (31, 36, -39, 40),  \nn\\
&& (-44, 52, 67),  (-31, 32, 33, 37),(38, 47, -90, 94),  (-31, 34, -35, 38),  (39, 55, -91, 96), (-32, -34, -36, 39), \nn\\
&&   (40, 70, -95, 97),(33, -35, 36, 40),  (41, -79, 80, -85),  (41, -42, 43, 48), (79, -86, 87),  (-41, 42, 46, -47),\nn\\
&&  (-26, 36, 80),  (-21, 31, 41, 43),  (-37, 98, -99),  (31, 44, -46, -48),(-27, 100, -101),  (21, 45, -47, 48),\nn\\
&&  (28, 46, \-86, \-98),  (\-21, 42, \-44, 46),  (38, 47, \-87, \-100), (\-31, \-42, \-45, 47),  (48, \-85, 99, \-101), (41, \-44, 45, 48),    \nn\\
&&  (49, -72, 73, 78), (49, -50, 51, 56), (72, -88, 89),  (-49, 50, 54, -55), (-16, 35, 73),  (-11, 32, 49, 51),  \nn\ee

\be
&& (37, -98, 99), (32, 52, -54, -56), (\-17, 102, \-103), (11, 53, \-55, 56), (18, 54, \-88, 98), (\-11, 50, \-52, 54),\nn\\
&& (39, 55, \-89, \-102), (\-32, \-50, \-53,55), (56, 78, -99, -103),  (49, -52, 53, 56),  (57, -65, 66, 71),  (57, -58, 59, 64),\nn\\
&&   (65, -90, 91),  (-57, 58, 62, -63), (-15, 25, 66),  (-12, 22, 57, 59),  (27, -100, 101),  (22, 60, -62, -64), \nn\\
&&  (17, -102, 103),  (12, 61, -63, 64),(19, 62, \-90, 100),  (\-12, 58, \-60, 62),  (29, 63, \-91, 102),  (\-22, \-58, \-61, 63),\nn\\
&&  (64, 71, \-101, 103),  (57, \-60, 61, 64),(-6, 34, 58),  (-1, -33, -57, 65),  (59, -92, 93),  (57, 66, -69, -70), \nn\\
&&   (37, -98, 99), (33, 67, -69, -71), (7, -104, -105),  (-1, 68, 70, 71),  (-8, 69, 92, 98),  (1, -66, -67, 69), \nn\\
&&  (40, 70, -93, -104),  (33, -66, 68, 70), (64, 71, -99, -105),  (57, -67, 68, 71),  (-5, 24, 50),  (-2, -23, -49, 72), \nn\\
&& (51, -94, 95),  (49, 73, -76, -77),(27, -100, 101),  (23, 74, -76, -78),  (-7, 104, 105),  (-2, 75, 77, 78),\nn\\
&&    (-9, 76, 94, 100),  (2, -73, 74, 76), (30, 77, -95, 104),  (23, -73, 75, 77),  (56, 78, -101, 105), (49, -74, 75, 78), \nn\\
&&  (-4, 14, 42),  (3, -13, -41, 79), (43, -96, 97), (41, 80, -83, 84),  (17, -102, 103),  (13, 81, -83, 85),  \nn\\
&& (-7, 104, 105),  (-3, 82, 84, 85),(-10, 83, 96, 102),  (-3, -80, -81, 83),  (-20, 84, 97, 104),  (-13, 80, 82, 84),  \nn\\
&&  (-48, 85, 103, 105),  (-41, 81, 82, 85), (8, -18, -46, 86),  (4, -14, -42),  (9, -19, -47, 87),  (-4, 14, 42), \nn\\
&&  (-8, -28, -54, 88),  (5, -24, -50), (10, -29, -55, 89),  (-5, 24, 50),  (-9, -38, -62, 90),  (6, -34, -58),  \nn\\
&& (-10, -39, -63, 91),  (-6, 34, 58), (-18, 28, 69, 92),  (15, -25, -66),  (20, -30, -70, 93),  (-15, 25, 66), \nn\\
&&  (-19, 38, 76, 94),  (16, -35, -73), (-20, -40, -77, 95),  (-16, 35, 73),  (-29, 39, 83, 96),  (26, -36, -80),  \nn\\
&& (-30, 40, 84, 97),  (-26, 36, 80), (-46, 54, 69, 98),  (44, -52, -67),  (48, -56, -71, 99),  (-44, 52, 67), \nn\\
&&  (-47, 62, 76, 100),  (45, -60, -74), (-48, -64, -78, 101),  (-45, 60, 74),  (-55, 63, 83, 102),  (53, -61, -81),  \nn\\
&& (-56, 64, 85, 103),  (-53, 61, 81), (-70, 77, 84, 104),  (-68, 75, 82),  (-71, 78, 85, 105),  (-68, 75, 82)\ \}
\labels{Elist}\ee}
\subsection{Solutions}
 The 210 $t$ equations in \eq{Elist} can be written in the matrix form $\tau\.t=\ol d$ as in \eq{teq6}.
 The $210\x105$ matrix $\tau$ turns out to have a rank of 79, thus possessing 131 left null vectors $u_x$
 so that $u_x\.\tau=0$. As a result, for \eq{Elist} to have a solution, the inhomogeneous terms $\ol d_i$
 must satisfy 131 sum rules of the form $R_x:=u_x\.\ol d=0$. Of the 131 sum rules, 40 of them have two terms,
 37 of them have four terms, eight of them have six terms, and the rest 46 of them have at least thirteen terms.
  It turns out that  sum rules are often
 satisfied when the number of terms are less or equal to six, but they are generally not satisfied when
 many terms are present. Hence
  the $t$ equaequationstions \eq{Elist}
 have no solutions.
 
 To illustrate this statement, let us pick four null vectors, with 2, 4, 6, 13 non-zero entries, for a detailed
 illustration. First, let us use
 the $t$ equations to  verify directly that these are indeed null vectors. Once verified, their respective sum rules are then checked. 
 
 Every null vector $u_x$ has 210 components. For those with very few non-zero entries, it is more economical
 to express $u_x$ by displaying the position of its non-zero entries, together with its coefficient (usually  $\-1$)
 if it is not 1. 
 
 \begin{enumerate}  
 \i $u_1=(13, 210)$.\quad This  null vector has only two non-zero entries, at position 13 and 210, both
 with value 1. Since $E_{13}= (68, -75, -82)$, meaning $t_{68}-t_{75}-t_{82}$, and $E_{210}=(-68, 75, 82)$,
 meaning  $-t_{68}+t_{75}+t_{82}$, (the left hand side of) these two equations obviously add up to zero,
 proving that $u_1=(13, 210)$ is indeed a null vector.
 \i $u_{131}=(-4,6,8,10)$.\quad The four non-zero entries are 
 $$-E_4=-(2, 5, 7, -9),\  E_6=(3, -4, 7, 10),\
 E_8=(-3, 4, 8, -9),\ E_{10}=(2, 5, -8, -10).$$
  These four equations do add up to be zero, so $u_{131}$ is indeed a null vector. 
\i $u_{126}=(-2,4,-6,-8,-14,20)$.\quad The six non-zero entries are 
\be  -E_2&=&-(1, 6, -7, -8), \quad E_4=(2, 5, 7, -9),\quad -E_6=-(3, -4, 7, 10),\nn\\
-E_8&=&-(-3, 4, 8, -9),\quad -E_{14}=-(-1, 2, 3, 7),\quad E_{20}=(3, -5, 6, 10).\nn\ee
They can be seen to add up to zero, so $u_{126}$ is a null vector.
\i $u_{127}=(1,-3,-5,-6,-7,-8,9,-11,13,14,15,-17,19)$. The thirteen non-zero entries are
\be
E_1&=&(1, -65, -68, 69), \quad -E_3=-(2, -72, -75, 76),\quad -E_5= -(3, 79, -82, -83), \nn\\
-E_6&=&-(3, -4, 7, 10),\quad -E_7= -(-79, 86, -87), \quad -E_8=-(-3, 4, 8, -9), \nn\\
E_9&=&(-72, 88, -89), \quad -E_{11}=-(-65, 90, -91), E_{13}=(68, -75, -82), \nn\\
E_{14}&=&(-1, 2, 3, 7), \quad E_{15}=(8, -69, 86, -88), \quad -E_{17}=-(9, -76, 87, -90), \nn\\
E_{19}&=&(10, -83, 89, -91).
\ee
Again, these 13 equations can be seen to add up to zero, proving that $u_{127}$ is a null vector.
 \end{enumerate}.
  
Having thus verified the correctness of the null vectors, let us now use the explicit expressions
of $(d_e, d'_e)$ in Table B6  to check  whether the corresponding
sum rules are satisfied.
\begin{enumerate} 
\i $ R_1=\ol d_{13}+\ol d_{210}=d_7+d'_{105}=\-T(126)T(354)\-T(612)T(345)=0$ because of the symmetries of $T$.
\i 
\be
R_{131}&=&-\ol d_4+\ol d_6+\ol d_8+\ol d_{10}=-d'_2+d'_3+d'_4+d'_5\nn\\
&=&\(-F(35649)+F(45693)+F(93456)+F(94356)\)T(\ol 912)\nn\\
&=&\(\-F(35496)\-F(54936)\-F(93546)\-F(49356)\)T(\ol 912)\nn\\
&=&\-D(35496)T(\ol 912)\simeq 0,\nn\ee
where \eq{F2}, \eq{D1}, and the numerator Slavnov-Taylor identity have been used.
\i
\be R_{126}&=&-\ol d_2+\ol  d_4-\ol d_6-\ol d_8-\ol d_{14}+\ol d_{20}=-d'_1+d'_2-d'_3-d'_4-d'_7+d'_{10}\nn\\
&=&\(-F(34659)+F(35649)-F(45693)-F(93456)-F(69543)+F(36549)\)T(\912)\nn\\
&=&\(F(34695)-F(53649)-F(45368)+F(93465)+F(69345)-F(36459)\)T(\912)\nn\\
&=&\(D(34695)-D(53649)\)T(\912)\simeq 0.\nn\ee
\i
\be R_{127}&=&\ol d_1-\ol d_3-\ol d_5-\ol d_6-\ol d_7-\ol d_8+\ol d_9-\ol d_{11}+\ol d_{13}+\ol d_{14}+\ol d_{15}-\ol d_{17}+\ol d_{19}\nn\\
&=&d_1- d_2-d_3- d'_3- d_4- d'_4+ d_5- d_{6}+ d_{7}+ d'_{7}+ d_{8}- d_{9}+ d_{10}\not=0\nn\ee
as can be seen from Table B6.
\end{enumerate}

\section{$\boldsymbol{q}$ equations for $\boldsymbol{N=6}$} 
This appendix provides supplementary material for Sec.~VIF. Please see the main text for some of the
definitions used below.

\subsection{Gauge constraint equations}
The following notations will be used for the $q$ parameters: $q_a\to a_1, q'_a\to a_2, q''_a\to a_3$.
In terms of these notations, the 144 constraint equations are given below by setting 
each entry to be zero. For example, \eq{qeq1} is given by the 6th entry below, and \eq{qeq2} is given by the
9th entry.

{\small\be
&&1_1\-2_1\-5_1\-6_1\+14_1,\quad \-1_1\+2_1\+8_1\-9_1\+15_1,\quad 1_2\-2_2\+79_2\+81_2\-93_1,\quad
1_3\+14_2\+79_3\+82_2,\nn\\
&&-1_3\+15_2\-79_3\+83_2,\quad 2_1\+3_1\+7_1\+8_1\+10_1 ,\quad
\-2_1\-3_1\+4_1\-5_1\+12_1,\quad 2_2\+3_2\+90_1\+91_1\+93_1,\nn\\
&&2_3\+8_3\+93_2\+105_2,\quad
\-2_3\-5_3\-93_2\-99_2,\quad 3_3\+10_3\+64_3\+66_3\+90_3,\quad \-3_3\+12_3\+73_3\+75_3\-90_3,\nn\\
&&\-4_1\+5_1\-8_1\+9_1\+13_1,\quad 4_2\-5_2\+70_2\+72_2\-99_1,\quad 4_3\+12_2\+70_3\+73_2,\quad
\-4_3\+13_2\-70_3\+74_2,\nn\\
&&5_1\+6_1\-7_1\-8_1\+11_1,\quad 5_2\+6_2\+96_1\+97_1\+99_1,\quad 5_3\-8_3\+99_2\-105_2,\quad 6_3\+11_3\+65_3\+68_3\+96_3,\nn\\
&&\-6_3\+14_3\+82_3\+84_3\-96_3,\quad 7_2\+8_2\+61_2\+63_2\+105_1,\quad 7_3\+10_2\+61_3\+64_2,\quad \-7_3\+11_2\-61_3\+65_2,\nn\\
&&\-8_2\+9_2\+102_1\+103_1\-105_1,\ 9_3\+13_3\+74_3\+77_3\+102_3,\ \-9_3\+15_3\+83_3\+86_3\-102_3,\ 16_1\-17_1\-20_1\-21_1\+29_1,\nn\\
&&\-16_1\+17_1\+23_1\-24_1\+30_1,\quad 16_2\-17_2\+71_1\-72_1\+78_1,\quad 16_3\+29_2\+71_3\+94_2,\quad \-16_3\+30_2\-71_3\+95_2\+17_1,\nn\\
&&18_1\+22_1\+23_1\+25_1,\quad \-17_1\-18_1\+19_1\-20_1\+27_1,\quad 17_2\+18_2\-75_1\-76_1\-78_1,\quad 17_3\+23_3\-78_2\-103_2,\nn\\
&&\-17_3\-20_3\+78_2\+87_2,\quad 18_3\+25_3\-64_3\-66_3\-75_3,\quad \-18_3\+27_3\+75_3\+88_3\-90_3,\quad \-19_1\+20_1\-23_1\+24_1\+28_1,\nn\\
&&19_2\-20_2\+80_1\-81_1\+87_1,\quad 19_3\+27_2\+80_3\+88_2,\quad \-19_3\+28_2\-80_3\+89_2,\quad 20_1\+21_1\-22_1\-23_1\+26_1,\nn\\
&&20_2\+21_2\-84_1\-85_1\-87_1,\quad 20_3\-23_3\-87_2\+103_2,\quad 21_3\+26_3\-65_3\-68_3\-84_3,\quad \-21_3\+29_3\+84_3\+94_3\-96_3,\nn\\
&&22_2\+23_2\-61_2\-63_2\-103_1,\quad 22_3\+25_2\-61_3\-64_2,\quad \-22_3\+26_2\+61_3\-65_2,
\quad \-23_2\+24_2\+103_1\+104_1\-105_1,\nn\\
&&24_3\+28_3\+89_3\+92_3\+104_3,\ 
\-24_3\+30_3\+95_3\+98_3\-104_3,\ 31_1\-32_1\-35_1\-36_1\+44_1,\quad  \-31_1\+32_1\+38_1\-39_1\+45_1,\nn\\
&& 31_2\-32_2\+62_1\-63_1\+69_1,\quad 31_3\+44_2\+62_3\+100_2,\quad\-31_3\+45_2\-62_3\+101_2,\quad
 32_1\+33_1\+37_1\+38_1\+40_1,\nn\\
&& \-32_1\-33_1\+34_1\-35_1\+42_1,\quad 32_2\+33_2\-66_1\-67_1\-69_1,\quad 32_3\+38_3\-69_2\-97_2,\quad \-32_3\-35_3\+69_2\+85_2\+33_3,\nn\\
&&40_3\-66_3\-73_3\-75_3,\quad \-33_3\+42_3\+66_3\-88_3\+90_3,\quad \-34_1\+35_1\-38_1\+39_1\+43_1,\quad 34_2\-35_2\-80_1\+81_1\+85_1,\nn\\
&& 34_3\+42_2\-80_3\-88_2,\quad \-34_3\+43_2\+80_3\-89_2,\quad 35_1\+36_1\-37_1\-38_1\+41_1,\quad 35_2\+36_2\-85_1\-86_1\-87_1,\nn\\
&& 35_3\-38_3\-85_2\+97_2,\quad 36_3\+41_3\-74_3\-77_3\-86_3,\quad \-36_3\+44_3\+86_3\+100_3\-102_3,\quad 37_2\+38_2\-70_2\+72_2\-97_1,\nn\\
&& 37_3\+40_2\-70_3\-73_2,\quad \-37_3\+41_2\+70_3\-74_2,\quad \-38_2\+39_2\+97_1\-98_1\+99_1,\quad 39_3\+43_3\-89_3\-92_3\-98_3,
\nn\ee

\be
&& \-39_3\+45_3\+98_3\+101_3\-104_3,\  46_1\-47_1\-50_1\-51_1\+59_1,\
\-46_1\+47_1\+53_1\-54_1\+60_1,\ 46_2\-47_2\-62_1\+63_1\+67_1,\nn\\
&& 46_3\+59_2\-62_3\-100_2,\quad \-46_3\+60_2\+62_3\-101_2,\quad 
47_1\+48_1\+52_1\+53_1\+55_1,\quad \-47_1\-48_1\+49_1\-50_1\+57_1,\nn\\
&& 47_2\+48_2\-67_1\-68_1\-69_1,\quad 47_3\+53_3\-67_2\-91_2,\quad 
\-47_3\-50_3\+67_2\+76_2,\quad 48_3\+55_3\-68_3\-82_3\-84_3,\nn\\
&& \-48_3\+57_3\+68_3\-94_3\+96_3,\quad\-49_1\+50_1\-53_1\+54_1\+58_1,\quad 
49_2\-50_2\-71_1\+72_1\+76_1,\quad 49_3\+57_2\-71_3\-94_2,\nn\\
&& \-49_3\+58_2\+71_3\-95_2,\quad 50_1\+51_1\-52_1\-53_1\+56_1,\quad
50_2\+51_2\-76_1\-77_1\-78_1,\quad 50_3\-53_3\-76_2\+91_2,\nn\\
&& 51_3\+56_3\-77_3\-83_3\-86_3,\quad \-51_3\+59_3\+77_3\-100_3\+102_3,\quad 
52_2\+53_2\-79_2\-81_2\-91_1,\quad 52_3\+55_2\-79_3\-82_2,\nn\\
&&\-52_3\+56_2\+79_3\-83_2,\quad \-53_2\+54_2\+91_1\-92_1\+93_1,\quad 
54_3\+58_3\-92_3\-95_3\-98_3,\quad \-54_3\+60_3\+92_3\-101_3\+104_3,\nn\\
&& 61_1\+63_1\+64_1\+66_1\+67_1,\-61_1\-63_1\+65_1\+68_1\+69_1,
62_2\-63_2\+100_1\-102_1\-103_1,\-62_2\+63_2\+101_1\-104_1\+105_1,\nn\\
&&63_3\+67_3\+91_3\+93_3\+105_3,\-63_3\+69_3\+97_3\+99_3\-105_3, \-63_3\-67_3\-76_3\-78_3\-103_3, 63_3\-69_3\-85_3\-87_3\+103_3,\nn\\
&& 66_2\+67_2\+90_2\+91_2,\quad \-66_2\-67_2\-75_2\-76_2,\quad 
\-67_3\-69_3\-72_3\-76_3\-97_3,\quad 67_3\+69_3\+81_3\+85_3\+91_3,\nn\\
&& 68_2\+69_2\+96_2\+97_2,\quad \-68_2\-69_2\-84_2\-85_2,\quad
70_1\+72_1\+73_1\+75_1\+76_1,\quad \-70_1\-72_1\+74_1\+77_1\+78_1,\nn\\
&& 71_2\-72_2\+94_1\-96_1\-97_1, \-71_2\+72_2\+95_1\+98_1\-99_1,
72_3\+76_3\-91_3\-93_3\-99_3, \-72_3\+78_3\+99_3\+103_3\-105_3,\nn\\
&& \-72_3\+78_3\+85_3\+87_3\-97_3,\quad 75_2\+76_2\-90_2\-91_2,\quad 
76_3\+78_3\-81_3\+87_3\-91_3,\quad 77_2\+78_2\+102_2\+103_2,\nn\\
&& \-77_2\-78_2\-86_2\-87_2,\quad 79_1\+81_1\+82_1\+84_1\+85_1,\quad
\-79_1\-81_1\+83_1\+86_1\+87_1,\quad 80_2\-81_2\+88_1\-90_1\-91_1,\nn\\
&& \-80_2\+81_2\+89_1\+92_1\-93_1,\ 81_3\+85_3\-93_3\-97_3\-99_3,\
\-81_3\+87_3\+93_3\-103_3\+105_3,\ 84_2\+85_2\-96_2\-97_2,\nn\\
&& 86_2\+87_2\-102_2\-103_2,\quad 92_2\-93_2\+104_2\-105_2,\quad
\-92_2\+93_2\-98_2\+99_2,\quad 98_2\-99_2\-104_2\+105_2.\labels{C1}
\ee}

\subsection{$\boldsymbol{\d\D_l=\bD_l}$ equations}
There are three groups of equations, $X, Y, Z$. Group $X$ equations consist of setting the $l$th member
of the following list to $g_l$, where $g_l$ is the $ss$ coefficient of $\bD_l$ in \eq{d1tod105}. For example,
\eq{dbd1a1} is given by the first member of $X$, and \eq{bdb4a1} is given by the fourth member of $X$.
A member of $X$ has three terms if $\bD_l$ belongs to $I_l$, and has four terms of $\bD_l$ belongs to $J_l$.
Altogether $X$ has 105 members.

{\small\be &&X=\{\nn\\
&&\-1_3,\- 2_3\+ 3_3,\quad  \-4_3\-5_3\+6_3,\quad
 7_ 3\- 8_ 3\- 9_ 3,\quad  \-7_ 2\+10_3\-11_ 3\-7_ 3, \quad  \-4_ 2\+12_ 3\-13_ 3\-4_ 3,\nn\\
&& \-1_ 2\+14_ 3\-15_ 3\-1_ 3,\quad 2_ 2\-5_ 2\-8_ 2\+2_ 3,\quad \- 3_ 2\+10_ 2\-12_ 2,\quad
 \-6_ 2\+11_ 2\-14_ 2,\quad  \-9_ 2\+13_ 2\-15_ 2,\nn\\
&&\-16_ 3\-17_ 
   3\+18_ 3,\quad\-19_ 3\-20_ 3\+21_ 3,\quad 22_ 
   3\-23_ 3\-24_ 3\-22_ 2,\quad 25_ 3\-26_ 
   3\-22_ 3,\quad \-19_ 2\+27_ 3\-28_ 3\-19_ 
   3,\nn\\
&&\-16_ 2\+29_ 3\-30_ 3\-16_ 3,\quad 17_
    2\-20_ 2\-23_ 2\+17_ 3,\quad \-18_ 2\+25_ 
   2\-27_ 2,\quad \-21_ 2\+26_ 2\-29_ 2,\quad \-24_ 
   2\+28_ 2\-30_ 2,\nn\\
&&\-31_ 3\-32_ 3\+33_ 
   3,\quad \-34_ 3\-35_ 3\+36_ 3,\quad 37_ 3\-38_ 
   3\-39_ 3,\quad \-37_ 2\+40_ 3\-41_ 3\-37_ 
   3, \quad \-34_ 2\+42_ 3\-43_ 3\-34_ 3, \nn\\
&&\-31_
    2\+44_ 3\-45_ 3\-31_ 3,\quad 32_ 2\-35_ 
   2\-38_ 2\+32_ 3,\quad \-33_ 2\+40_ 2\-42_ 
   2,\quad \-36_ 2\+41_ 2\-44_ 2,\quad \-39_2\+43_
    2\-45_ 2,\nn\\
&& \-46_ 3\-47_ 3\+48_ 3,\quad \-49_
    3\-50_ 3\+51_ 3,\quad 52_ 3\-53_ 3\-54_ 
   3,\quad \-52_ 2\+55_ 3\-56_ 3\-52_ 3,\quad \-49_
    2\+57_ 3\-58_ 3\-49_ 3,\nn\\ 
&& \-46_ 2\+59_ 
   3\-60_ 3\-46_ 3,\quad 47_ 2\-50_ 2\-53_ 
   2\+47_ 3,\quad \-48_ 2\+55_ 2\-57_ 2,\quad \-51_ 
   2\+56_ 2\-59_ 2,\quad \-54_ 2\+58_ 2\-60_ 
   2,\nn\\
&&61_ 3\-62_ 3\-63_ 3,\  \-61_ 2\+64_
    3\-65_ 3\-61_ 3,\  31_ 1\-46_ 1\-62_ 
   2\+31_ 3,\  \-47_ 1\+66_ 3\-67_ 3\-47_ 
   3, \  \-32_ 1\+68_ 3\-69_ 3\-32_ 3,\nn\\
&&\-33_
    1\+64_ 2\-66_ 2,\quad \-48_ 1\+65_ 2\-68_ 
   2,\quad \-63_ 2\+67_ 2\-69_ 2,\quad 70_ 3\-71_ 
   3\-72_ 3,\quad \-70_ 2\+ 73_ 3\-74_ 3\-70_ 
   3,\nn\\
&&16_ 1\-49_ 1\-71_ 2\+16_ 3,\quad \-50_ 
   1\+75_ 3\-76_ 3\-50_ 3,\quad \-17_ 1\+77_ 
   3\-78_ 3\-17_ 3,\quad \-18_ 1\+73_ 2\-75_ 
   2,\quad \-51_ 1\+74_ 2\-77_ 2,\nn\\  
&&\-72_ 2\+76_
    2\-78_ 2,\quad 79_ 3\-80_ 3\-81_ 3,\quad\-79_
    2\+82_ 3\-83_ 3\-79_ 3,\quad 19_ 1\-34_ 
   1\-80_ 2\-19_ 3,\quad \-35_ 1\+84_ 3\-85_ 
   3\-35_ 3,\nn\\
&&\-20_ 1\+86_ 3\-87_ 3\-20_ 
   3,\quad \-21_ 1\+82_ 2\-84_ 2,\quad \-36_ 1\+83_
    2\-86_ 2,\quad \-81_ 2\+85_ 2\-87_ 2, \quad 1_ 
   1\-52_ 1\-79_ 1\+1_ 3,\nn\\
&&\-80_ 1\+88_ 3\-89_ 3\-80_ 3,\quad \-53_ 1\+90_ 3\-91_ 
   3\-53_ 3,\quad 2_ 1\-92_ 3\-93_ 3\+2_ 3, 
 \quad 3_ 1\-88_ 2\-90_ 2, \quad \-54_ 1\+89_ 2\-92_ 2,\nn\\
&&\-81_ 1\+91_ 2\-93_ 2, \quad 4_ 1\-37_ 1\-70_ 1\+4_ 3, \quad \-71_ 1\+94_ 3\-95_ 3\-71_ 3, \quad 
\-38_ 1\+96_ 3\-97_ 
   3\-38_ 3,\quad 5_ 1\-98_ 3\-99_ 3\+5_ 3,\nn\\
&&6_ 1\-94_ 2\-96_ 2,\quad \-39_ 1\+95_ 2\-98_ 2,\quad \-72_ 1\+97_ 2\-99_ 2, \quad 7_ 1\-22_ 1\-61_ 1\+7_ 3,\quad \-62_ 1\+100_ 3\-101_ 3\-62_ 3,\nn\\
&& \-23_ 1\+102_ 3\-103_ 
   3\-23_ 3, \quad 8_ 1\-104_ 3\-105_ 3\+8_ 
   3, \quad 9_ 1\-100_ 2\-102_ 2, \quad 24_ 1\-101_
    2\-104_ 2, \quad 63_ 1\-103_ 2\-105_ 
   2,\nn\\
&&10_ 1\-25_ 1\-64_ 1\+10_ 2,\  11_ 
   1\-26_ 1\- 65_ 1\+11_ 2, \  12_ 1\-40_ 
   1\-73_ 1\+12_ 2, \  13_ 1\-41_ 1\-74_ 
   1\+13_ 2, \  14_ 1\-55_ 1\-82_ 1\+14_ 2,\nn\\
&&15_ 1\-56_ 1\-83_ 1\+15_ 2, \ 27_ 1\-42_ 1\-88_ 1\+27_ 2, \  28_ 1\-43_ 
   1\-89_ 1\+28_ 2, \  29_ 1\-57_ 1\-94_ 
   1\+29_ 2, \  30_ 1\-58_ 1\-95_ 1\+30_ 2,\nn\\
 &&44_ 1\-59_ 1\-100_ 1\+44_ 2, \ 45_ 1\-60_ 1\-101_ 1\+45_ 2, \ 66_ 1\-75_ 
   1\-90_ 1\+66_ 2, \ 67_ 1\-76_ 1\-91_ 
   1\+67_ 2, \ 68_ 1\-84_ 1\-96_ 1\+68_ 2, \nn\\
&& 69_ 1\-85_ 1\-97_ 1\+69_ 2, \ 77_ 1\-86_ 1\-102_ 1\+77_ 2, \ 78_ 1\-87_ 
   1\-103_ 1\+78_ 2, \ 92_ 1\-98_ 1\-104_ 
   1\+92_ 2, \ 93_ 1\-99_ 1\-105_ 1\+93_ 
   2\nn\\
&& \}\labels{dddl}\ee}

Group $Y$ consists of 60 pairs of equations, namely 120 equations. They are listed below. The three quantities
within each pair of parentheses are equal. These equations occur only for $\bD_l$ in class $J_l$. For example,
\eq{bdb4a3} is given by the  first member of $Y$.  
{\small\be
Y=\{&&(\-7_ 3,10_ 2,\-11_ 2), (\-4_ 3,12_ 2,\-13_  2), (\-1_ 3,14_ 2,\-15_ 2), (2_ 3, \-5_  3, \8_ 3), (\-2_ 3,25_ 2, \-26_ 2),\nn\\ 
&&(\-19_ 3,27_ 2, \-28_ 2), (\-16_ 3, 29_ 2,\-30_ 2), (17_ 3, \-20_ 3, \-23_ 3), (\-37_ 3, 40_ 2, \-41_ 2), 
(\-34_ 3,42_ 2, \-43_ 2), \nn\\
&&(\-31_3, 44_ 2, \-45_ 2), (32_ 3,\-35_ 3,\-38_ 3), (\-52_ 3, 55_ 2, \-56_ 2), 
(\-49_ 3,57_2, \-58_ 2), (\-46_ 3, 59_ 2, \-60_ 2),\nn\\
&& (47_3, \-50_ 3, \-53_ 31), (-61_ 3, 64_ 2, \-65_ 2), 
(31_ 3,\-46_ 3,\-62, 3), (-47_ 3, 66_2,\-67_2), (-32_3, 68_ 2, \-69_ 2),\nn\\
&& (\-70_3, 73_ 2, \-74_ 2), (16_ 3, \-49_ 3, \-71_  3), (\-50_ 3, 75_ 2,\-76_ 2), (-17_ 3, 77_ 2, \-78_ 2), (-79_ 3, 82_ 2, \-83_ 2),\nn\\
&&(19_ 3, \-34_ 3, \-80_ 3), (\-35_ 3, 84_ 2, \-85_2), (-20_ 3, 86_ 2, \-87_ 2), 
(1_ 3, \-52_ 3,\-79_ 3), (-80_ 3, 88_ 2, \-89_ 2),\nn\\
&& (\-53_3, 90_ 2, \-91_ 2), (2_ 3, \-92_ 2, \-93_ 2), (4_ 3,\-37_ 3,\-70_ 31), (\-71_ 3, 94_ 2,\-95_ 2), (-38_ 3, 96_ 2, \-97_ 2),\nn\\
&& (5_ 3, \-98_ 2, \-99_ 2), (7_ 3, \-22_ 3, \-61_ 3), 
(\-62_ 3, 100_ 2, \-101_ 2), (-23_ 3, 102_ 2, \-103_ 2), (8_ 3, \-104_ 2, \-105_2),\nn\\ 
&&(10_ 2, \-25_ 2, \-64_ 2), (11_ 2, \-26_ 
   2,\-65_ 2), (12_ 2, \-40_ 2, \-73_ 2), ( 13_
    2, \-41_ 2, \-74_ 2), ( 14_ 2, \-55_ 2, \-82_ 
   2),\nn\\
   && ( 15_ 2,\-56_ 2, \-83_ 2), ( 27_ 2,\-42_ 
   2, \-88_ 2), ( 28_ 2,\-43_ 2, \-89_ 2), ( 29_
    2, \-57_ 2, \-94_ 2), ( 30_ 2, \-58_ 2, \-95_ 
   2), \nn\\
   &&( 44_ 2, \-59_ 2, \-100_ 2), ( 45_ 2, \-60_
    2, \-101_ 2), ( 66_ 2, \-75_ 2, \-90_ 
   2), ( 67_ 2, \-76_ 2, \-91_ 2), ( 68_ 2, \-84_ 
   2, \-96_ 2),\nn\\
   &&( 69_ 2, \-85_ 2,\-97_ 2), ( 77_
    2, \-86_ 2, \-102_ 2), ( 78_ 2, \-87_ 2, \-103_
    2), ( 92_ 2, \-98_ 2, \-104_ 2), ( 93_ 2, 
   \-99_ 2, \-105_ 2)\ \}.\nn\\
\labels{Yeq}\ee}   

Group $Z$ consists of the following 225 $q$ parameters which are zero.
{\small\be
Z=\{&&
1_ 2, 1_ 1, 2_ 2, 2_ 1, 3_ 3, 3_ 2, 3_ 1, 4_ 2, 4_ 1, 5_ 2, 5_ 1, 6_ 3, 6_ 2, 6_ 1, 7_ 2,
 7_ 1, 8_ 2, 8_ 1, 9_ 3, 9_ 2,\nn\\
 && 9_ 1, 10_ 3, 10_ 1, 11_ 3, 11_ 1, 
12_ 3, 12_ 1, 13_ 3, 13_ 1, 14_ 3, 14_ 1, 15_ 3, 15_ 1, 16_ 2, 16_ 1, 17_ 2, 17_ 1, 
18_ 3, 18_ 2, 18_ 1, \nn\\
&&19_ 2, 19_ 1, 20_ 2, 20_ 1, 21_ 3, 21_ 2, 21_ 1, 22_ 
2, 22_ 1, 23_ 2, 23_ 1, 24_ 3, 24_ 2, 24_ 1, 25_ 3, 
25_ 1, 26_ 3, 26_ 1, 27_ 3,27_ 1, \nn\\
&& 28_ 3, 28_ 1, 29_ 
3, 29_ 1, 30_ 3, 30_ 1, 31_ 2, 31_ 1, 32_ 2, 32_ 1, 
33_ 3, 33_ 2, 33_ 1, 34_ 2, 34_ 1, 35_ 2, 35_ 1, 36_ 
3, 36_ 2, 36_ 1,\nn\\
&& 37_ 2, 37_ 1, 38_ 2, 38_ 1, 39_ 3, 
39_ 2, 39_ 1, 40_ 3, 40_ 1, 41_ 3, 41_ 1, 42_ 3, 42_ 
1, 43_ 3, 43_ 1, 44_ 3, 44_ 1, 45_ 3, 45_ 1, 46_ 2,\nn\\
&& 
46_ 1, 47_ 2, 47_ 1, 48_ 3, 48_ 2, 48_ 1, 49_ 2, 49_ 
1, 50_ 2, 50_ 1, 51_ 3, 51_ 2, 51_ 1, 52_ 2, 52_ 1, 
53_ 2, 53_ 1, 54_ 3, 54_ 2, 54_ 1,\nn\\
&& 55_ 3, 55_ 1, 56_ 
3, 56_ 1, 57_ 3, 57_ 1, 58_ 3, 58_ 1, 59_ 3, 59_ 1, 
60_ 3, 60_ 1, 61_ 2, 61_ 1, 62_ 2, 62_ 1, 63_ 3, 63_ 
2, 63_ 1, 64_ 3,\nn\\
&&64_ 1, 65_ 3, 65_ 1, 66_ 3, 66_ 1, 
67_ 3, 67_ 1, 68_ 3, 68_ 1, 69_ 3, 69_ 1, 70_ 2, 70_ 
1, 71_ 2, 71_ 1, 72_ 3, 72_ 2, 72_ 1, 73_ 3, 73_ 1,\nn\\
&& 
74_ 3, 74_ 1, 75_ 3, 75_ 1, 76_ 3, 76_ 1, 77_ 3, 77_ 
1, 78_ 3, 78_ 1, 79_ 2, 79_ 1, 80_ 2, 80_ 1, 81_ 3, 
81_ 2, 81_ 1, 82_ 3, 82_ 1, 83_ 3, \nn\\
&&83_ 1, 84_ 3, 84_ 
1, 85_ 3, 85_ 1, 86_ 3, 86_ 1, 87_ 3, 87_ 1, 88_ 3, 
88_ 1, 89_ 3, 89_ 1, 90_ 3, 90_ 1, 91_ 3, 91_ 1, 92_ 
3, 92_ 1, 93_ 3,\nn\ee}
{\small
\be
&& 93_ 1, 94_ 3, 94_ 1, 95_ 3, 95_ 1, 
96_ 3, 96_ 1, 97_ 3, 97_ 1, 98_ 3, 98_ 1, 99_ 3, 99_ 
1, 100_ 3, 100_ 1, 101_ 3, 101_ 1, 102_ 3, 102_ 1, 103_ 
3, \nn\\
&&103_ 1, 104_ 3, 104_ 1, 105_ 3, 105_ 1\}.\labels{Zeq}\ee}
For example, \eq{dbd1a2} corresponds to members $1_1, 1_2, 2_1, 2_2, 3_1, 3_2$ of $Z$, and \eq{bdb4a2} corresponds to members $7_1, 10_1, 11_1$ of $Z$.

\subsection{Solutions}
The $3\x105=315$ $q$ parameters must satisfy the 144 gauge constraint equations \eq{C1}, 
the 105 equations of group $X$, the 120 equations of group $Y$, and the 225 equations of group $Z$.
With the equations numbers so much larger than the number of $q$ parameters, there would be no solutions
unless these equations are highly degenerate.

To show that there is really no solution, one can proceed as follows. By substituting $Z$ into $Y$, more $q$ parameters must vanish. Substituting these vanishing parameters into $X$, one gets 105 equations relating
$q$'s on the left hand side to  $g_l$ on the right hand side.  Since $g_l=0$ for the 60 $l$'s with $l\in J_l$,
the combination of $q$ on the left must be zero as well. It turns out that for those $l$, there is always a single
$q$ term left on the left, so in this way one gets even more zero $q$ parameters. With all these vanishing
$q$ parameters substituted into  the left hand side of the remaining 45 $X$ equations for $l\in I_l$, where
$g_l\not=0$, there can be no solutions if any of the left hand side of the 45 equations is identically zero,
namely, has no non-vanishing $q$ parameters left. This turns out to be indeed the case for 
$l=7, 8, 13, 14, 19, 20, 37, 40, 43, 44, 45$. Thus even without ever using the 144 gauge constraint equations \eq{C1}, one sees that the $X,Y,Z$ solutions have no solutions as long as $g_l\not=0$ for $l\in J_l$.

\end{document}